\documentclass[12pt,a4paper]{article}
\usepackage{epsfig}
\usepackage{verbatim}
\usepackage{graphicx}
\usepackage[utf8]{inputenc} 
\usepackage{url}

\newcommand{\code}{\texttt}
\newcommand{\Rin}[1]{\code{>\ {#1}}}
\newcommand{\Rout}[1]{{\,\code{[1]\ {#1}}}}

\begin{document}

\newcommand{\bm}[1]{\mbox{\boldmath$#1$}}

\def\mvec#1{{\bm{#1}}}   

\title{Bertrand `paradox' reloaded\\
(with details on transformations of variables, \\
an introduction to Monte Carlo simulation\\
and an inferential variation of the problem)\footnote{Note 
based on lectures 
to PhD students in Rome. The Android app  mentioned in the text is available at {\tt http://www.roma1.infn.it/\~\,dagos/prob+stat.html\#bpr}.}
}

\author{G.~D'Agostini \\
Universit\`a ``La Sapienza'' and INFN, Roma, Italia \\
{\small (giulio.dagostini@roma1.infn.it,
 \url{http://www.roma1.infn.it/~dagos})}
}

\date{}

\maketitle

\begin{abstract}
This note is mainly to point out, if needed, that
uncertainty about models and their parameters
has little to do with a `paradox'. The proposed 
`solution' is to formulate practical questions instead
of seeking refuge into abstract principles. 
(And, in order to be concrete, some details on how to calculate
the probability density functions of the chord lengths are 
provided, together with some comments on simulations
and an appendix on the inferential aspects of the problem.)
\end{abstract}
{\small
\begin{flushright}
{\sl ``On trace {\it au hasard} une corde dans un cercle.} \\
{\sl Quelle est la probabilité pour qu'elle soit plus petite }\\
{\sl que le c\^oté du triangle équilatéral inscrit?} \\
\ldots  \\
{\sl Entre ces trois réponses, quelle est la véritable?}\\
{\sl Aucune des trois n'est fausse, aucune n'est exacte,} \\
{\sl la question est mal posée.''}\\
{(Joseph Bertrand)}\\
\mbox{}\\
{\sl ``Probability is either referred to real cases} \\
{\sl or it is nothing'' } \\
{(Bruno de Finetti)}\\
\mbox{}\\
{\sl ``As far as the laws of mathematics refer to reality, } \\
{\sl they are not certain,  } \\
{\sl and as far as they are certain,  } \\
t{\sl hey do not refer to reality.'' } \\
{(Albert Einstein)}
\end{flushright}
}

\thispagestyle{empty}

\newpage
\section{Introduction}
The question asked by Joseph Bertrand in his 1889 book
{\em Calcul des probabilités}~\cite{Bertrand-CDP}
is about the probability that a chord drawn `at random'
is smaller that the side of the equilateral 
triangle inscribed in the same
circle (see e.g. \cite{Wiki-Bertrand_paradox}). Obviously 
the question can be restated asking about the probability that 
the chord  will be larger or smaller than the radius, or whatever segment 
you like, upper to the diameter (for which the solution
is trivially 100\%). The reason of the original choice 
of the side of such a triangle is that
the calculation is particularly easy, 
{\em under the hypotheses} Bertrand originally considered, as we shall see.

The question can be restated in  more general terms, i.e. that  
of finding the probability distribution of the length $l$ of a chord.
Indeed, as well known, our uncertainty about the value
a continuous variable can assume can be described by a {\em probability
density function}, hereafter `pdf', $f(l)$. This
should be written, more precisely, 
as $f(l\,|\,I_s(t))$, where $I_s(t)$ is the {\em Information} available
to the {\em subject} $s$ at the {\em time} $t$. In fact, 
      \begin{quote}
      ``{\sl
      Since the knowledge may be different with different persons
      or with the same person at different times, they may anticipate
      the same event with more or less confidence, and thus different numerical
      probabilities may be attached to the same event}.''\,\cite{Schrodinger}
      \end{quote}
And, hence,  probability is always conditional 
      probability, as again well stated by Schr\"odinger\,\cite{Schrodinger},
      \begin{quote}
      ``{\sl
      Thus whenever we speak loosely of `the probability of an event,'
      it is always to be understood: probability with regard to a certain 
      given state of knowledge.''
      }
      \end{quote}
These quotes, which are 100\% in tune with common sense, 
definitely rule out the use of the appellative of `paradox' 
for the problem of the chords. 
In other words, Bertrand's `paradox' belongs
to a completely different class than e.g. Bertrand Russell's 
{\em barber paradox}. 
Absurd is instead the positions of those who maintain that
the problem should have a unique solution once it is 
{\em ``well posed''}`\cite{Jaynes_Bertrand}, or that 
they have found the {\em ``conclusive answer''}~\cite{DiPorto}.

In my point of view the question proposed by Bertrand can be 
only answered if framed in a given contest and `asked' somehow,
either to human beings, or -- and hence the quote marks -- 
to Nature by performing suitable experiments 
(but making a particular simulation, of the kind of that proposed
in Ref.~\cite{DiPorto}, is the same as asking human beings
-- and even making an experiment the result will depend on the 
experimental setup!). 

For example we can ask suddenly  students, 
without any apparent reason (for them), 
to draw a chord in a circle printed on a sheet of paper. 
And to give more sense (and fun) to the `experiment' 
we can make a bet among us on the resulting length
(the bet could be even more detailed, concerning 
for example the orientation of the chord -- for example
it never happened to me that
a student drawn a vertical one, but perhaps Japanese students might
have higher tendency to draw segments top down!).
Or we can ask students to write, with the their preferred 
programming language and plotting package, 
a `random chord generator'. In this
case our bet will be about the length 
that will result from a certain
extraction, e.g. the first, or the 100th (if we were not
informed about all or some of the previous 99 results, because 
this information might change
our odds about the outcome of the following ones,
as we shall see in Appendix C). 

Indeed I have done experiments of this kind since several years,
and I have formed my opinion on how students react, 
depending on the class they attend
and how they are skilled in mathematics, and \ldots even in games
(yes! in the answer there is even a flavor of game theory, because
smart students unavoidable try to guess the reason of the question
and try to surprise you!) 
For example unsophisticated 
students draw `typical' chords, of the kind you can get
searching for the keywords ``{\tt chord geometry}'' 
in Google Images.\footnote{
\url{https://www.google.it/search?q=chord+geometry&source=lnms&tbm=isch}}
$[$Hence, for example,  a possible \underline{real} 
`well posed' question could be the following:
``what is the length of the chord (in units of the radius) that will appear
in e.g. the 27-th (from left to right, top to bottom) 
image returned by the search engine?'']
When instead I propose the question to students of advanced
years I have quite some expectation that one or more of them
will draw a diameter (just a maximum chord) or even a tiny 
segment almost tangent to the circumference.

Essentially this is all what I have to say about this
so called `paradox'. The rest of the note has been written 
for didactic purposes in order to show how to evaluate
the probability distributions of interest and how to make 
the simulations. The paper is supplemented by three appendices.
Appendix A is a kind of extra exercise on transformation
of variables. Appendix B show a simple way to write
an `infinite number' of chord generators
(indeed a generator characterized by two {\em hyperparameters}). 
Appendix C is finally 
an inferential variation of Bertrand problem, in which 
we use `experimental data' to guess the generator used
and to predict the length of other chords
produced with the same generator.

\section{Basic (`classic') geometric solutions}
Let us now start going through the `classical' solutions
of the problem, i.e. those analyzed by Bertrand and which
typically appear in the literature and on the web.
The adjective {\em geometric} is to distinguish then from a more `physical'
one, based on a kind of realistic game that, as we shall see 
in section \ref{sec:mikado}. 
 
\subsection{Endpoints uniformly chosen on the circumference}
\begin{figure}
\begin{tabular}{ccc}
\epsfig{file=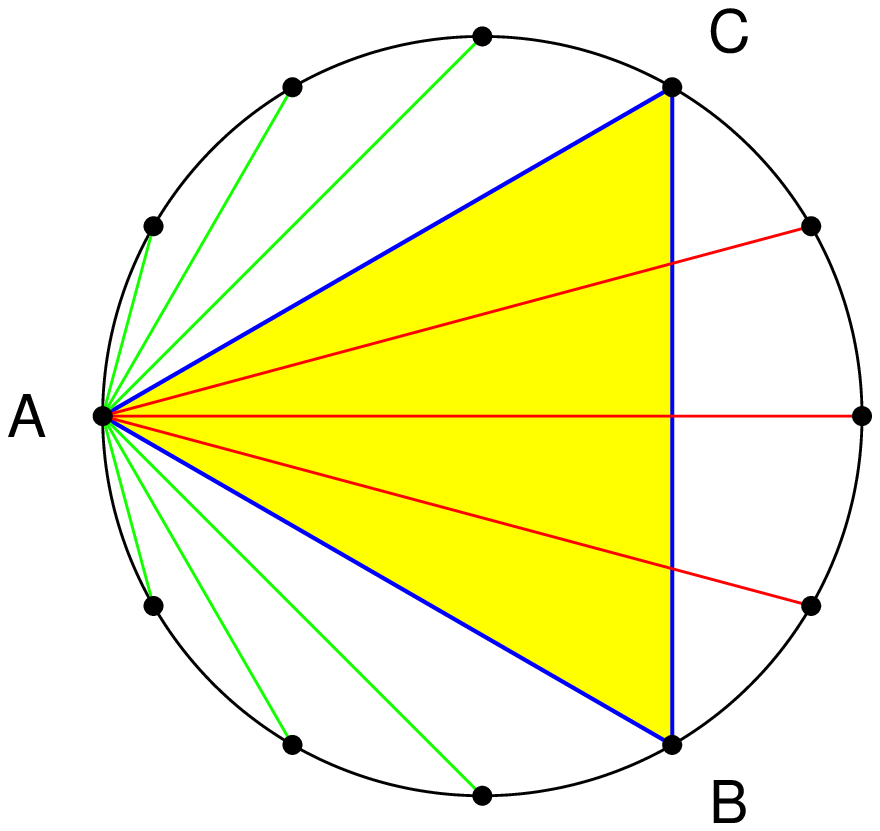,clip=,width=0.475\linewidth} & &
\epsfig{file=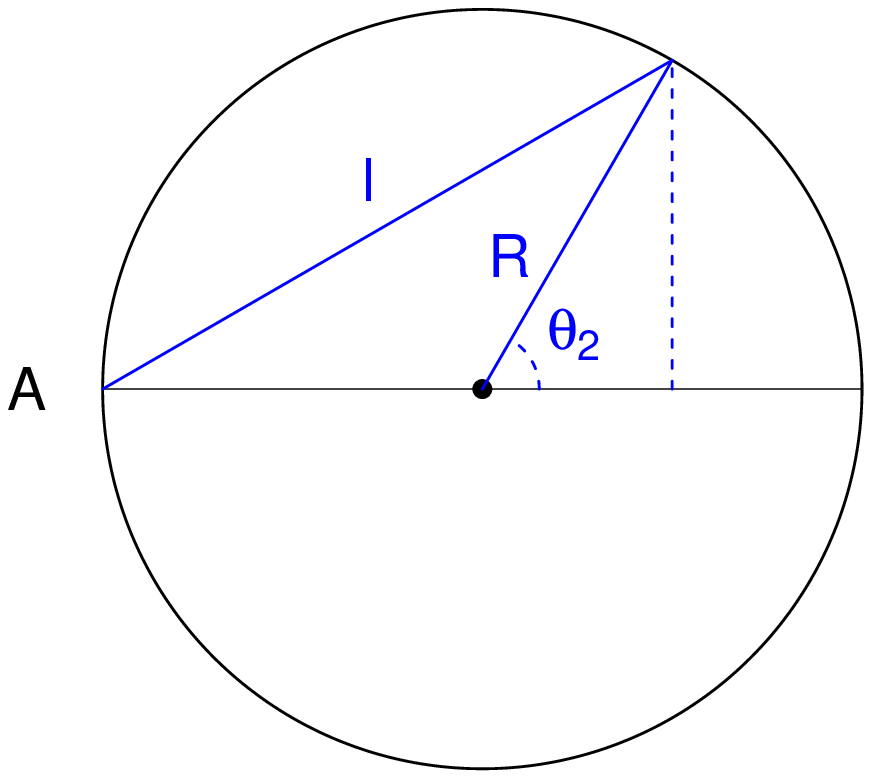,clip=,width=0.475\linewidth} \\
\end{tabular}
\caption{\small \sf Left: circle with inscribed equilateral triangle and
chords with one end in $A$ and the other distributed uniformly 
around the circumference. Right: geometric construction to 
show how to evaluate the length $l$ from $\theta_2$, with 
$0\le \theta_2\pi$ (see text for details).} 
\label{fig:bertrand_meth1}
\end{figure}

A way to draw `at random' a chord is to choose two points
on the circumference and to join them with a segment.
If we indicate the first point with $A$, corresponding to a vertex
of the equilateral triangle, as shown in 
Fig.~\ref{fig:bertrand_meth1},
the chords smaller that the side of the triangle are those 
with the other end either in the arc between $A$ and $B$
or in that between $A$ and $C$. The resulting probability
is thus simply {\bf 2/3}.

A more complete information about our beliefs that the
length falls in any given interval 
is provided by the pdf $f(l\,|\,{\cal M}_1)$, where
${\cal M}_1$ stands for `Model 1'. Since there is a one to one 
correspondence
between a point on the circle and the angle between the radius to that
point and the $x$-axis (according to usual trigonometry convention),
we can turn our extraction method into two angles, $\theta_1$ and $\theta_2$, 
uniformly distributed between 0 and $2 \pi$. If we are only interested
in the length of the chords and not in their position inside the 
circle we can fix  $\theta_1$ at $\pi$ and consider 
 $\theta_2$ in the interval between 0 and $\pi$. The corresponding 
chord will have a 
length of (see right plot in Fig.~\ref{fig:bertrand_meth1})
\begin{eqnarray}
l & = & \sqrt{(R+R \cos\theta_2)^2+R^2\sin^2\theta_2} \nonumber \\
  & = & R\,  \sqrt{(1+\cos\theta_2)^2+\sin^2\theta_2} \nonumber \\
& = & R\,  \sqrt{2+2 \cos\theta_2}\,,
\end{eqnarray}
or, more conveniently, the normalized length $\lambda$ will be
\begin{eqnarray}
\lambda = \frac{l}{R} & = & \sqrt{2+2 \cos\theta_2}\,.
\end{eqnarray}
The problem is thus how to calculate the pdf\,\footnote{Obviously, the pdf's 
$f(\theta_2\,|\,{\cal M}_1)$ and
$f(\lambda\,|\,{\cal M}_1)$ are usually expressed by 
different mathematical 
functions, a point very clear among physicists. Mathematics
oriented guys like to clarify it, thus writing e.g.
$f_{\Lambda}(\lambda\,|\,{\cal M}_1) $,  $f_{\Theta_2}(\theta_2\,|\,{\cal M}_1)$, and so on.
I will add a proper subscript only if it is not clear from the context
what is what.
}  
 $f(\lambda\,|\,{\cal M}_1)$ from 
\begin{eqnarray}
f(\theta_2\,|\,{\cal M}_1) &=& \frac{1}{\pi}\hspace{0.7cm} 
(0\le \theta_2 \le \pi)\,.
\end{eqnarray}
We shall use the general 
rule\,\footnote{An alternative way is to use 
the `text book' transformation rule, valid for a monotonic
function $y=g(x)$ that relates the generic variable $X$
to the variable $Y$:
\begin{eqnarray*}
f_Y(y) &=& f_X (g^{-1}(y))\left|\frac{d}{dy}g^{-1}(y)\right|\,,
\hspace{1.5cm}(A)
\end{eqnarray*} 
which can be derived in the following way for the general variables 
$X$ and $Y$ [capital letters indicates the variable, small letters the possible values
-- now it becomes important to make clear the different pdf's and we shall then
use the notation $f_X()$ and $f_Y()$]:
\begin{description}
\item[\fbox{$g'(x) \ge 0$}]
 If $g()$ is {\em non-decreasing} in the range of $X$ we have
$$F_Y(y) \equiv P(Y \le y) =  P(g(X) \le y) = P(X \le g^{-1}(y)) \equiv 
 F_X(g^{-1}(y))\,.$$
Making use of the rules of calculus we have then
$$f_Y(y) = \frac{d}{dy}\,F_Y(y) = \frac{d}{dy}\,F_X(g^{-1}(y)) =  f_X((g^{-1}(y))\cdot 
 \frac{d}{dy}\,g^{-1}(y) $$
\item[\fbox{$g'(x) \le 0$}] If, instead,  $g()$ is {\em non-increasing} we have
$$F_Y(y) \equiv P(Y \le y) =  P(g(X) \le y) = P(X \ge g^{-1}(y)) \equiv 
1-  F_X(g^{-1}(y))\,,$$
and then
$$f_Y(y) = \frac{d}{dy}\,F_Y(y) = - \frac{d}{dy}\,F_X(g^{-1}(y)) =  - f_X((g^{-1}(y))\cdot 
 \frac{d}{dy}\,g^{-1}(y)
 = f_X((g^{-1}(y)) \cdot \left[\!- \frac{d}{dy}\,g^{-1}(y)\!\right] $$
where the factorization in the last step is due to the fact 
that $f_X()$ cannot be negative, and for this reason the absolute
value in (A) is only on the second factor.
\end{description}
Equation (A) takes then into account the two possibilities and, 
let us stress once more,
it is valid for monotonic transformations. We shall use it, to double 
check our results
in footnotes \ref{fn:TrasformazioneTextbook_1},
\ref{fn:TrasformazioneTextbook_2} and \ref{fn:TrasformazioneTextbook_3}.
\label{fn:TrasformazioneTextbook}
}
\begin{eqnarray}
f(y) &=& \int_{-\infty}^{+\infty}
\!\delta\left(y - g(x)\right)\cdot f(x) \,dx\,,
\label{eq:GeneraRuleTransf_xy}
\end{eqnarray}
in which $Y$ is related to the $X$ by  $Y=g(X)$, with 
 $g()$ a generic function and $\delta()$ the 
Dirac delta.
 In our case we have then
\begin{eqnarray}
f(\lambda\,|\,{\cal M}_1) &=& \int_0^\pi
\!\delta\left(\lambda -  \sqrt{2+2 \cos\theta_2}\right)\cdot f(\theta_2\,|\,{\cal M}_1) 
\,d\theta_2\,.
\label{eq:GeneraRuleTransf}
\end{eqnarray}
Eq. (\ref{eq:GeneraRuleTransf})
has the very simple interpretation of `summing up' 
all infinitesimal probabilities `$f(\theta_2\,|\,{\cal M}_1) \,d\theta_2$'
that contribute to the `same value' of $\lambda$
(the quote marks are due to the fact that we are dealing
with continuous quantities and hence we have use the rules of 
calculus). This interpretation is very useful to estimate 
$f(\lambda\,|\,{\cal M}_1)$ by simulation: extract values 
of $\theta_2$ according to $f(\theta_2\,|\,{\cal M}_1)$; 
for each value of $\theta_2$ calculate the corresponding
$\lambda$; summarize the result by suitable 
statistical indicators and visualize it with an histogram.  

\begin{figure}
\begin{tabular}{c}
 {\large $[\,{\cal M}_1\,]$ }\\
\epsfig{file=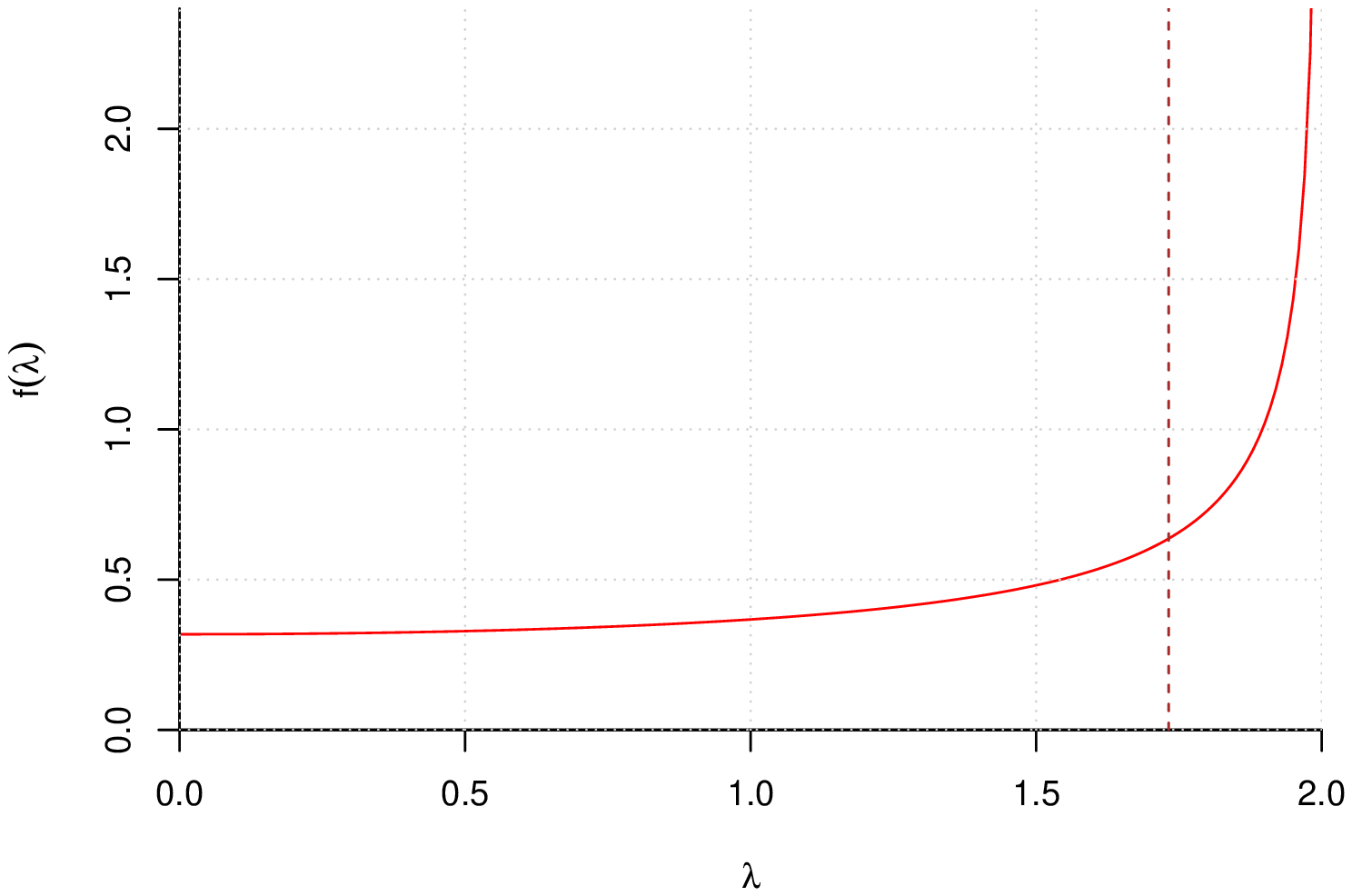,clip=,width=\linewidth} \\
\epsfig{file=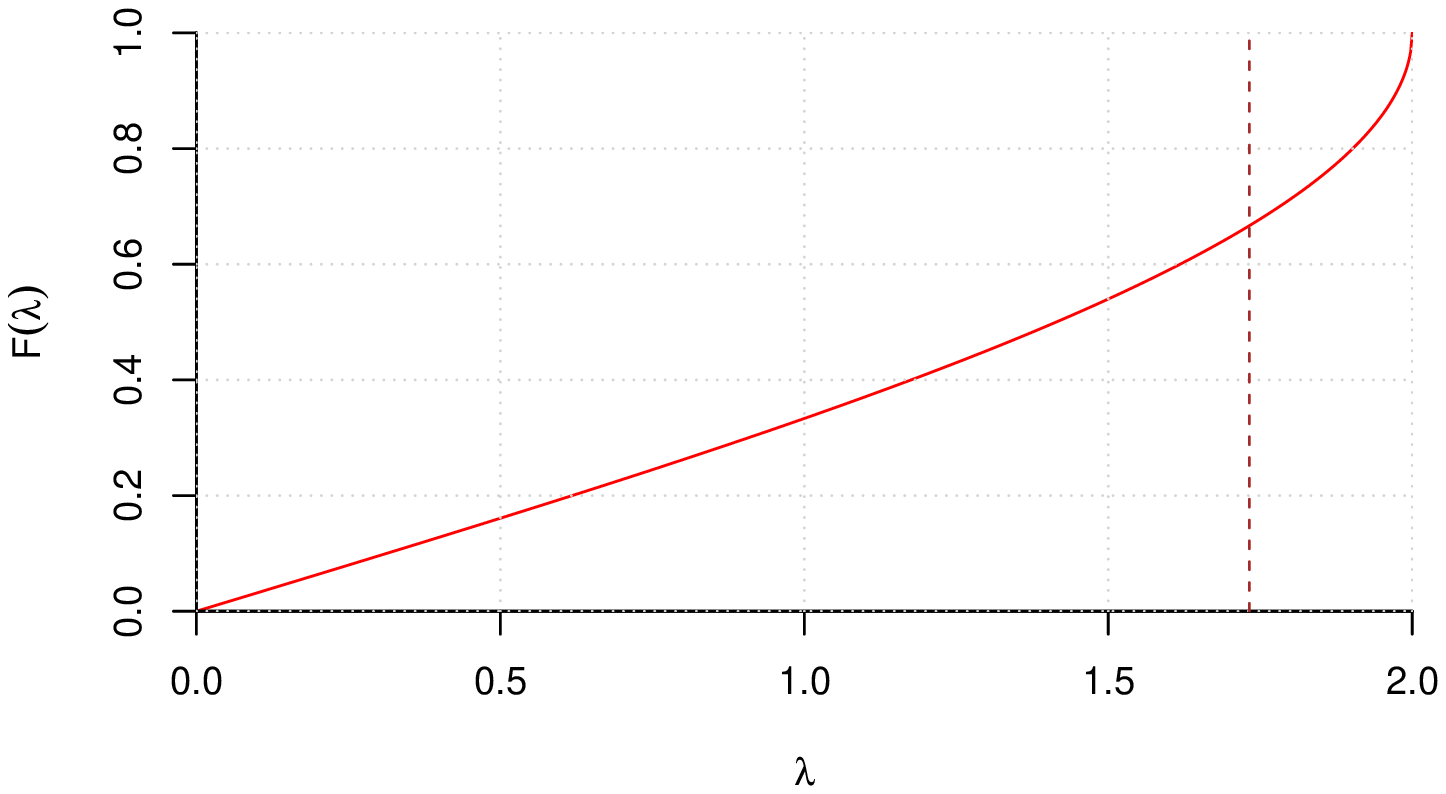,clip=,width=\linewidth}
\end{tabular}
\caption{\small \sf Probability distribution of $\lambda=l/R$ of the chords
generated with Method 1. The dashed vertical line indicates $\lambda=\sqrt{3}$.}
\label{fig:f_F_meth1}
\end{figure}

Making use of the properties of the Dirac delta and taking 
into account that $\lambda(\theta_2)$ decreases monotonically
we get\,\footnote{In the step from 
Eq.~(\ref{eq:IntegraleTrasformazione_0}) 
to  Eq.~(\ref{eq:IntegraleTrasformazione}) 
we are making use of the famous property (at least among physicists) 
of the Dirac delta
\begin{eqnarray*}
\delta(g(x)) &=& \sum_i \frac{\delta(x-x_i)}{|g'(x_i)|}\,,
\end{eqnarray*}
where $x_i$ are the real roots of $g(x)$.
In our case we have a single root, which we write as $x^*$, and hence 
we get
\begin{eqnarray*}
\delta(g(x)) &=& \frac{\delta(x-x^*)}{|g'(x^*)|}\,.
\end{eqnarray*}
If we apply it to the 
general transformation rule (\ref{eq:GeneraRuleTransf_xy}) we obtain
\begin{eqnarray*}
f_Y(y) &=& \int_{-\infty}^{\infty}\!\delta(y-g(x))\cdot f_X(x)\,dx \\
&=&  \int_{-\infty}^{\infty}\! \frac{\delta(x-x^*)}{|g'(x^*)|}  \cdot f_X(x)\,dx\\
&=& \frac{1}{|g'(x^*)|}\cdot f_X(x^*)\,, 
\end{eqnarray*}
in which we recognize Eq.~(A) of footnote \ref{fn:TrasformazioneTextbook},
if we note that $f_X(x^*)$ is $f_X(g^{-1}(y))$ and 
the derivative of $g(x)$ w.r.t. $x$ calculated in $x^*$ is 
the inverse of the derivative of the inverse function $g^{-1}$ 
(w.r.t. $y$!)
calculated in $y=g(x^*)$ [for the latter observation just think at
the Leibniz notation $dy/dx = 1/(dx/dy)$]. 

Anyway, in order to avoid confusion  
the denominator in Eq.~(\ref{eq:IntegraleTrasformazione}) has
been written in the most unambiguous way.
}
\begin{eqnarray}
f(\lambda\,|\,{\cal M}_1) &=& \int_0^\pi
\!\delta\left(\lambda -  \sqrt{2+2 \cos\theta_2}\right)\cdot \frac{1}{\pi} 
\,d\theta_2 \label{eq:IntegraleTrasformazione_0}\\
&=& \int_0^\pi \! \frac{\delta(\theta_2-\theta_{2}^*)}
  {\left|\left.\left(\frac{d}{d\theta_2}(\lambda -  
       \sqrt{2+2 \cos\theta_2})\right)\right|_{\theta_2=\theta_{2}^*}\right|}
\cdot \frac{1}{\pi} 
\,d\theta_2\,,\label{eq:IntegraleTrasformazione}
\end{eqnarray} 
where $\theta_{2}^*$ is the solution of the equation
\begin{eqnarray}
\lambda -  \sqrt{2+2 \cos\theta_2} &=& 0\,,
\end{eqnarray} 
that is
\begin{eqnarray}
\theta_{2}^* &=& \arccos\left(\frac{\lambda^2}{2}-1\right)\,.
\end{eqnarray} 
Being the derivative in Eq.~(\ref{eq:IntegraleTrasformazione})
\begin{eqnarray}
\left.\left(\frac{d}{d\theta_2}(\lambda -  
       \sqrt{2+2 \cos\theta_2})\right)\right|_{\theta_2=\theta_{2}^*}
&=& \left.\frac{\sin\theta_2}{\sqrt{2+2\cos\theta_2}}\right|_{\theta_2=\theta_{2}^*}\\
&=& 1-\left(\frac{\lambda}{2}\right)^2
\end{eqnarray}
the pdf of interest is\,\footnote{As an exercise, we can check the result 
with that obtainable using the `text book' transformation rule 
described
in footnote \ref{fn:TrasformazioneTextbook}. In our case, using 
the general symbol $g()$ introduced there, we have 
$\lambda=g(\theta_2) = \sqrt{2+2\cos{\theta_2}}$
and $\theta_2=g^{-1}(\lambda) = \arccos\left(\lambda^2/2-1\right)$. 
Applying Eq.~(A) of footnote \ref{fn:TrasformazioneTextbook}
we have
\begin{eqnarray*}
f_\Lambda(\lambda) &=& f_{\Theta_2}(g^{-1}\left(\lambda)\right) \cdot 
\left|\frac{d}{d\lambda} g^{-1}(\lambda)\right| \\
 &=& \frac{1}{\pi} \cdot \left|\frac{d}{d\lambda} \arccos\left(\lambda^2/2-1\right))\right| \\
 &=&  \frac{1}{\pi} \cdot \left|-\frac{\lambda}{\sqrt{1-(\lambda^2/2-1)^2}}\right|\,, 
\end{eqnarray*}
also yielding Eq.~(\ref{eq:l_lambda_M1}). (The minus sign resulting from the derivation
is because $\lambda$ decreases as $\theta_2$ increases, as clear from 
Fig~\ref{fig:bertrand_meth1}.) 
\label{fn:TrasformazioneTextbook_1}
}
\begin{eqnarray}
f(\lambda\,|\,{\cal M}_1) &=& \frac{1}{\pi}\cdot 
\frac{1}{\sqrt{1-(\lambda/2)^2}}
\hspace{0.7cm} (0 \le \lambda \le 2)\,,
\label{eq:l_lambda_M1}
\end{eqnarray}
from which we can also calculate the cumulative function 
\begin{eqnarray}
F(\lambda\,|\,{\cal M}_1) \equiv \int_0^\lambda
f(\lambda'\,|\,{\cal M}_1)\,d\lambda' &=& \frac{2}{\pi}\cdot
\arcsin\left(\frac{\lambda}{2}\right)\,,
\end{eqnarray}
both shown in Fig.~\ref{fig:f_F_meth1}.

\begin{figure}
\begin{tabular}{|c|c|}
\hline
\epsfig{file=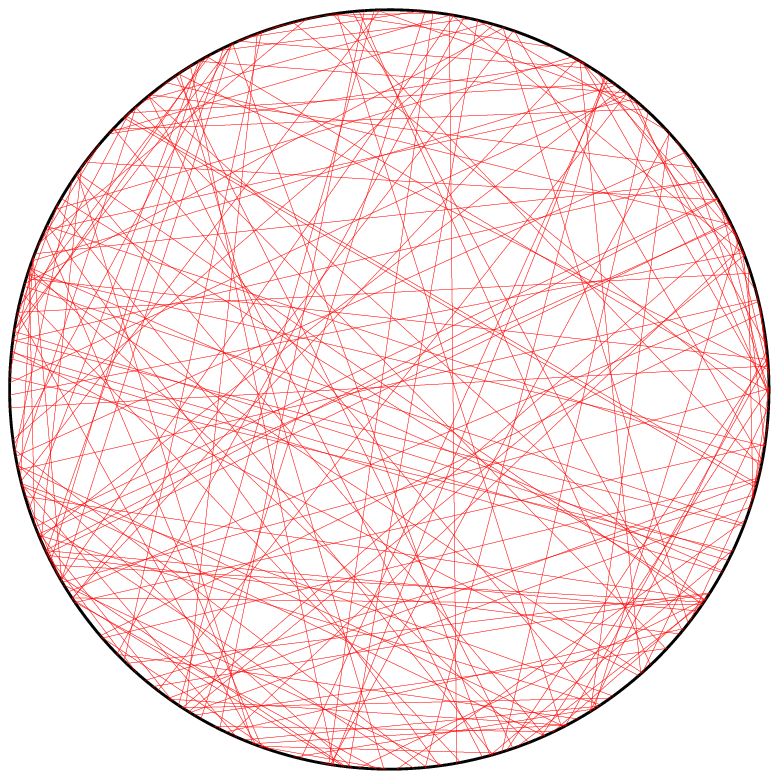,clip=,width=0.46\linewidth} &
\epsfig{file=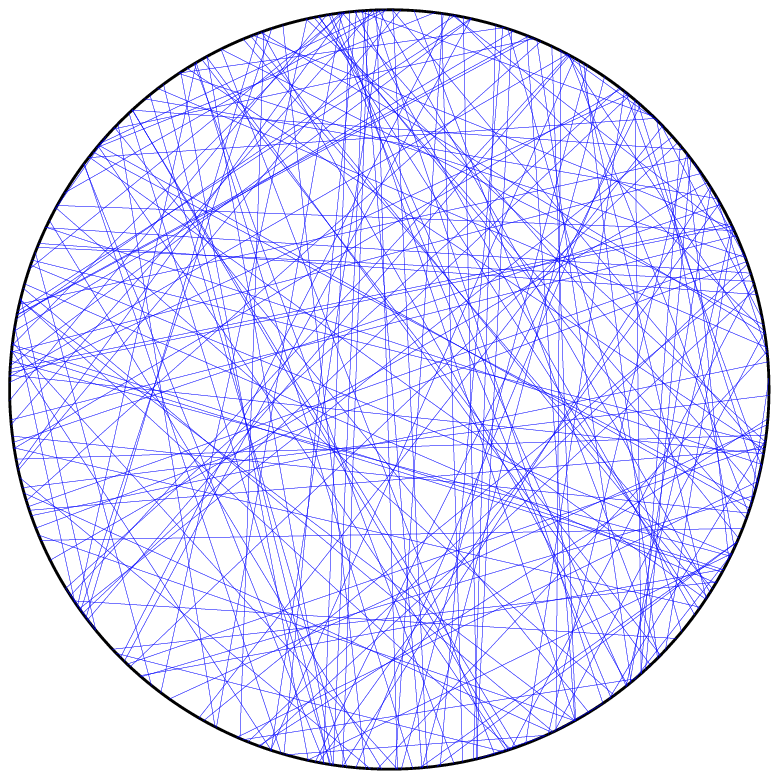,clip=,width=0.46\linewidth} \\
\hline
\epsfig{file=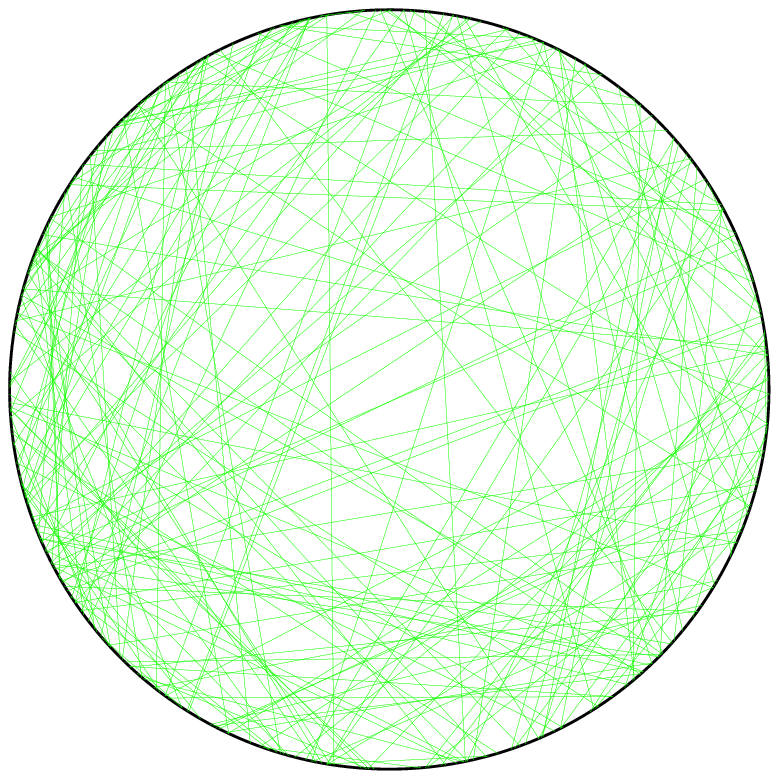,clip=,width=0.46\linewidth} &
\epsfig{file=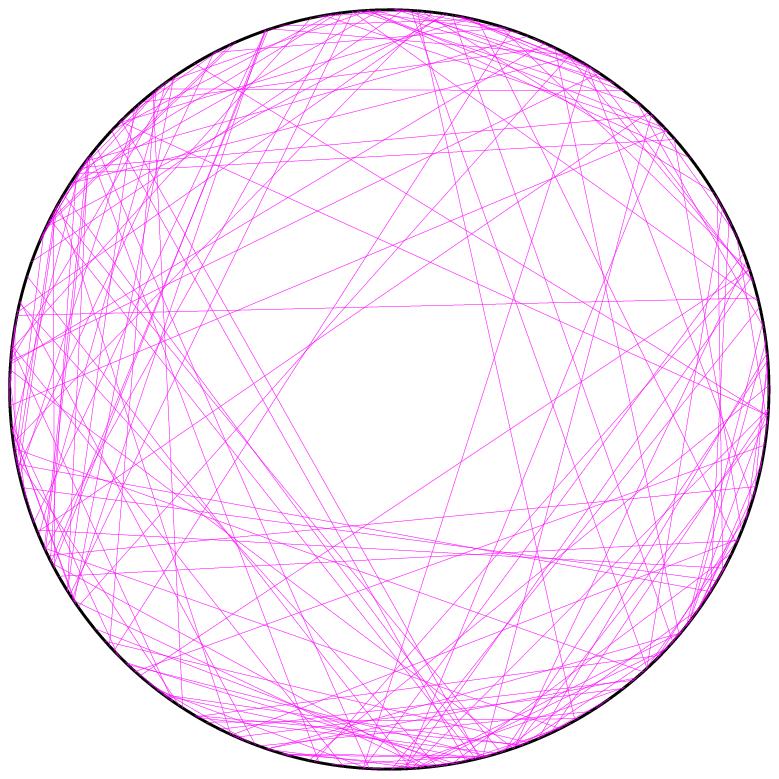,clip=,width=0.46\linewidth}\\
\hline
\epsfig{file=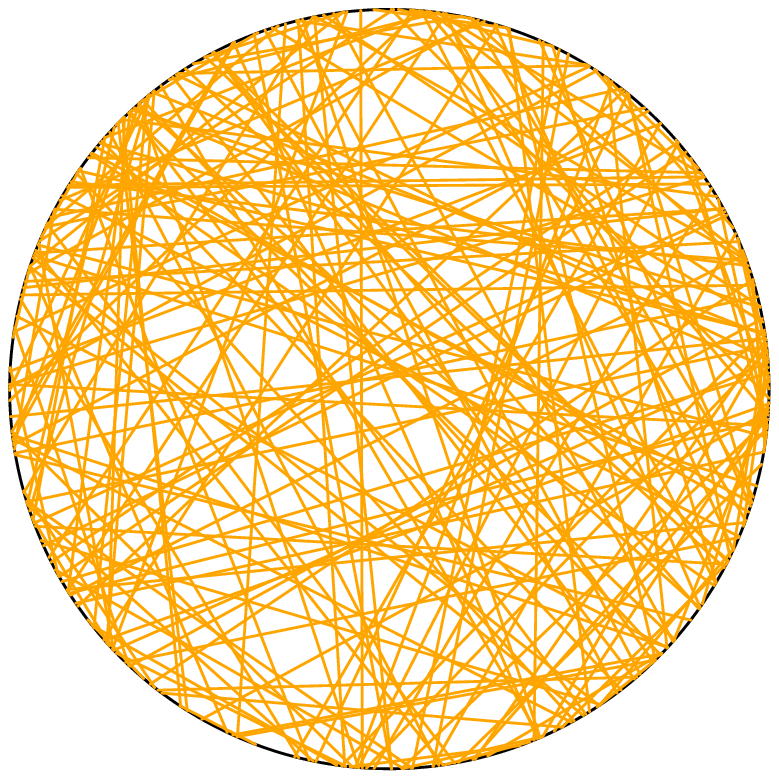,clip=,width=0.46\linewidth} &
\epsfig{file=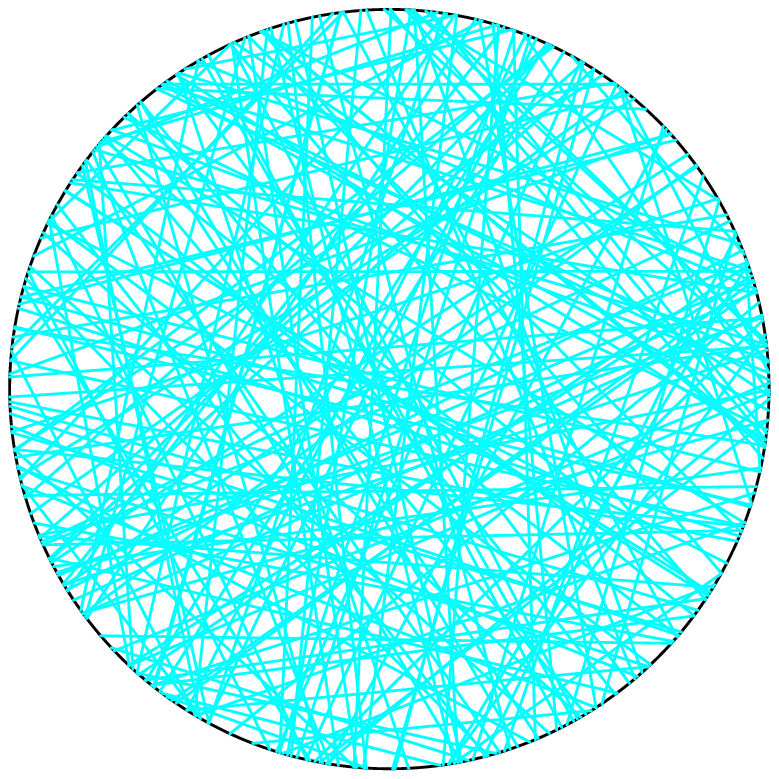,clip=,width=0.46\linewidth}\\
\hline
\end{tabular}
\caption{\small \sf Samples of chords generated the various methods.}
\label{fig:corde}
\end{figure}

\begin{figure}
\begin{tabular}{|c|c|}
\hline
\epsfig{file=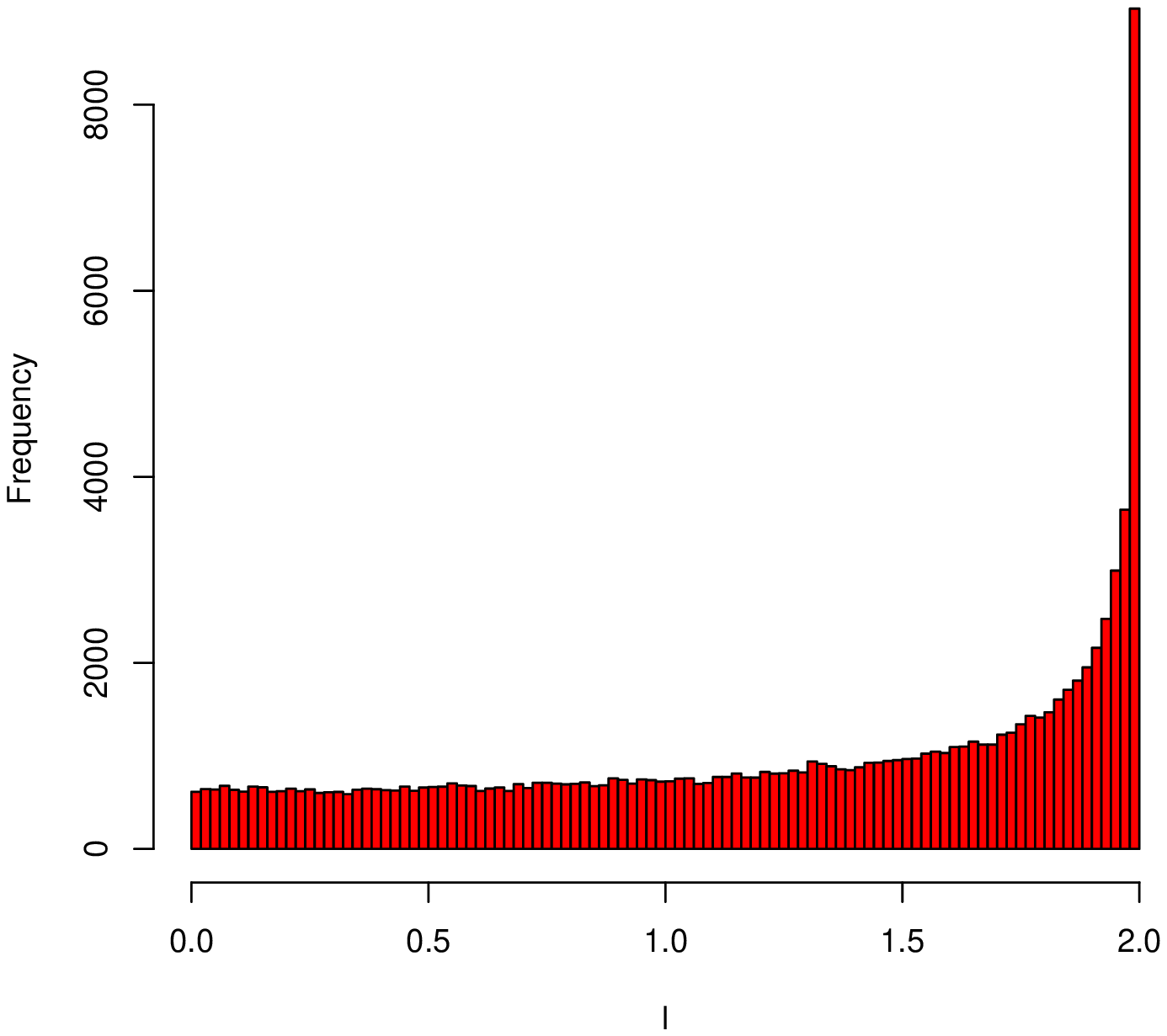,clip=,width=0.46\linewidth} &
\epsfig{file=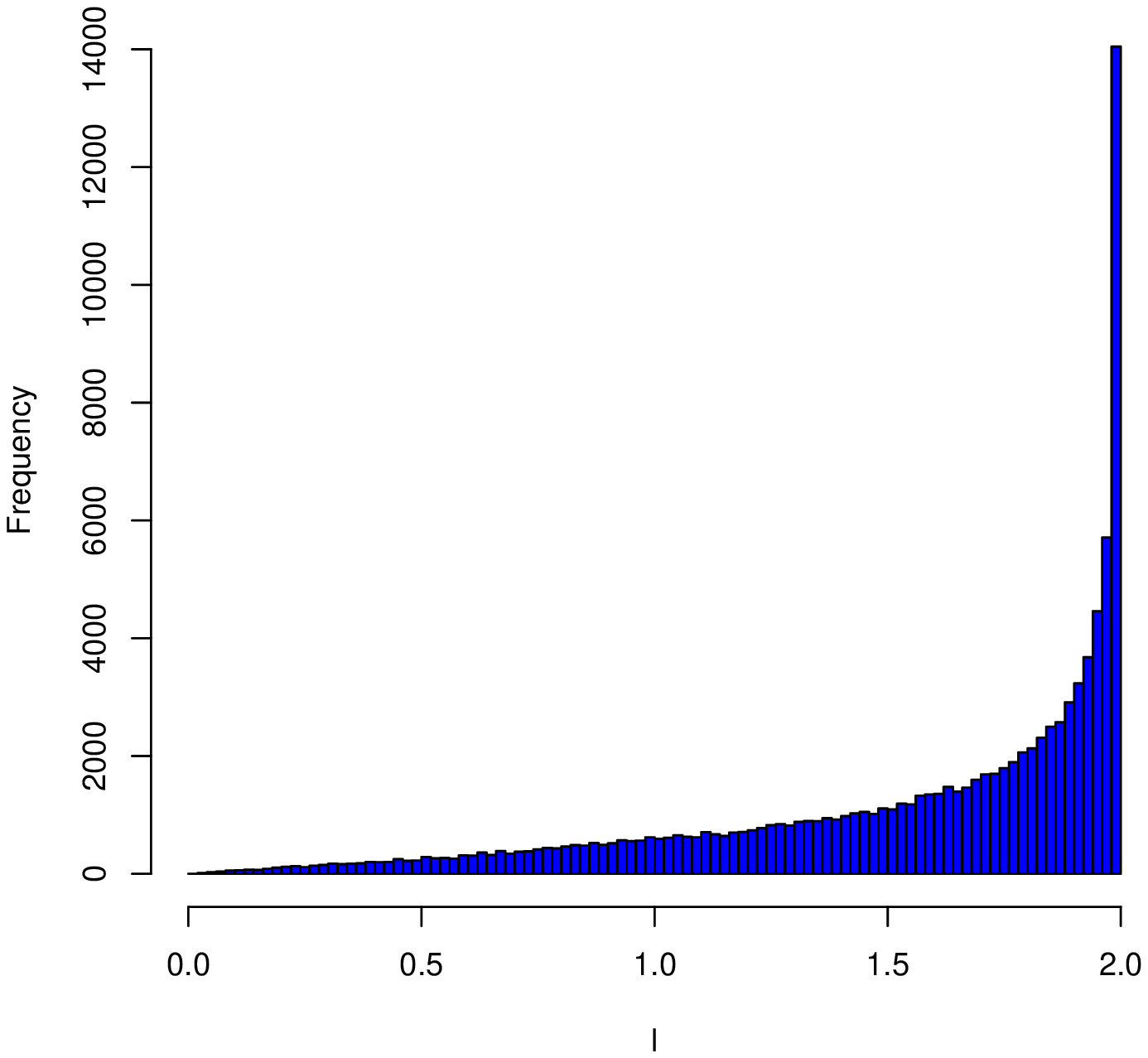,clip=,width=0.46\linewidth} \\
\hline
\epsfig{file=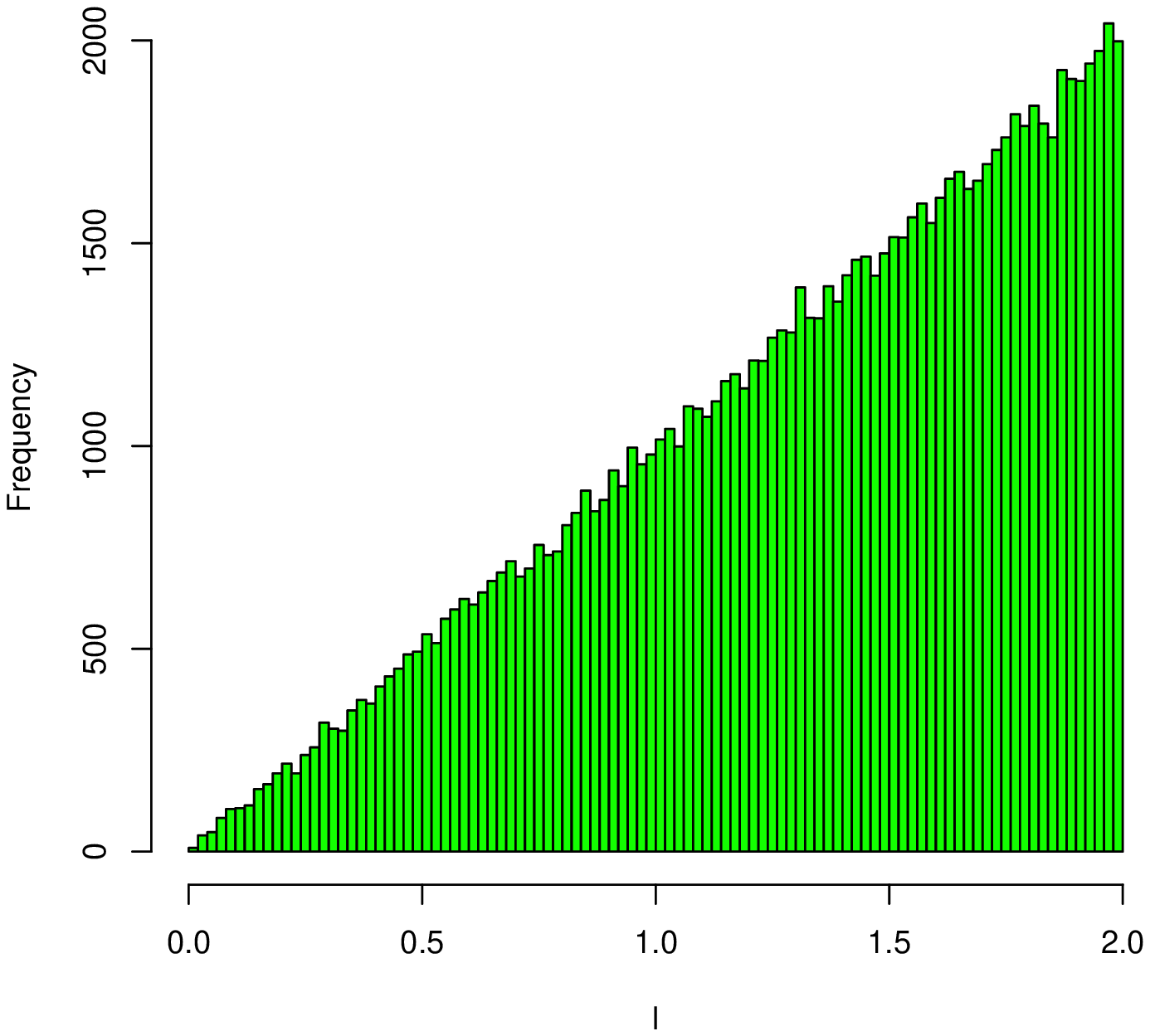,clip=,width=0.46\linewidth} &
\epsfig{file=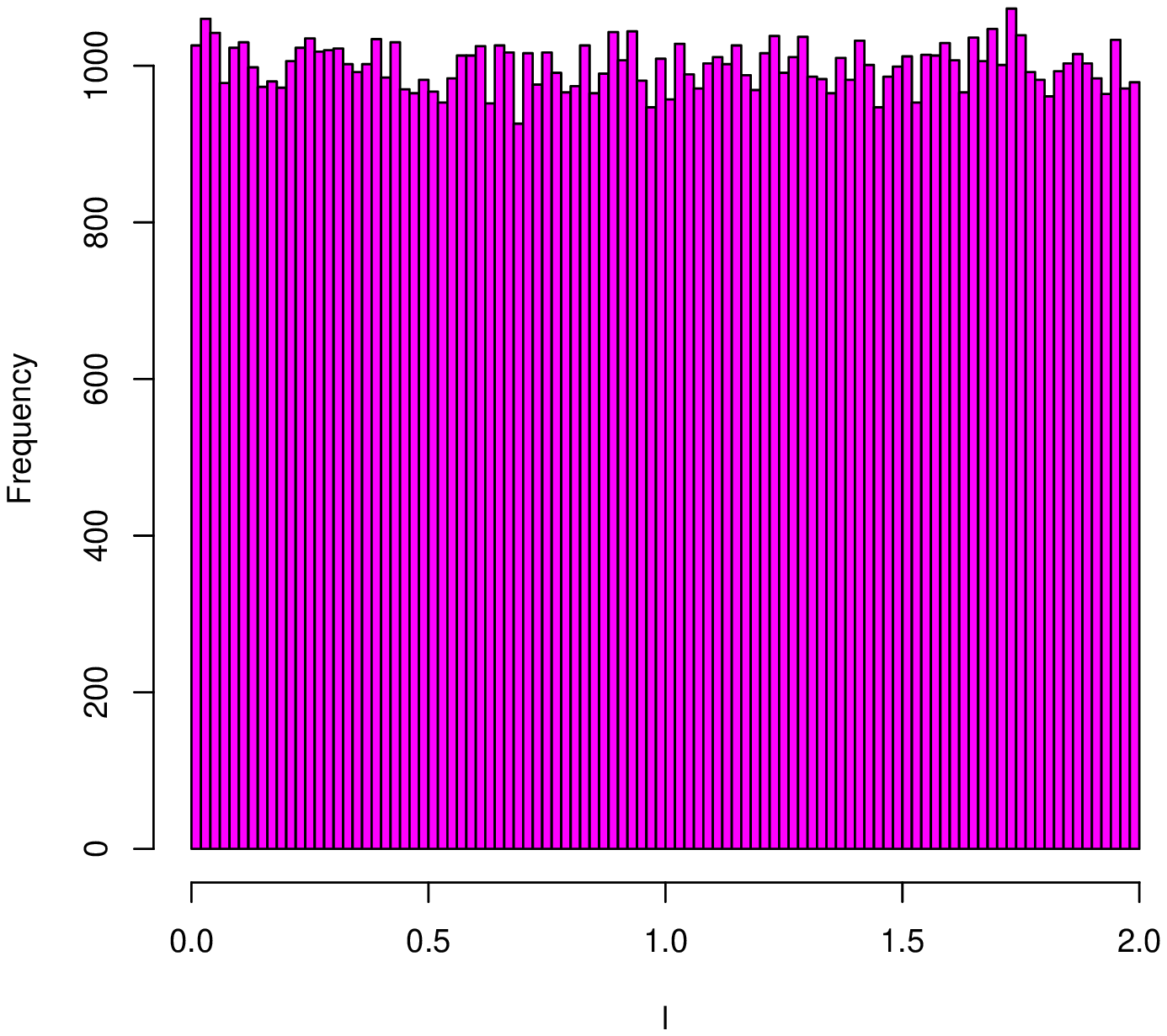,clip=,width=0.46\linewidth}\\
\hline
\epsfig{file=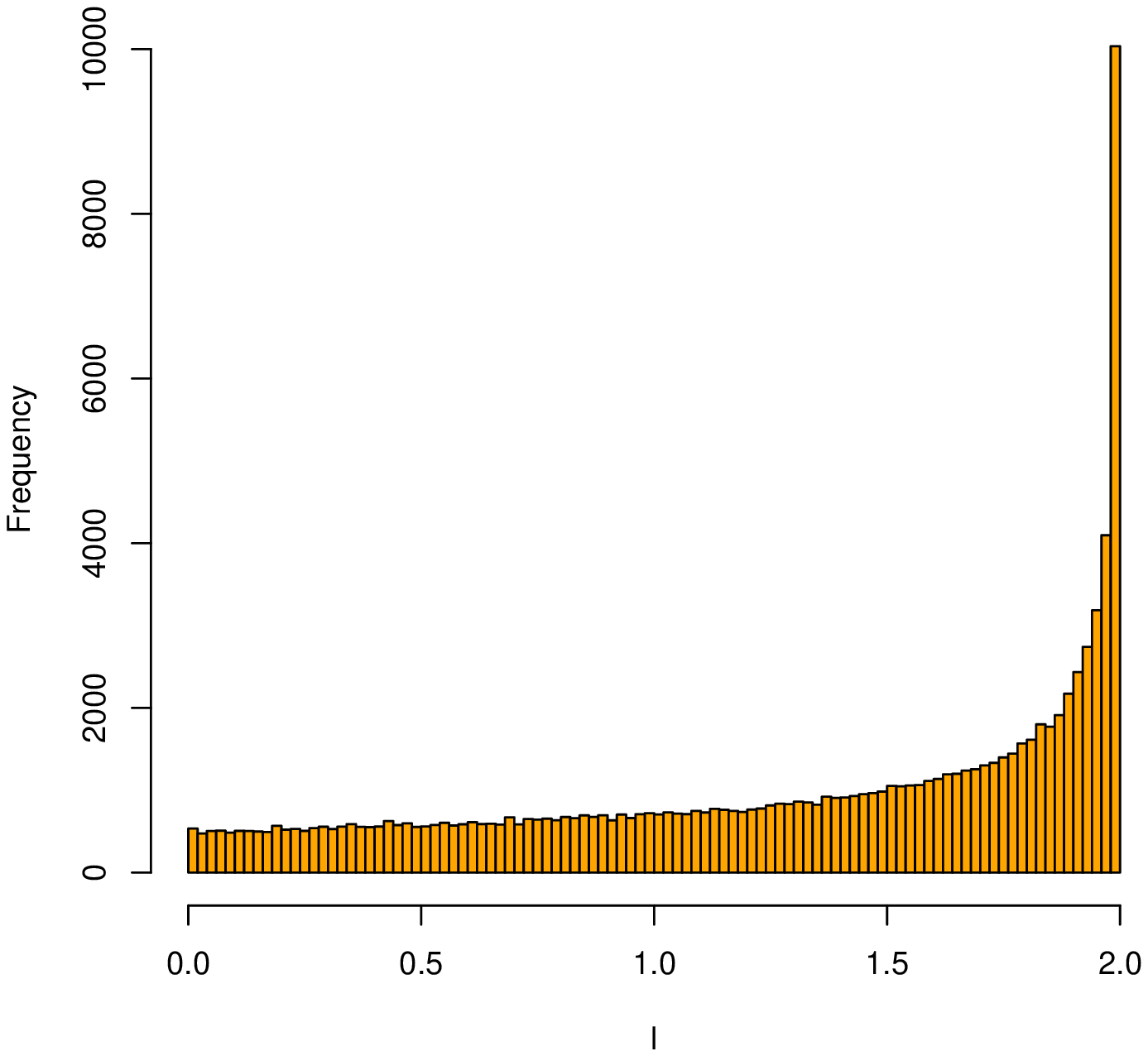,clip=,width=0.46\linewidth} &
\epsfig{file=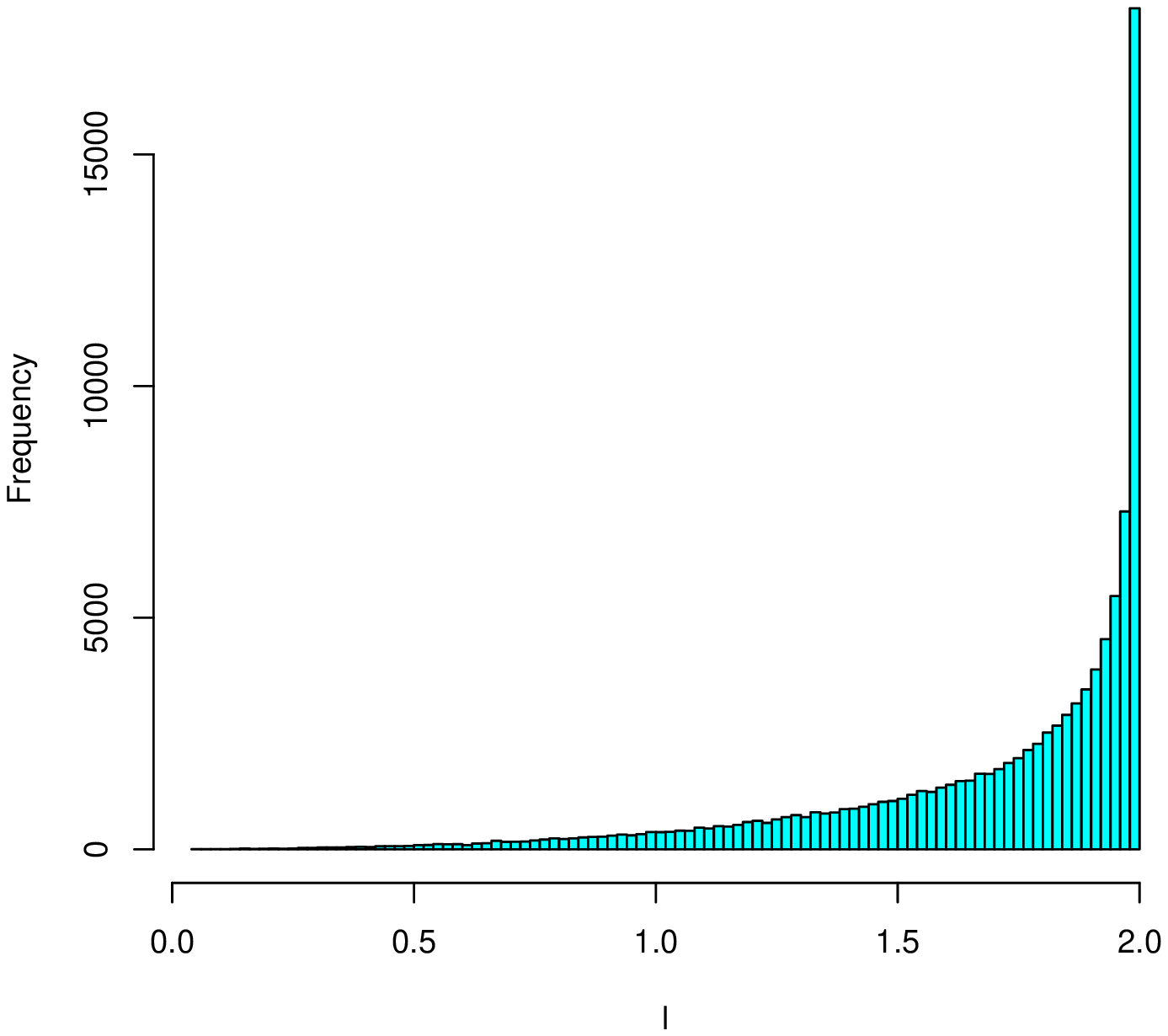,clip=,width=0.46\linewidth}\\
\hline
\end{tabular}
\caption{\small \sf Distribution of the length of the chords
in sample produced with the various methods. }
\label{fig:l}
\end{figure}

\begin{figure}
\begin{tabular}{|c|c|}
\hline
\epsfig{file=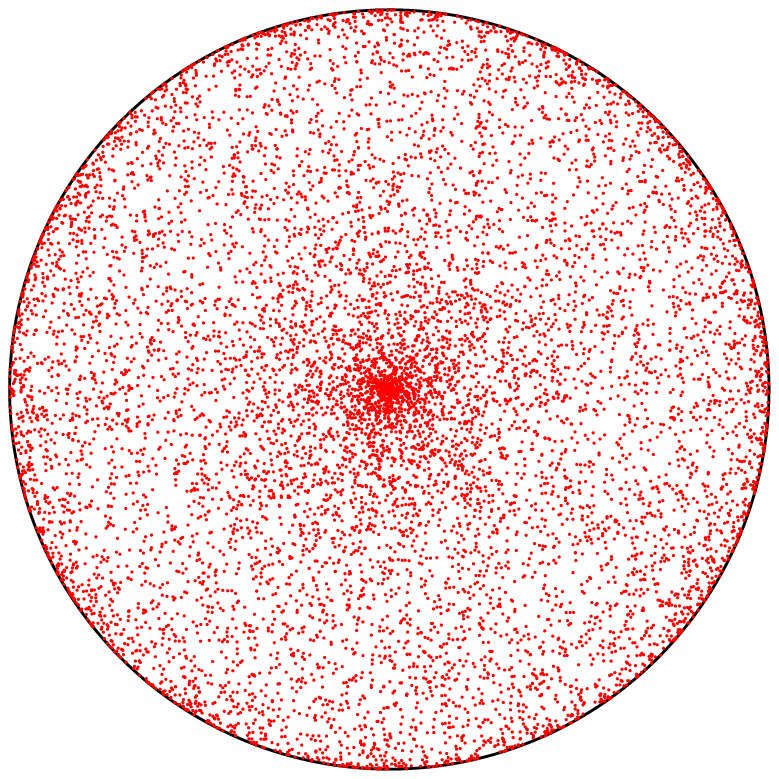,clip=,width=0.46\linewidth} &
\epsfig{file=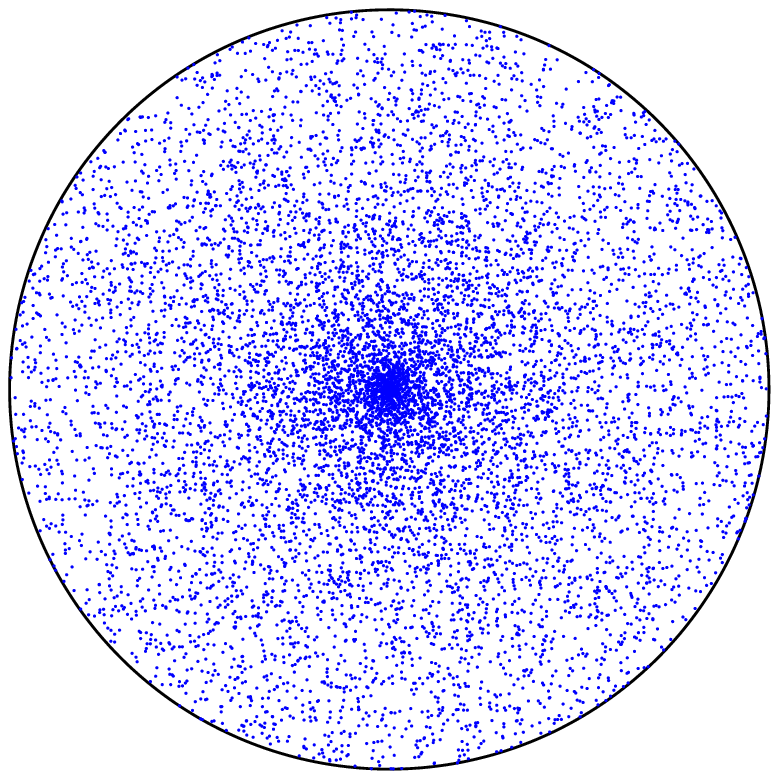,clip=,width=0.46\linewidth} \\
\hline
\epsfig{file=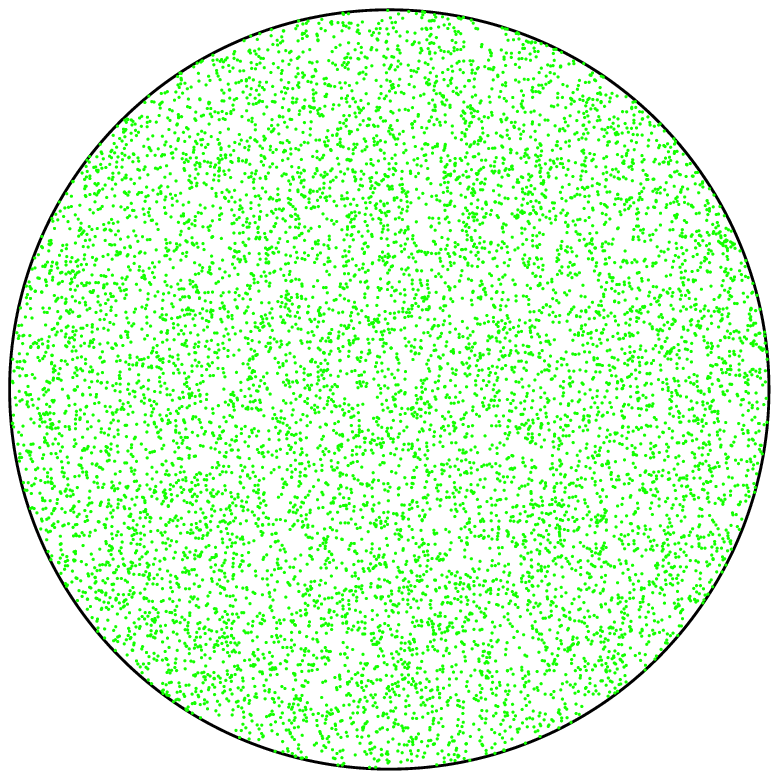,clip=,width=0.46\linewidth} &
\epsfig{file=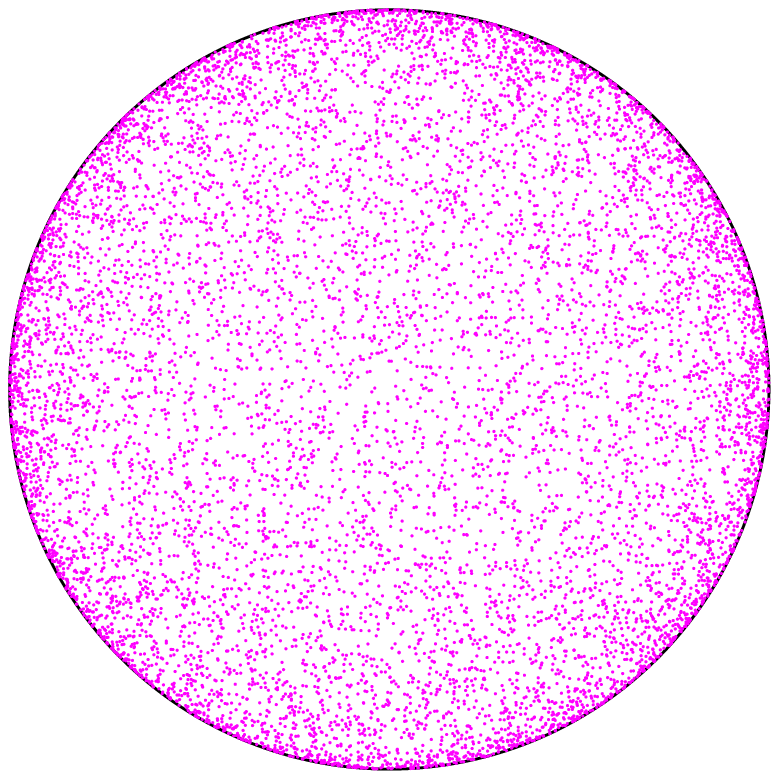,clip=,width=0.46\linewidth}\\
\hline
\epsfig{file=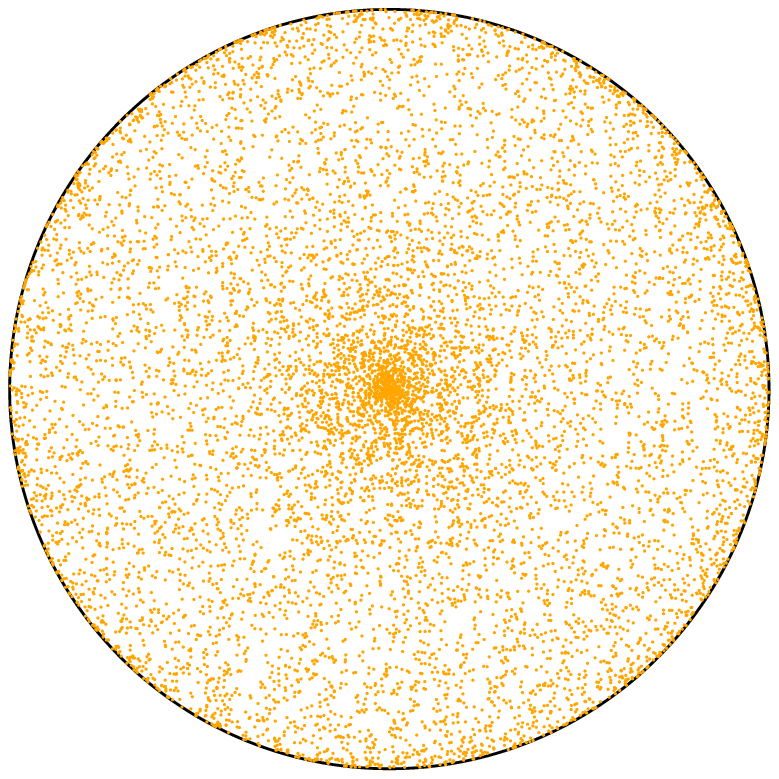,clip=,width=0.46\linewidth} &
\epsfig{file=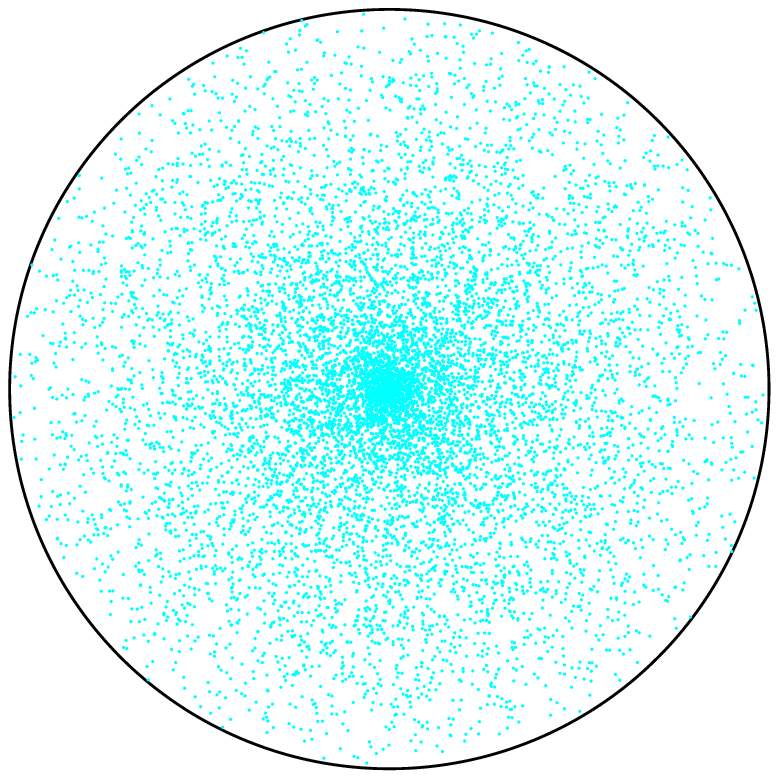,clip=,width=0.46\linewidth}\\
\hline
\end{tabular}
\caption{\small \sf Sample of centers of the chords generated
with the various methods}
\label{fig:centri}
\end{figure}

\begin{figure}
\begin{tabular}{|c|c|}
\hline
\epsfig{file=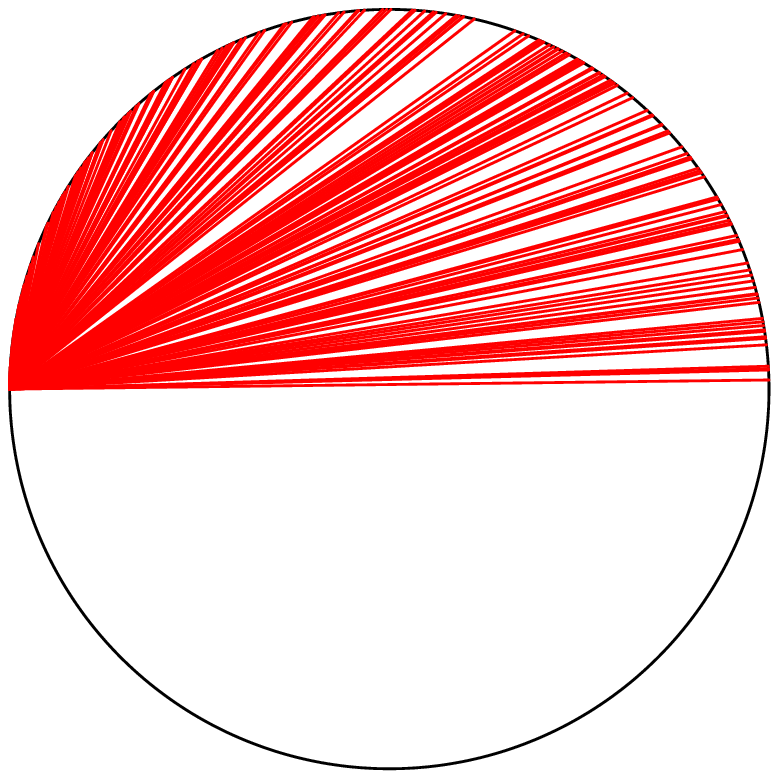,clip=,width=0.46\linewidth} &
\epsfig{file=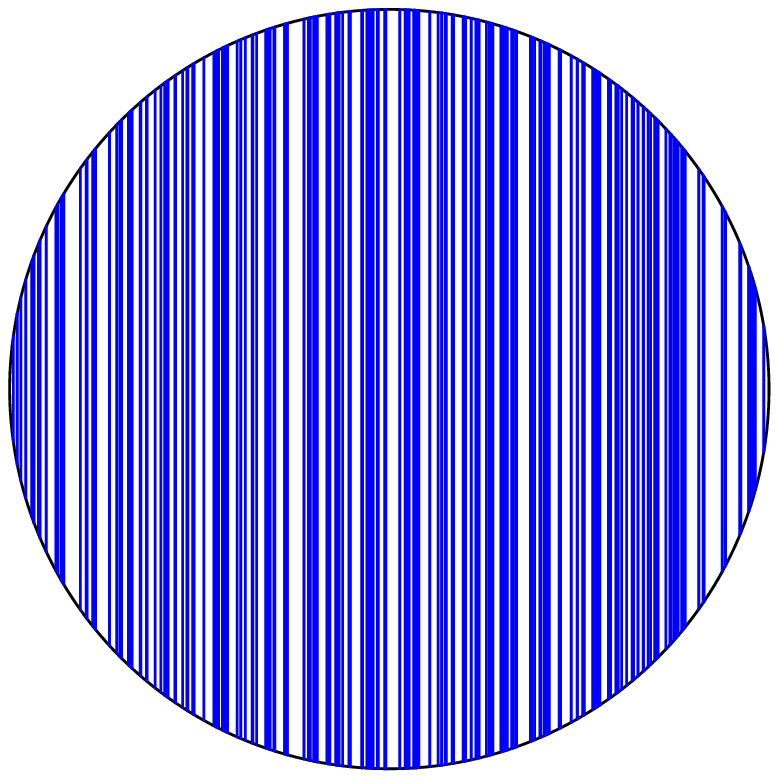,clip=,width=0.46\linewidth} \\
\hline
\epsfig{file=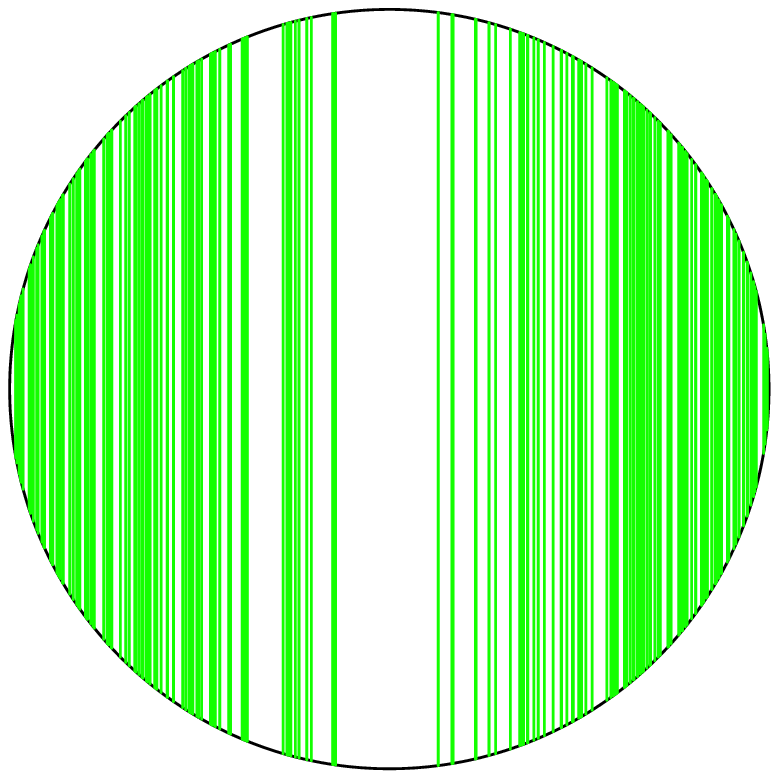,clip=,width=0.46\linewidth} &
\epsfig{file=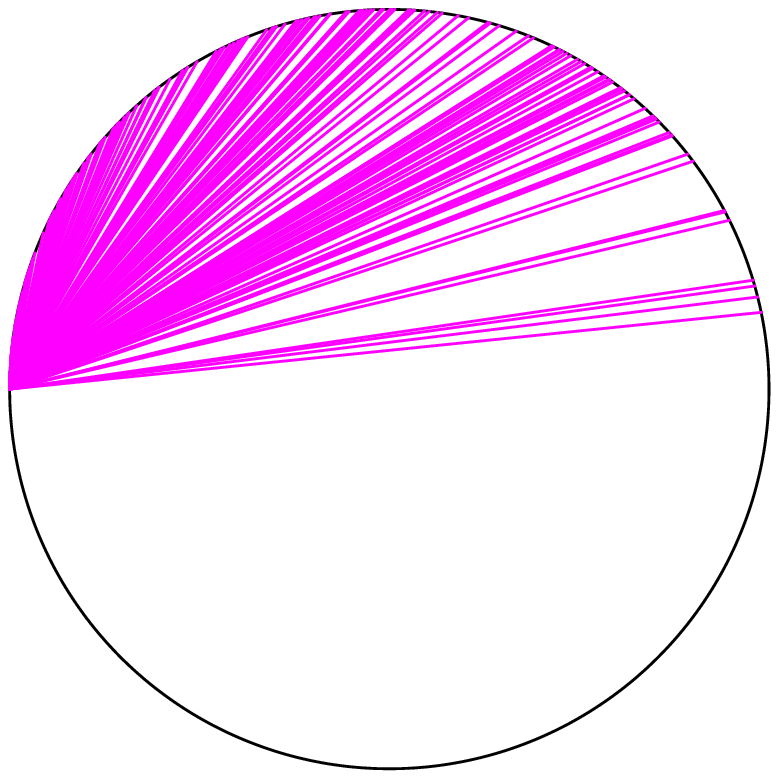,clip=,width=0.46\linewidth}\\
\hline
\epsfig{file=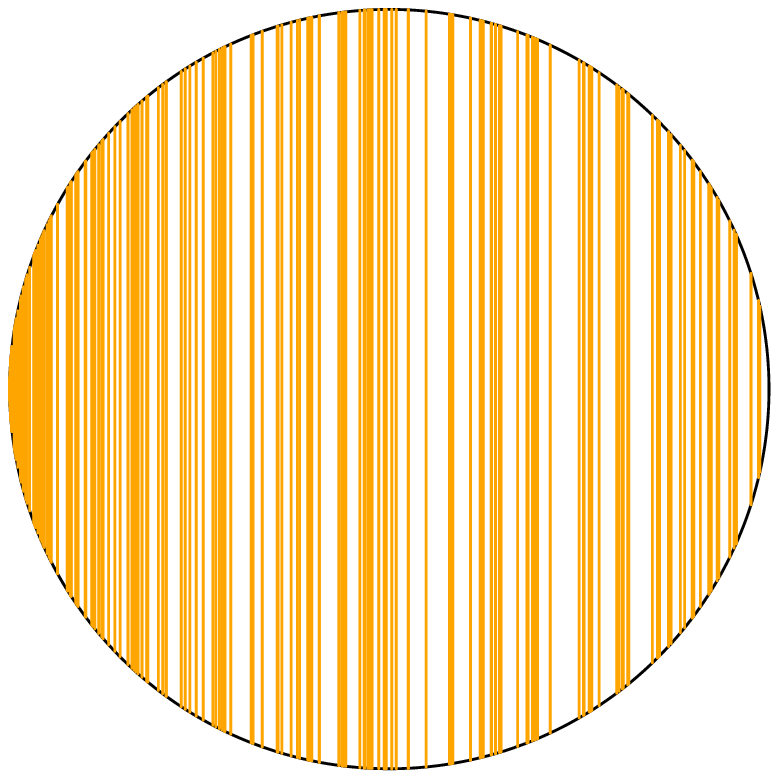,clip=,width=0.46\linewidth} &
\epsfig{file=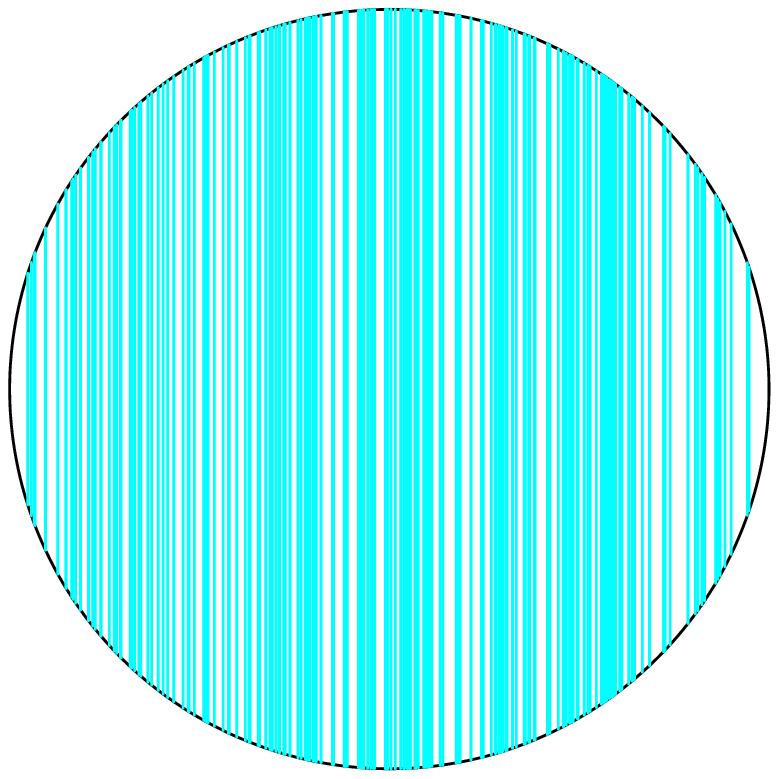,clip=,width=0.46\linewidth}\\
\hline
\end{tabular}
\caption{\small \sf Samples of chords generated the various methods
with respect to a preferred axis (``without rotation'' -- see text). }
\label{fig:corde_norot}
\end{figure}

We can finally check the probability of interest, and also calculate 
the probability of a chord to be smaller than the radius of the circle:
\begin{eqnarray*}
P(\lambda \le \sqrt{3}\,|\, {\cal M}_1) =
F(\sqrt{3}\,|\, {\cal M}_1)
 & = &  
\frac{2}{\pi} \cdot \arcsin\left(\frac{\sqrt{3}}{2}\right) =
\frac{2}{\pi} \cdot \frac{\pi}{3} = \mathbf{\frac{2}{3}} \\
P(\lambda \le 1\,|\, {\cal M}_1) =
F(1\,|\, {\cal M}_1)
& = &  
\frac{2}{\pi} \cdot \arcsin\left(\frac{1}{2}\right) =
\frac{2}{\pi} \cdot \frac{\pi}{6} = \frac{1}{3} 
\end{eqnarray*}
Simulations of chords with this methods are 
reported in the top left plot of Figures 
\ref{fig:corde}-\ref{fig:corde_norot} 
at the end of the paper:
 Fig.~\ref{fig:corde} shows a sample of random
chords; Fig.~\ref{fig:centri} the position of their center.
The distribution of the lengths in units of $R$ 
are shown in the histograms of Fig.~\ref{fig:l},
while those  of Fig.~\ref{fig:r} show the distribution
of the distance of the chords from the center of the circle,
about which we shall say more in Appendix A. 
Finally, Fig.~\ref{fig:corde_norot} is like Fig.~\ref{fig:corde}, but 
`without rotation', the meaning of this espression becoming
clearer as we go through the various methods.
 In the case of the first method this corresponds
to the way we have followed to derive $f(\lambda\,|\, {\cal M}_1)$
in this subsection, that is fixing $\theta_1$ at $\pi$ and 
extracting $\theta_2$ between 0 and $\pi$.

\newpage
\begin{figure} 
\begin{center}
\epsfig{file=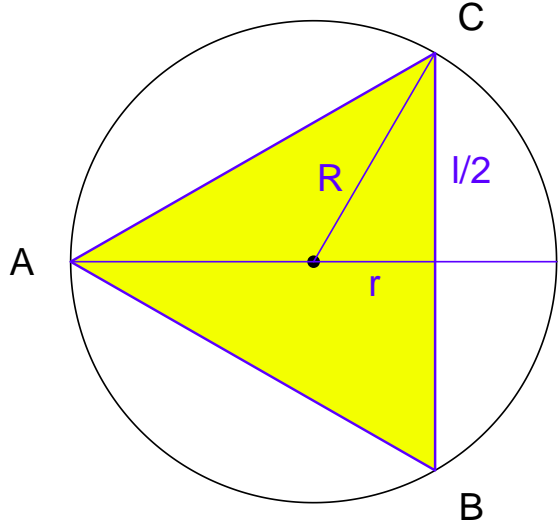,clip=,width=0.475\linewidth}
\end{center}
\caption{\small \sf Construction of a chord orthogonal to a radius
and distant $r$ from the center of the circle.}
\label{fig:bertrand_meth2}
\end{figure}
\subsection{Chords orthogonal to a radius, with center 
uniformly distributed along it}
\begin{figure}
\begin{tabular}{c}
 {\large $[\,{\cal M}_2\,]$ }\\
\epsfig{file=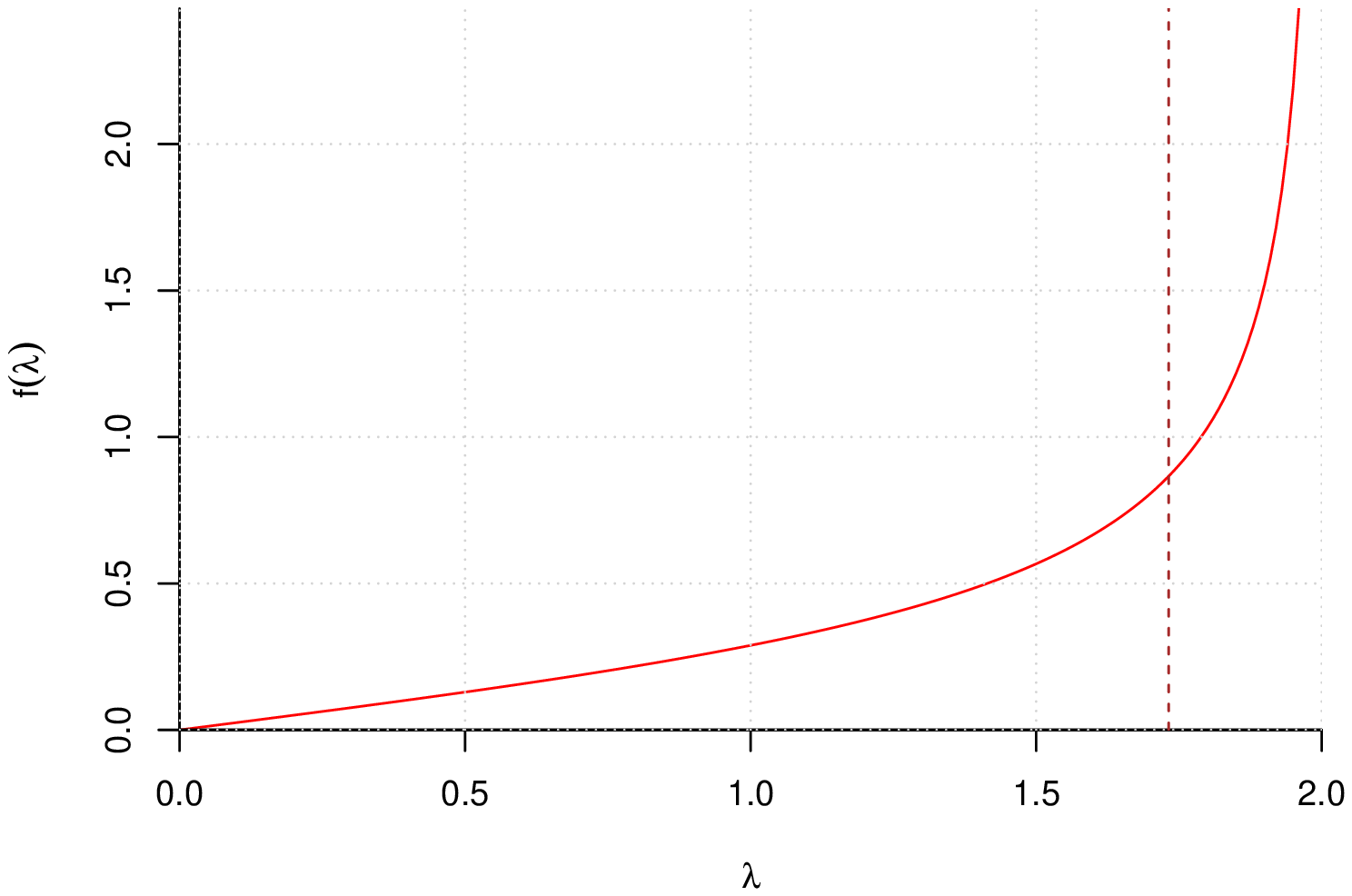,clip=,width=\linewidth} \\
\epsfig{file=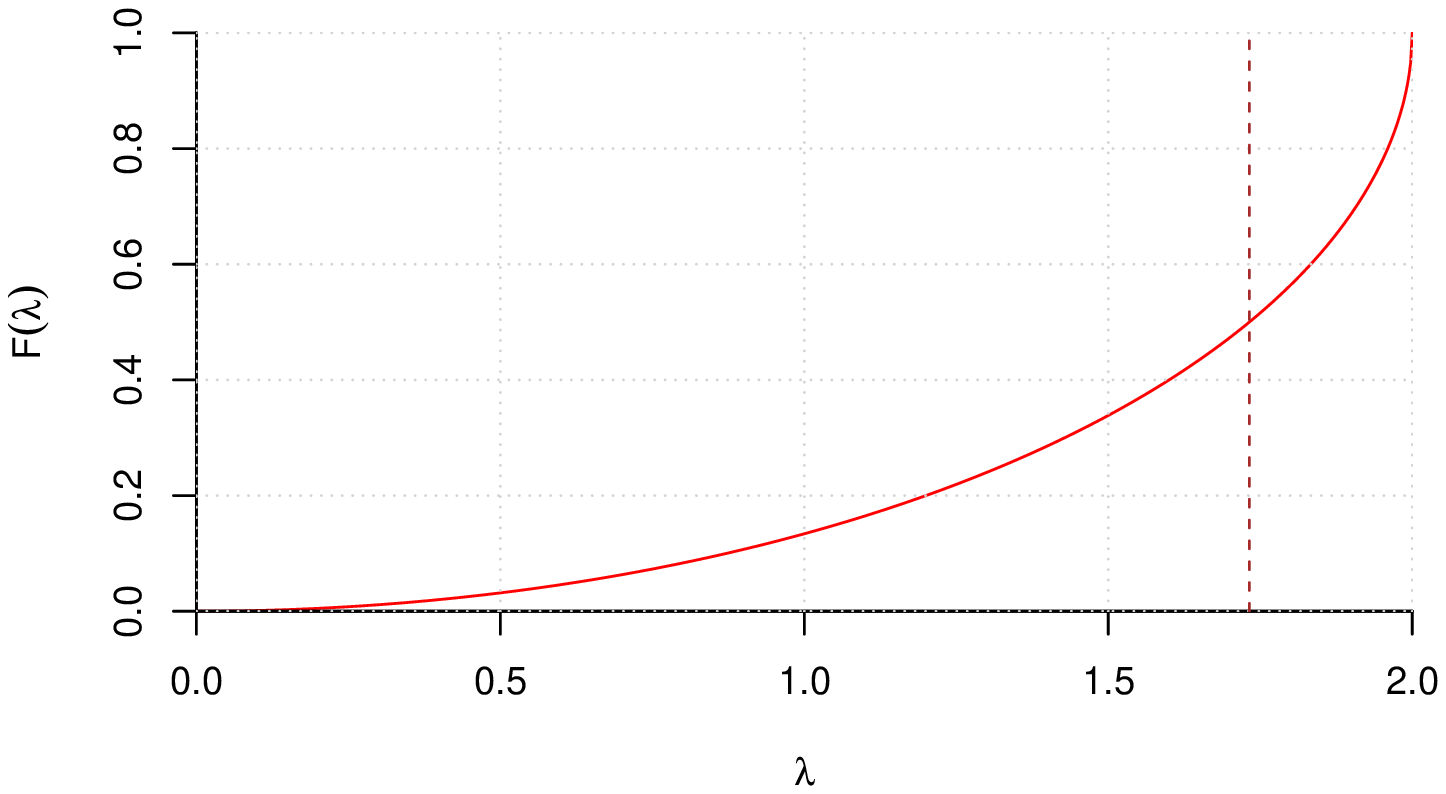,clip=,width=\linewidth}
\end{tabular}
\caption{\small \sf Probability distribution of $\lambda=l/R$ of the chords
generated with Method 2. The dashed vertical line indicates $\lambda=\sqrt{3}$.}
\label{fig:f_F_meth2}
\end{figure}

The second `classical' algorithm consists in choosing chords
orthogonal to a radius with its center uniformly distributed
along it. As we easily see from 
Fig.~\ref{fig:bertrand_meth2},
the condition for a chord to be smaller than the side
of the triangle ($l_T$) is the same as requiring
that its distance from the center of the circle, 
indicated by $r$ in the figure, is above $R/2$. That is 
\begin{eqnarray}
P(l \le l_T\,|\,{\cal M}_2) &=& P(r \ge R/2\,|\,{\cal M}_2) = \mathbf{\frac{1}{2}}\,.
\end{eqnarray}
Let us repeat the exercise of evaluating the probability distribution
of the lengths of the chords obtained with this method. 
Using $\rho$ to indicate $r/R$, in analogy to $\lambda=l/R$, 
we have (see figure)
\begin{eqnarray}
\lambda &=& 2\,\sqrt{1-\rho^2}\,,
\end{eqnarray}
with
\begin{eqnarray}
f(\rho\,|\,{\cal M}_2) &=& 1 \hspace{0.7cm}  (0\le \rho \le 1)\,.
\end{eqnarray}
Then, the pdf of interest will be given by
\begin{eqnarray}
f(\lambda\,|\,{\cal M}_2) &=& \int_0^1
\!\delta\left(\lambda -  2 \sqrt{1-\rho^2}\right)\cdot 1 \,d\rho 
\label{eq:trasf_M2_inizio}\\
&=& \int_0^1 \! \frac{\delta(\rho-\rho^*)}
  {\left|\left.\left(\frac{d}{d\rho}(\lambda -  
       2 \sqrt{1-\rho^2})\right)\right|_{\rho=\rho^*}\right|}
\,d\rho\,,\label{eq:trasf_M2} \\
&=& \frac{1}{2\rho^*/\sqrt{1-{\rho^*}^2}}
\end{eqnarray}
with 
\begin{eqnarray}
\rho^* &=& \sqrt{1-(\lambda/2)^2}\,. \label{eq:rho*M2}
\end{eqnarray}
The pdf and the cumulative distribution of interest
are then\,\footnote{Let us 
repeat the exercise of using Eq.~(A) of footnote \ref{fn:TrasformazioneTextbook}
also in this case, starting now from $\lambda=2\sqrt{1-\rho^2} \equiv g(\rho)$, 
$\rho = \sqrt{1-(\lambda/2)^2} \equiv g^{-1}(\lambda)$ and $f_{P}(\rho) = 1$:
\begin{eqnarray*}
f_\Lambda(\lambda) &=& f_{P}( g^{-1}(\lambda)) 
\cdot \left| \frac{d}{d\lambda} \sqrt{1-(\lambda/2)^2} \right| \\
&=& 1\cdot \frac{\lambda}{4\,\sqrt{1-(\lambda/2)^2}}\,,
\end{eqnarray*}
that is precisely Eq.~(\ref{eq:f_lambda_meth2}).
\label{fn:TrasformazioneTextbook_2}
}
\begin{eqnarray}
f(\lambda\,|\,{\cal M}_2) &=& \frac{\lambda}{4\sqrt{1-(\lambda/2)^2}} 
\hspace{0.7cm}  (0\le \lambda \le 2)\label{eq:f_lambda_meth2}\\
F(\lambda\,|\,{\cal M}_2) &=& 1 - \sqrt{1-(\lambda/2)^2}\,,
\end{eqnarray}
plotted in  Fig.~\ref{fig:f_F_meth2} and from which we can calculate the
probabilities of interest:
\begin{eqnarray}
F(\sqrt{3}\,|\,{\cal M}_2) &=& \mathbf{\frac{1}{2}} \\
F(1\,|\,{\cal M}_2) &=& 1-\frac{\sqrt{3}}{2} \approx 13.4\%.
\end{eqnarray}
Simulations of chords with this methods are 
reported in the plots of Figs.~\ref{fig:corde}-\ref{fig:corde_norot}. 

\begin{figure}[h]
\begin{center}
\epsfig{file=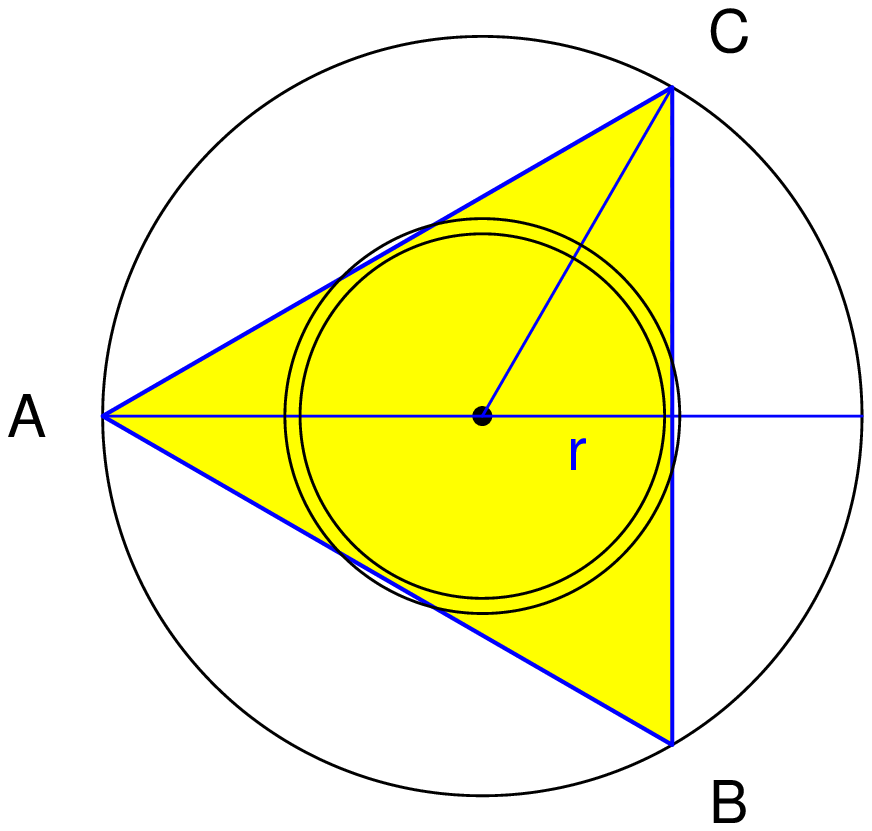,clip=,width=0.45\linewidth}
\end{center}
\vspace{-0.5cm}
\caption{\small \sf As Fig.~\ref{fig:bertrand_meth2}, but 
with the annulus of infinitesimal  width $dr$ drawn at $r$ 
to show that in Model 3 the infinitesimal probability
$dP$ that the distance of a chord from the center is between
$r$ and $r+dr$ is given by
$dP \equiv dF(r) =2\pi r dr /(\pi R^2)= (2r/R^2)\,dr$, 
and hence $dF(\rho)=2\rho\,d\rho$, 
or, in our notation, $f(\rho\,|\,{\cal M}_3) = 2\rho$, 
with $\rho=r/R$.}
\label{fig:bertrand_meth3}
\end{figure}

\vspace{-0.3cm}
\subsection{Center of chords uniformly chosen inside the 
circle, with chords orthogonal to radius}\label{ss:Meth3}
The third method is a variant of the second, in which the centers
of the chord, instead of being generated uniformly along a radius,
are generated uniformly inside the circle. 
The centers of the chords fall inside the circle of radius $R/2$
with probability 1/4, and than 
(see again Fig.~\ref{fig:bertrand_meth2})
\vspace{-0.2cm}
\begin{eqnarray}
P(l \le l_T\,|\,{\cal M}_3) &=& P(r\ge \frac{R}{2}\,|\,{\cal M}_3) = \mathbf{\frac{3}{4}}\,.
\end{eqnarray}
Let us repeat once more the exercise of calculating the
probability distribution of $\lambda$. The difference with 
respect to the previous case is that now the pdf of the center of
the chords is proportional to $r$, since $dP\propto (2\pi r) dr$ 
(see Fig.~\ref{fig:bertrand_meth3}). 
The pdf of $\rho$ is then, after normalization,
\begin{eqnarray}
f(\rho\,|\,{\cal M}_3) &=& 2 \rho \hspace{0.7cm}  (0\le \rho \le 1)\,.
\end{eqnarray}
and the equivalent of 
Eqs.~(\ref{eq:trasf_M2_inizio})-(\ref{eq:trasf_M2}) and sequel are now
\begin{eqnarray}
f(\lambda\,|\,{\cal M}_3) &=& \int_0^1
\!\delta\left(\lambda -  2 \sqrt{1-\rho^2}\right)\cdot 2 \rho \,d\rho 
\label{eq:trasf_M3_inizio}\\
&=& \int_0^1 \! \frac{\delta(\rho-\rho^*) \cdot 2 \rho }
  {\left|\left.\left(\frac{d}{d\rho}(\lambda -  
       2 \sqrt{1-\rho^2})\right)\right|_{\rho=\rho^*}\right|}
\,d\rho\,,\label{eq:trasf_M3} \\
&=& \frac{2\,\rho^*}{2\rho^*/\sqrt{1-{\rho^*}^2}} = \sqrt{1-{\rho^*}^2}
\end{eqnarray}
with the same $\rho^*$ of Eq.~(\ref{eq:rho*M2}), 
thus leading to\,\footnote{Let us 
repeat once more the exercise done in footnotes 
\ref{fn:TrasformazioneTextbook_1} and \ref{fn:TrasformazioneTextbook_2}, 
since in this case the starting pdf is not a constant, being
  $f_{P}(\rho) = 2 \rho$. All the rest is like in
footnote  \ref{fn:TrasformazioneTextbook_2}.  Here it is:
\begin{eqnarray*}
f_\Lambda(\lambda) &=& f_{P}( g^{-1}(\lambda)) 
\cdot \left| \frac{d}{d\lambda} \sqrt{1-(\lambda/2)^2} \right| \\
&=& 2\sqrt{1-(\lambda/2)^2}\cdot 
\frac{\lambda}{4\sqrt{1-(\lambda/2)^2}} = \frac{\lambda}{2}\,.
\end{eqnarray*}
\label{fn:TrasformazioneTextbook_3}
}
\begin{eqnarray}
f(\lambda\,|\,{\cal M}_3) &=& 
\sqrt{1-\left(1-(\lambda/2)^2\right)} 
= \frac{\lambda}{2}\,.
\label{eq:trasf_M3a}
\end{eqnarray}
\begin{figure}
\begin{tabular}{c}
 {\large $[\,{\cal M}_3\,]$ }\\
\epsfig{file=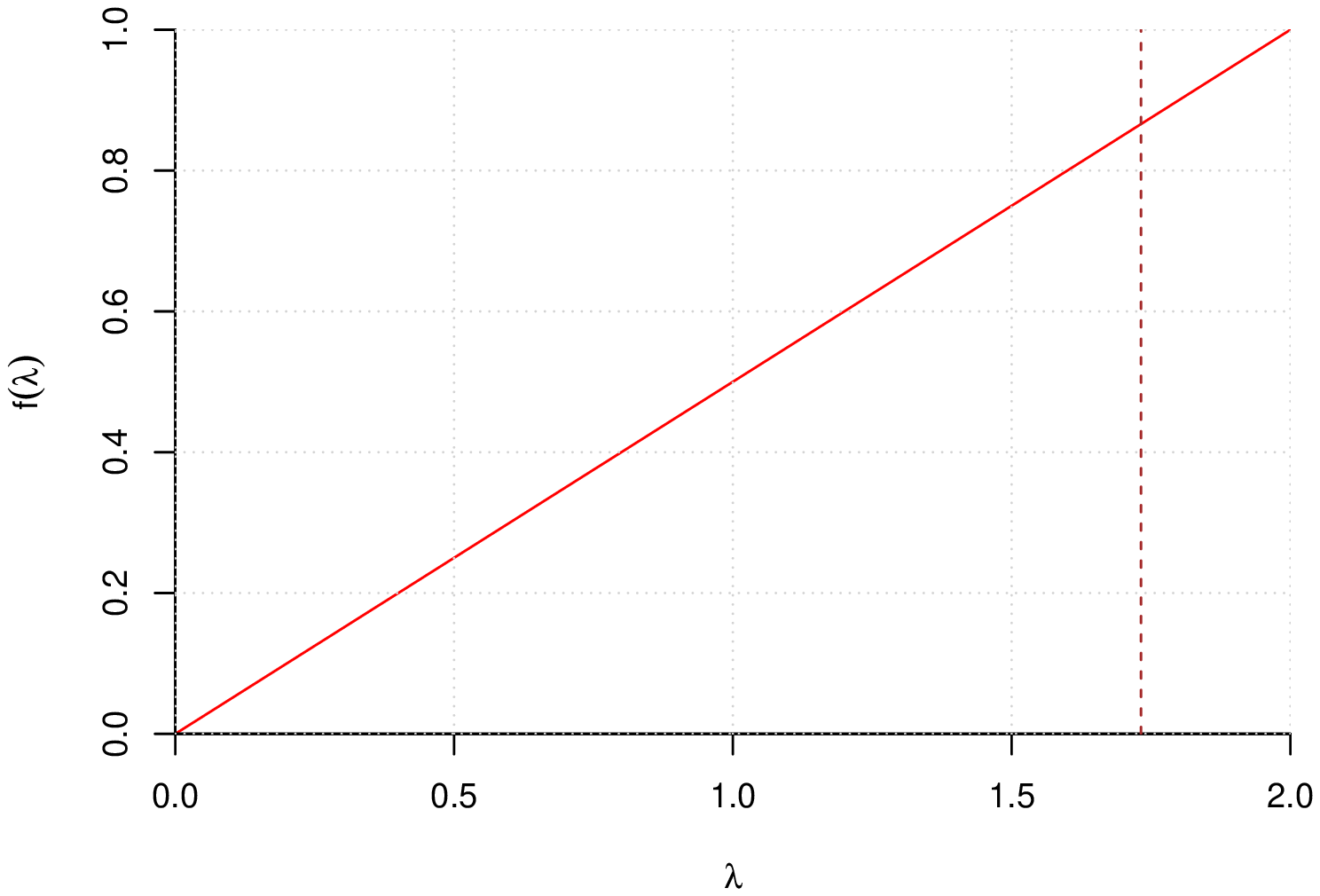,clip=,width=\linewidth} \\
\epsfig{file=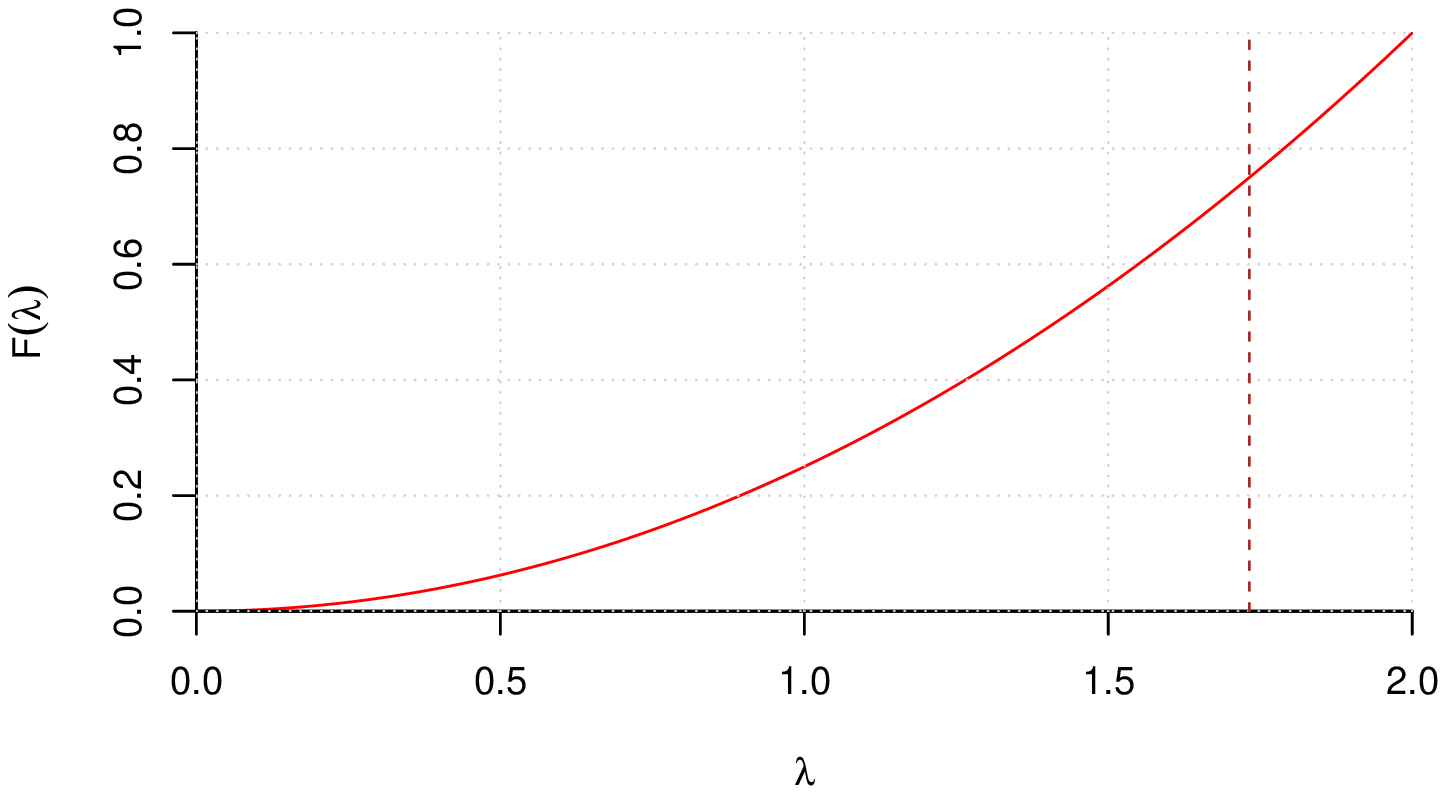,clip=,width=\linewidth}
\end{tabular}
\caption{\small \sf Probability distribution of $\lambda=l/R$ of the chords
generated with Method 3. The dashed vertical line indicates $\lambda=\sqrt{3}$.}
\label{fig:f_F_meth3}
\end{figure}
The result can be then summarized as
\begin{eqnarray}
f(\lambda\,|\,{\cal M}_3) &=& \frac{\lambda}{2}
\hspace{0.7cm}  (0\le \lambda \le 2)\\
F(\lambda\,|\,{\cal M}_3) &=&  \frac{\lambda^2}{4}\,,
\end{eqnarray}
from which we obtain
\begin{eqnarray}
F(\sqrt{3}\,|\,{\cal M}_3) &=& \mathbf{\frac{3}{4}} \\
F(1\,|\,{\cal M}_3) &=& \frac{1}{4}.
\end{eqnarray}
Simulations of chords with this methods are 
reported in the plots of Figs.~\ref{fig:corde}-\ref{fig:corde_norot}. 

\section{Some ``ruler and compass'' methods}
Let us now see another two `geometric' methods in which
chords a drawn by opening at random 
an ideal compass (`ideal' because the minimum distance
between the end points is taken to be zero) 
from a reference point along the circumference.
(By `opening at random' we mean that the endpoints of the 
compass will define, uniformly, segments up to the diameter.
Changing the opening method we can then define two classes
of chord generators.)

\begin{figure}[!t]
\begin{center}
\epsfig{file=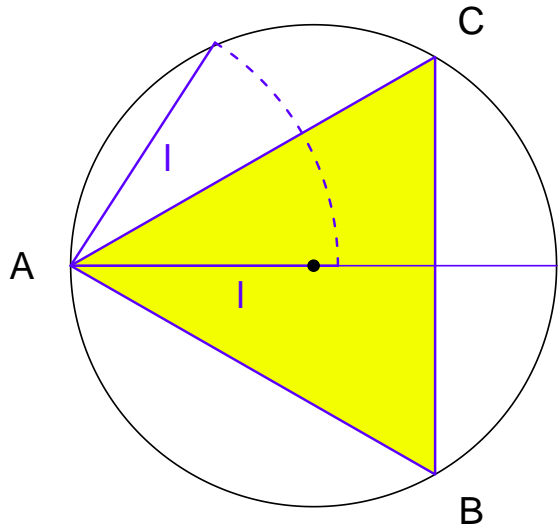,clip=,width=0.475\linewidth}
\end{center}
\caption{\small \sf A way to draw chords with lengths
uniformly distribute between 0 and $2R$.} 
\label{fig:bertrand_meth4}
\end{figure}
\subsection{Chords with length uniformly distributed between 0 and $2 R$}
It is curious that the method of simply taking chords with 
length uniformly distributed up to the length of the diameter
is usually not taken into account, although the idea 
is not bizarre at all. Indeed this could even be the {\em natural}
procedure to someone used to operate in classical geometry 
with ruler and compass: place the needle of the compass in a point of
the circumference (e.g. $A$ in Fig.~\ref{fig:bertrand_meth4}) 
to define one endpoint of the chord;
then place the pencil lead along the diameter impinging the 
circumference in $A$;
finally rotate the compass in either direction 
(anticlockwise in the figure) and find the second endpoint of 
the chord.

\begin{figure}
\begin{tabular}{c}
 {\large $[\,{\cal M}_4\,]$ }\\
\epsfig{file=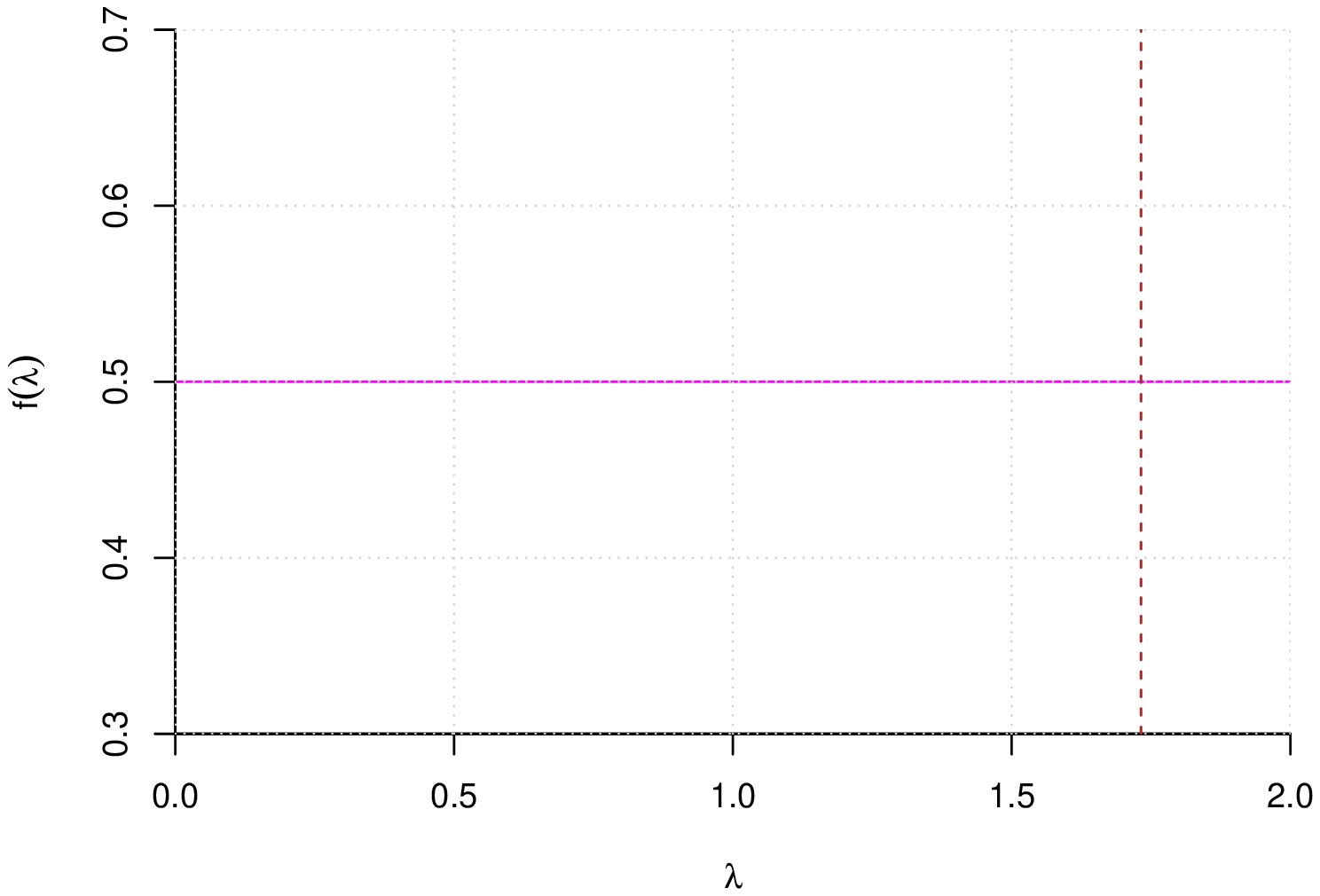,clip=,width=\linewidth} \\
\epsfig{file=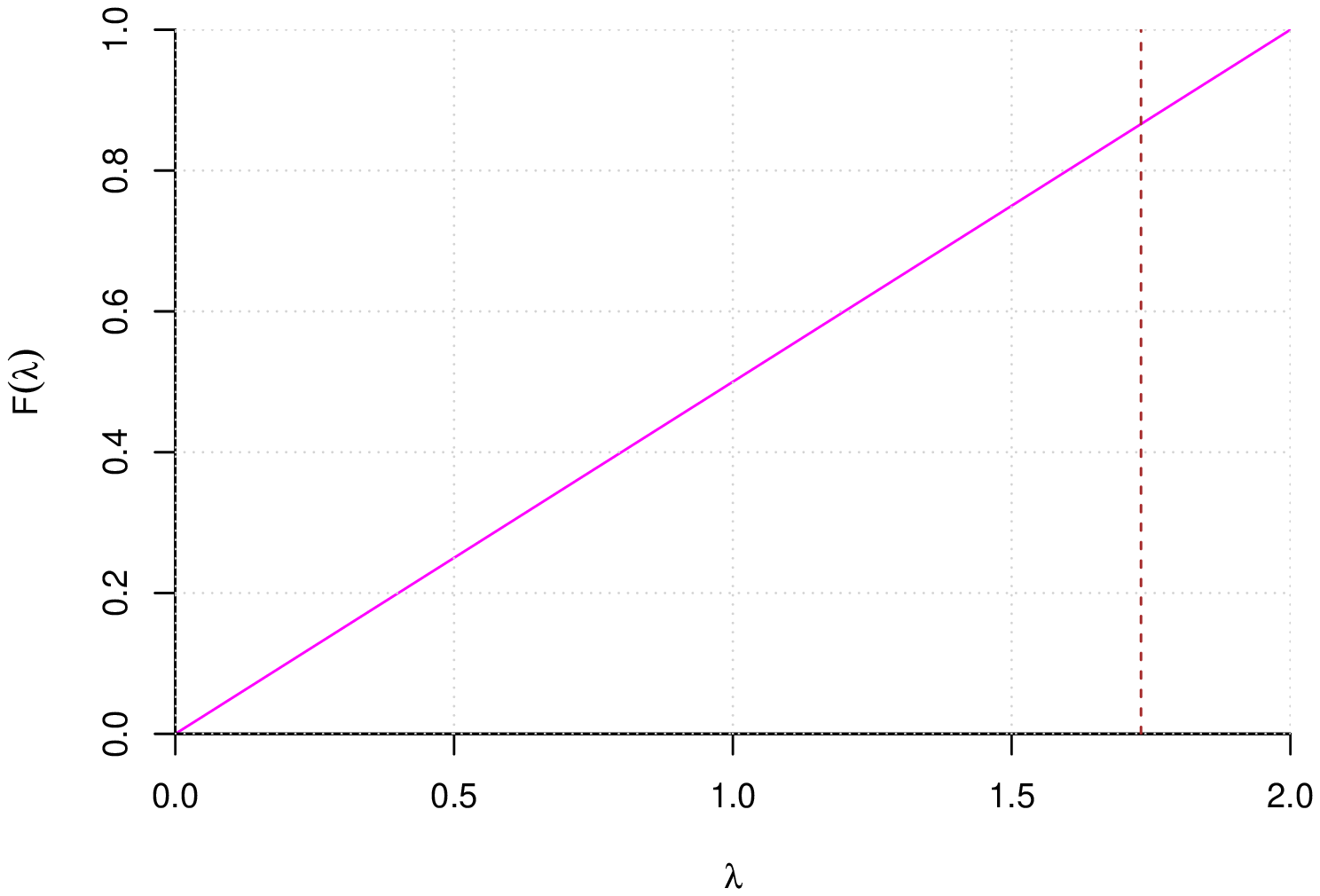,clip=,width=\linewidth}
\end{tabular}
\caption{\small \sf Probability distribution of $\lambda=l/R$ of the chords
generated with Method 4. The dashed vertical line indicates $\lambda=\sqrt{3}$.}
\label{fig:f_F_meth4}
\end{figure}

The probability distribution of $\lambda$ (see Fig.~\ref{fig:f_F_meth4})
as well as the 
probabilities of interests are in this case really trivial:
\begin{eqnarray}
f(\lambda\,|\,{\cal M}_4) &=& \frac{1}{2}
\hspace{0.7cm}  (0\le \lambda \le 2)\\
F(\lambda\,|\,{\cal M}_4) &=&  \frac{\lambda}{2}\\
F(\sqrt{3}\,|\,{\cal M}_4) &=& \mathbf{\frac{\sqrt{3}}{2}} \\
F(1\,|\,{\cal M}_4) &=& \frac{1}{2}.
\end{eqnarray}
Simulations of chords with this methods are 
reported in the plots of Figs.~\ref{fig:corde}-\ref{fig:corde_norot}. 

Nevertheless, although the results from this extraction model are
very easy, if we ask someone to make a computer 
program to really \underline{draw}
cords `at random', I would not bet 8.7 to 1.3 
(that is $\approx \sqrt{3}/2$
to $1-\sqrt{3}/2$) 
that a chord will be smaller than the side of the triangle! 
This point will become clear in
section~\ref{sec:drawing_program}.

\subsection{A variant of Method 4}
\begin{figure}
\begin{tabular}{cc}
\epsfig{file=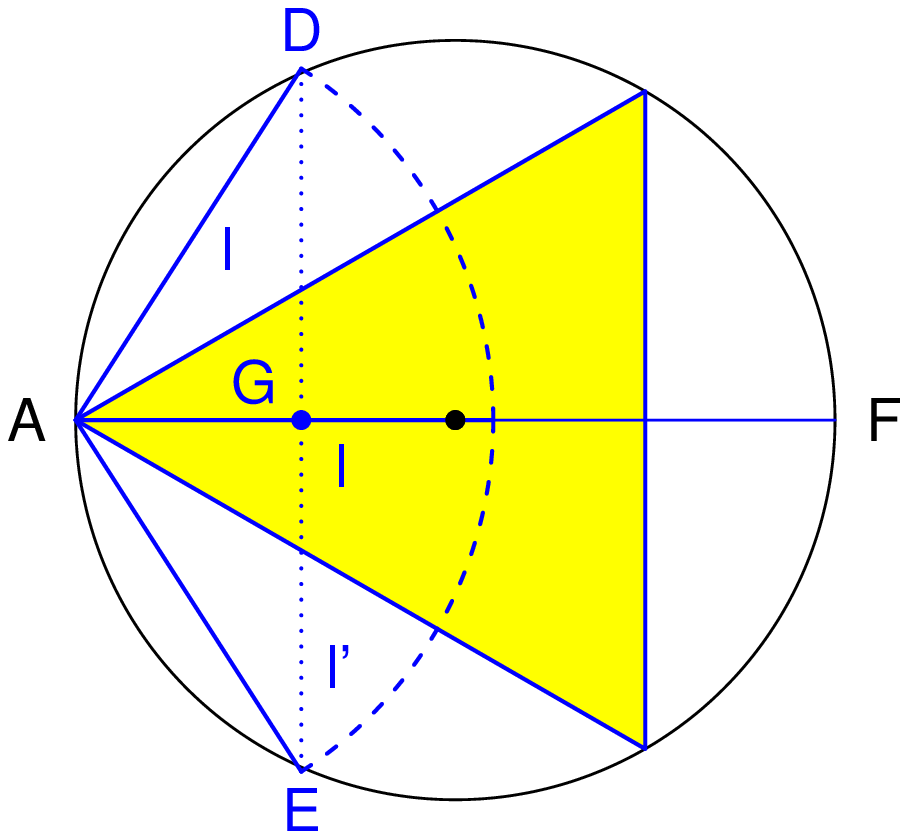,clip=,width=0.42\linewidth}\ \ & 
\epsfig{file=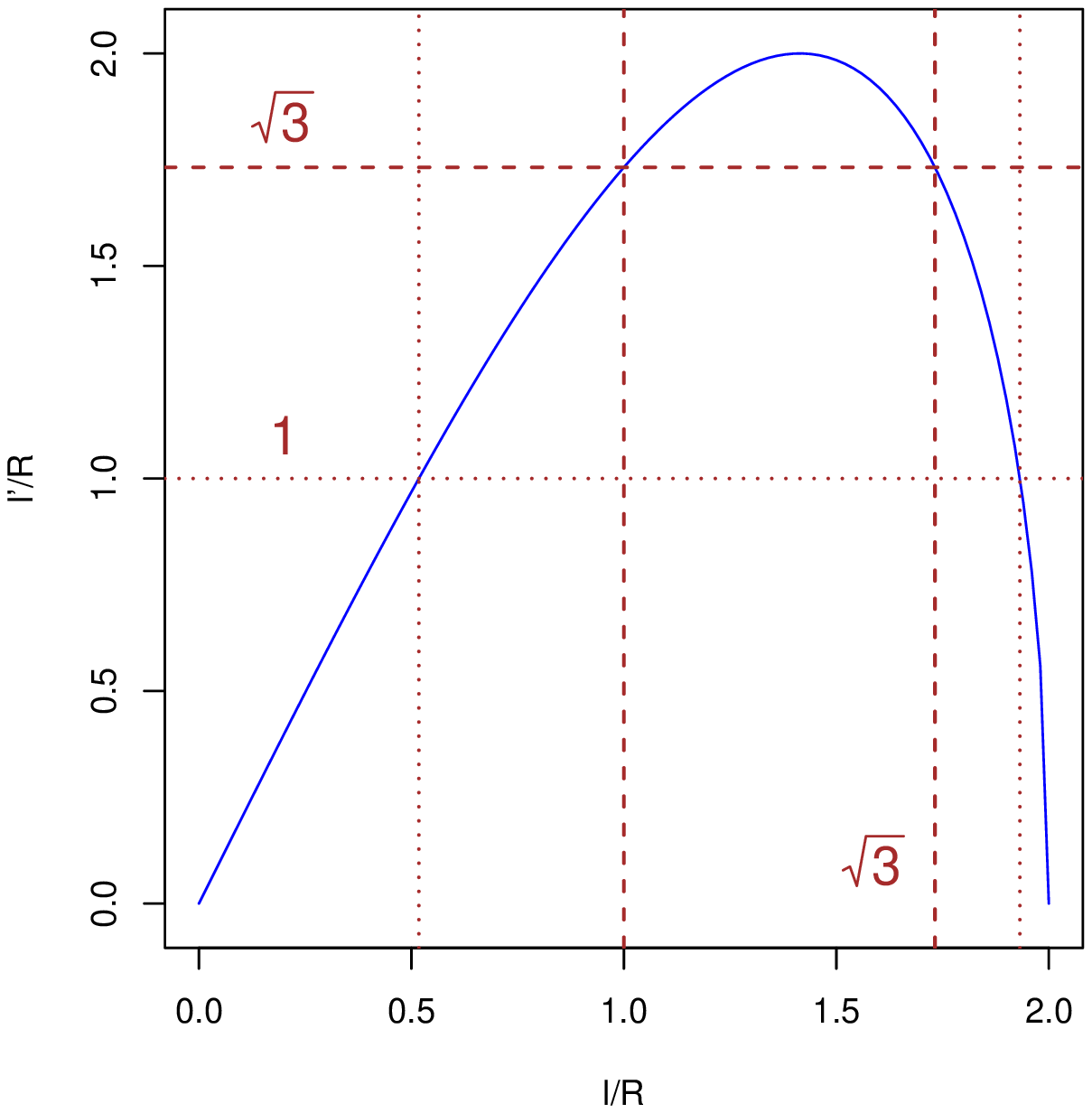,clip=,width=0.56\linewidth}
\end{tabular}
\caption{\small \sf  An alternative way to drawn chords
with ruler and compass. Left: construction of the chord. Right: 
relation between the $l$ and $l'$ (both in units of the radius).}
\label{fig:bertrand_meth4a}
\end{figure}

When using ruler and compass it is almost {\em automatic} that,
after having drawn an arc in one direction to intercept 
the circumference,
one rotates the tool the other direction, thus identifying the two
points $E$ and $D$ of Fig.~\ref{fig:bertrand_meth4a},
which then become the {\em natural} endpoints of a chord
(the reader understands that at this point the use of the adjective `natural'
is at limit of being sarcastic). The length $l'$ of this chord
is related to the segment $l$ defined\footnote{The points
$ADF$ define a rectangular triangle. The length of the 
segment $\overline{AG}$ is then equal to $l^2/2R$,
while the square of the length of $\overline{DG}$ is equal to
the product of the lengths of  $\overline{AG}$ and of  $\overline{GF}$.
It follows then
\begin{eqnarray*}
l'&=& 2\times l\sqrt{l^2/2R\cdot \left(2R-l^2/2R\right)}\,.
\end{eqnarray*}
} by the opening of the
compass by $l'=2\,l\,\sqrt{1-(l/2R)^2}$, or, in units
of the radius,
\begin{eqnarray}
\lambda' &=& 2\lambda\sqrt{1-(\lambda/2)^2}.
\end{eqnarray}
This relation is shown in the right plot of Fig.~\ref{fig:bertrand_meth4a},
from which we can calculate the usual probability that $\lambda'$ is 
smaller that $\sqrt{3}$. This is equal to the probability
that $l$ is smaller than 1, that is $1/2$, plus the probability
that it is larger that $\sqrt{3}$, that is $(2-\sqrt{3})/2$. 
The result is then $\mathbf{(3-\sqrt{3})/2} \approx 63.4$\%,
that do not correspond to none of the previous methods. 
Similarly, the probability that the chord is smaller than the radius,
or that $\lambda'<1$, is equal to the probability
that $\lambda$ is smaller than $\sqrt{2-\sqrt{3}}$, plus the 
probability that it is larger than  $\sqrt{2+\sqrt{3}}$,
that is\,\footnote{It is remarkable that $\sqrt{2+\sqrt{3}}-
\sqrt{2-\sqrt{3}} = \sqrt{2}$, an identity used to rewrite 
Eq.~(\ref{eq:P_lambda'_<=1}) in a more compact form.} 
\begin{eqnarray}
\frac{\sqrt{2-\sqrt{3}}}{2}+ \frac{2-\sqrt{2+\sqrt{3}}}{2} 
&=&1-\frac{1}{\sqrt{2}}
\approx 29.3\,\%\,. \label{eq:P_lambda'_<=1}
\end{eqnarray}
(This second calculation is important in order to double check the
result that we shall get later, being the resulting formulae
not very `pretty'.)

\begin{figure}
\begin{tabular}{c}
 {\large $[\,{\cal M}_5\,]$ }\\
\epsfig{file=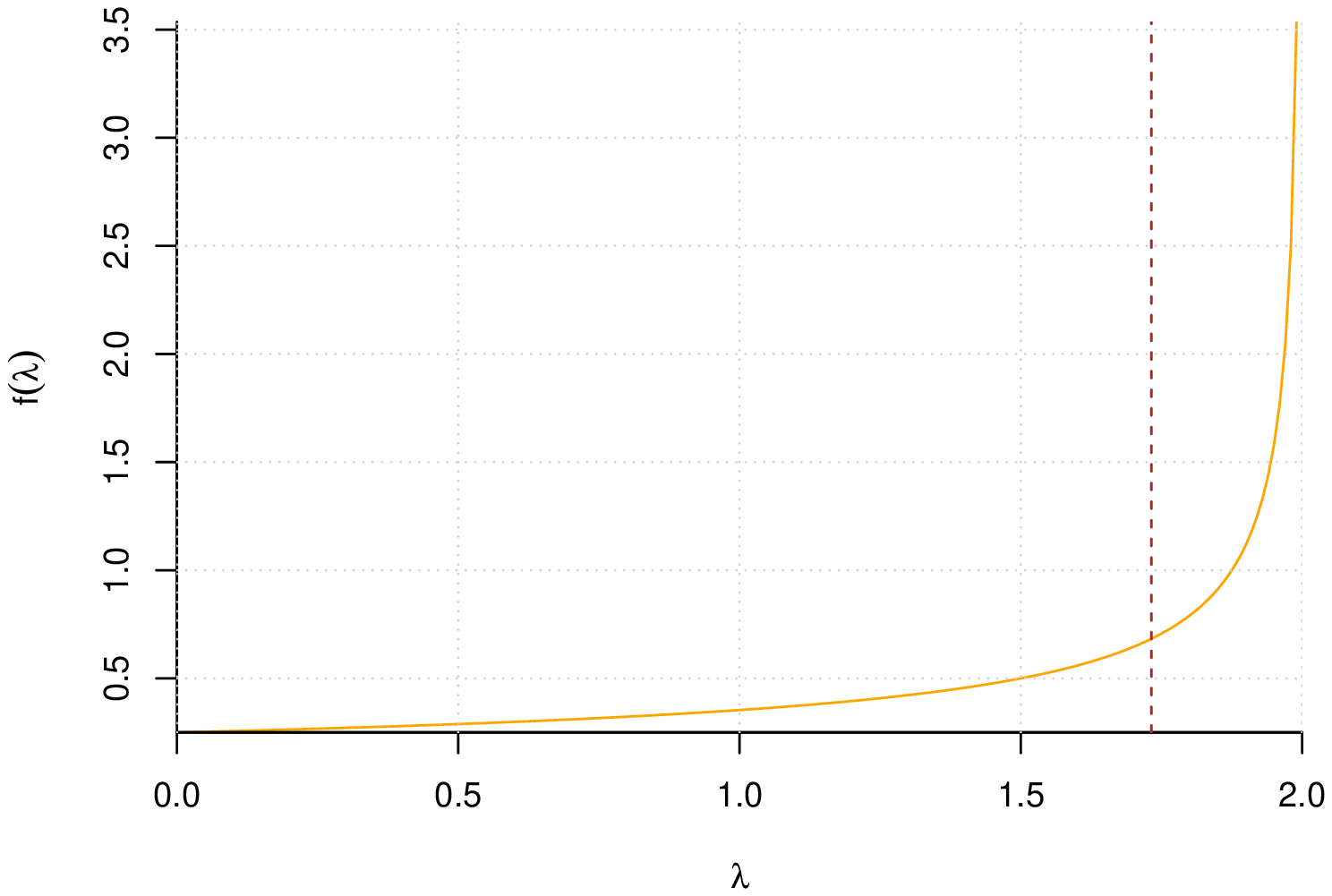,clip=,width=\linewidth} \\
\epsfig{file=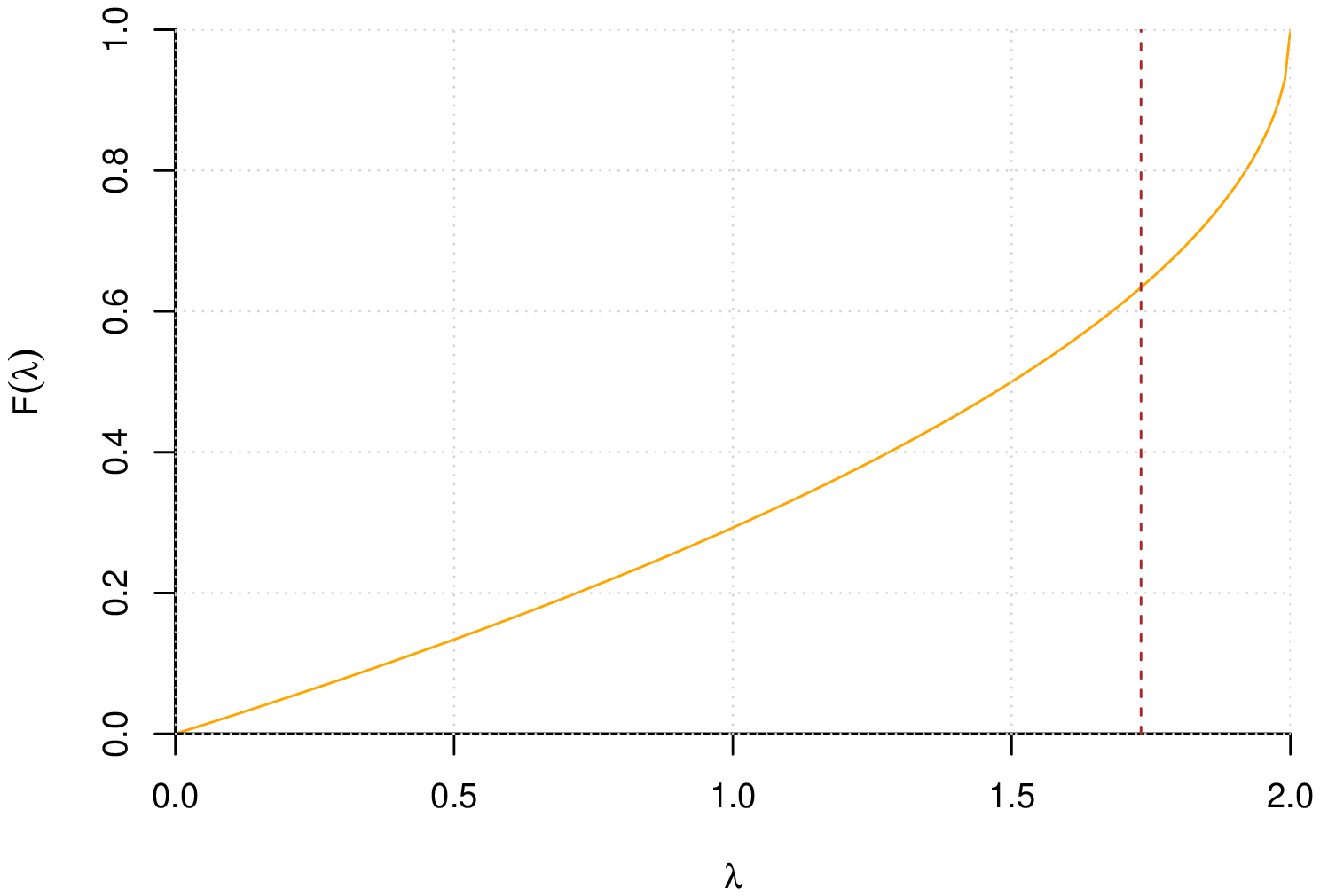,clip=,width=\linewidth}
\end{tabular}
\caption{\small \sf Probability distribution of $\lambda=l/R$ of the chords
generated with Method 5. The dashed vertical line indicates $\lambda=\sqrt{3}$.}
\label{fig:f_F_meth5}
\end{figure}

Also in this case it is possible to arrive to a closed
expression of the pdf. We do here the exercise mainly to show
a case in which the property of the Dirac delta needed
to make the transformation involves more than one root,
as evident from the non-monotonic relation shown in 
Fig.~\ref{fig:bertrand_meth4a}. Here are the details:
\begin{eqnarray*}
f(\lambda'\,|\,{\cal M}_5) &=& \int_0^2\!
\delta\left(\lambda'-2\lambda\sqrt{1-(\lambda/2)^2}\right)\cdot
\frac{1}{2}\,d\lambda \\
 &=&  \int_0^2\!\left(
\frac{\delta(\lambda-\lambda^*_1)}{|\frac{d}{d\lambda}(-2\lambda\sqrt{1-(\lambda/2)^2} )|_{\lambda=\lambda_1^*}} +
\frac{\delta(\lambda-\lambda^*_2)}{|\frac{d}{d\lambda}(-2\lambda\sqrt{1-(\lambda/2)^2} )|_{\lambda=\lambda_2^*}} 
\right)\cdot\frac{1}{2}\,d\lambda \\
\end{eqnarray*}
with
\begin{eqnarray}
\lambda_1^* &=& \sqrt{2-\sqrt{4-\lambda'^2}} \\
\lambda_2^* &=& \sqrt{2+\sqrt{4-\lambda'^2}}\,.
\end{eqnarray}
The result is, continuing to indicate in this subsection the chord of interest with $\lambda'$,
\begin{eqnarray}
f(\lambda'\,|\,{\cal M}_5) &=& 
\frac{\sqrt{2+\sqrt{4-\lambda'^2}} +\sqrt{2-\sqrt{4-\lambda'^2}} }
     {4\sqrt{4-\lambda'^2}}\,. \label{eq:flM5_da_Delta}
\end{eqnarray}
\newpage
The cumulative distribution can be obtained making the usual 
integral, although the integrand is in this case particularly 
nasty.\footnote{Wolfram Mathematica provides the following result: 
\begin{eqnarray*}
F(\lambda'\,|\,{\cal M}_5) &=& \frac{1}{2\lambda'}\left(
2\lambda' -2\sqrt{2-\sqrt{4-\lambda'^2}} - \sqrt{(4-\lambda'^2)\cdot
 (2-\sqrt{4-\lambda'^2})}\right.  \\
&& \left.
+2\sqrt{2+\sqrt{4-\lambda'^2}} - 
 \sqrt{(4-\lambda'^2)\cdot (2+\sqrt{4-\lambda'^2})}
\right)\,,
\end{eqnarray*}
that, although apparently quite different 
from Eq.~(\ref{eq:LlM5_da_Geometria}), can be checked to be
numerically equivalent to it.
\label{fn:IntegraleMathematica}
}. 
In reality the cumulative distribution can be calculated
extending the reasoning followed above to calculate the probability
that the chord is smaller than the side of the triangle and 
than the radius. In fact, remembering the transformation from $\lambda$
to $\lambda'$ plotted in Fig.~\ref{fig:bertrand_meth4a} and using
capital letters to distinguish the variables from their values, 
we have
\begin{eqnarray}
P(\Lambda' \le \lambda') &=& 
P\left(\Lambda \le \sqrt{2-\sqrt{4-\lambda'^2}}\right)
+ P\left(\sqrt{2+\sqrt{4-\lambda'^2}} \le \Lambda \le 2\right) \\
&=& \frac{\sqrt{2-\sqrt{4-\lambda'^2}}}{2}
   +  \frac{2-\sqrt{2+\sqrt{4-\lambda'^2}}}{2}\,,
\end{eqnarray}
from which it follows
\begin{eqnarray}
F(\lambda'\,|\,{\cal M}_5) &=& \frac{1}{2}\,\left(2 + 
\sqrt{2-\sqrt{4-\lambda'^2}} - \sqrt{2+\sqrt{4-\lambda'^2}}\,\right)\,,
\label{eq:LlM5_da_Geometria}
\end{eqnarray}
that, derivated, reproduces Eq.~(\ref{eq:flM5_da_Delta}). Pdf 
and cumulative functions are shown in Fig.~\ref{fig:f_F_meth5}.
The two probabilities of interest are 
then
\begin{eqnarray}
P(\lambda' \le \sqrt{3}\,|\,{\cal M}_5) &=& \mathbf{\frac{3-\sqrt{3}}{2}}
\approx 0.634 \\
P(\lambda' \le 1\,|\,{\cal M}_5) &=& 1-\frac{1}{\sqrt{2}} \approx 0.293\,.
\label{eq:Plambda'le1}
\end{eqnarray}
Results based on simulations are shown in 
Figs.~\ref{fig:corde}-\ref{fig:corde_norot}.

\subsection{Summary of the results obtained by the various methods}
\begin{figure}
\epsfig{file=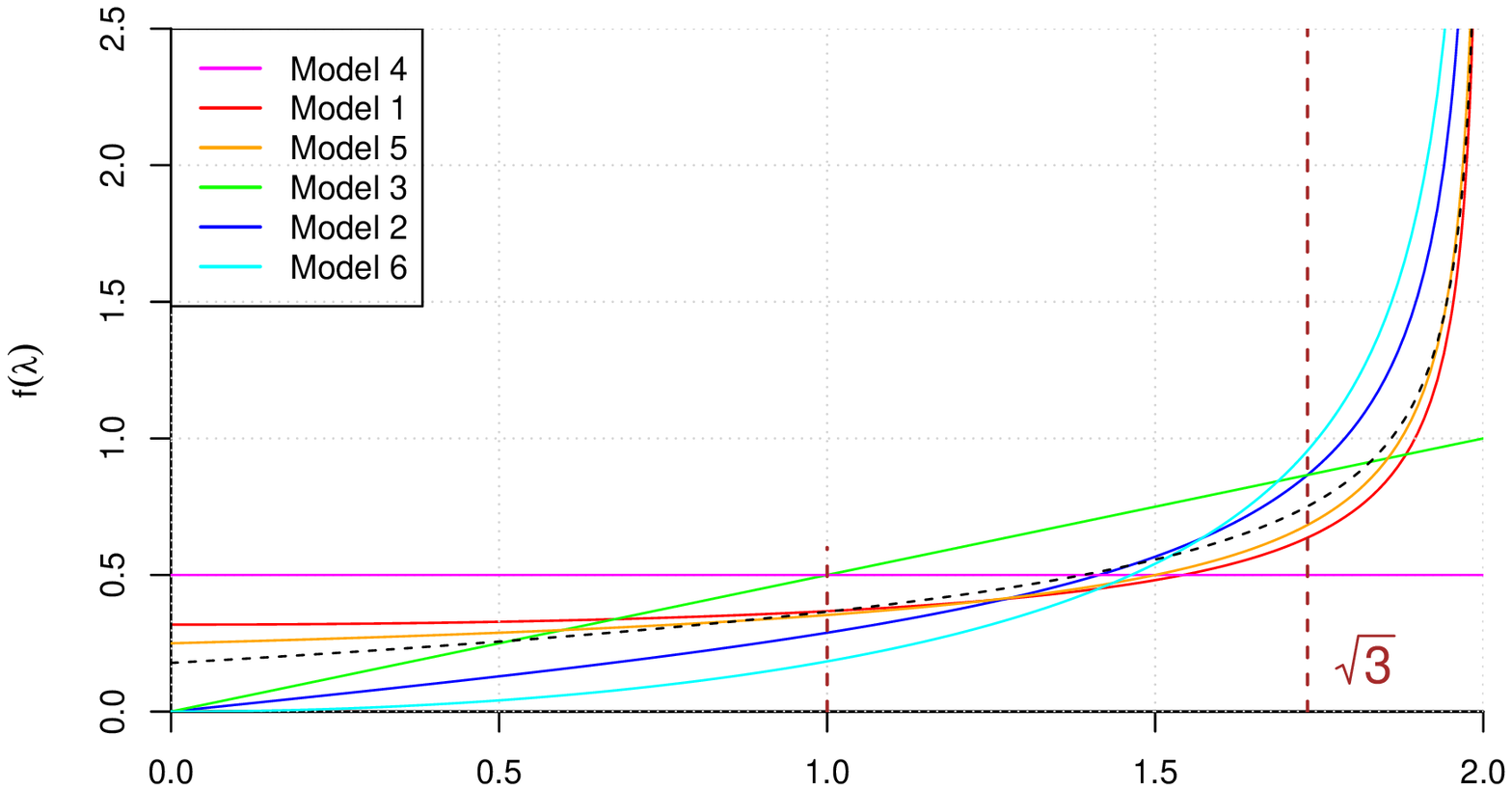,clip=,width=\linewidth}\\
\mbox{}\\
\epsfig{file=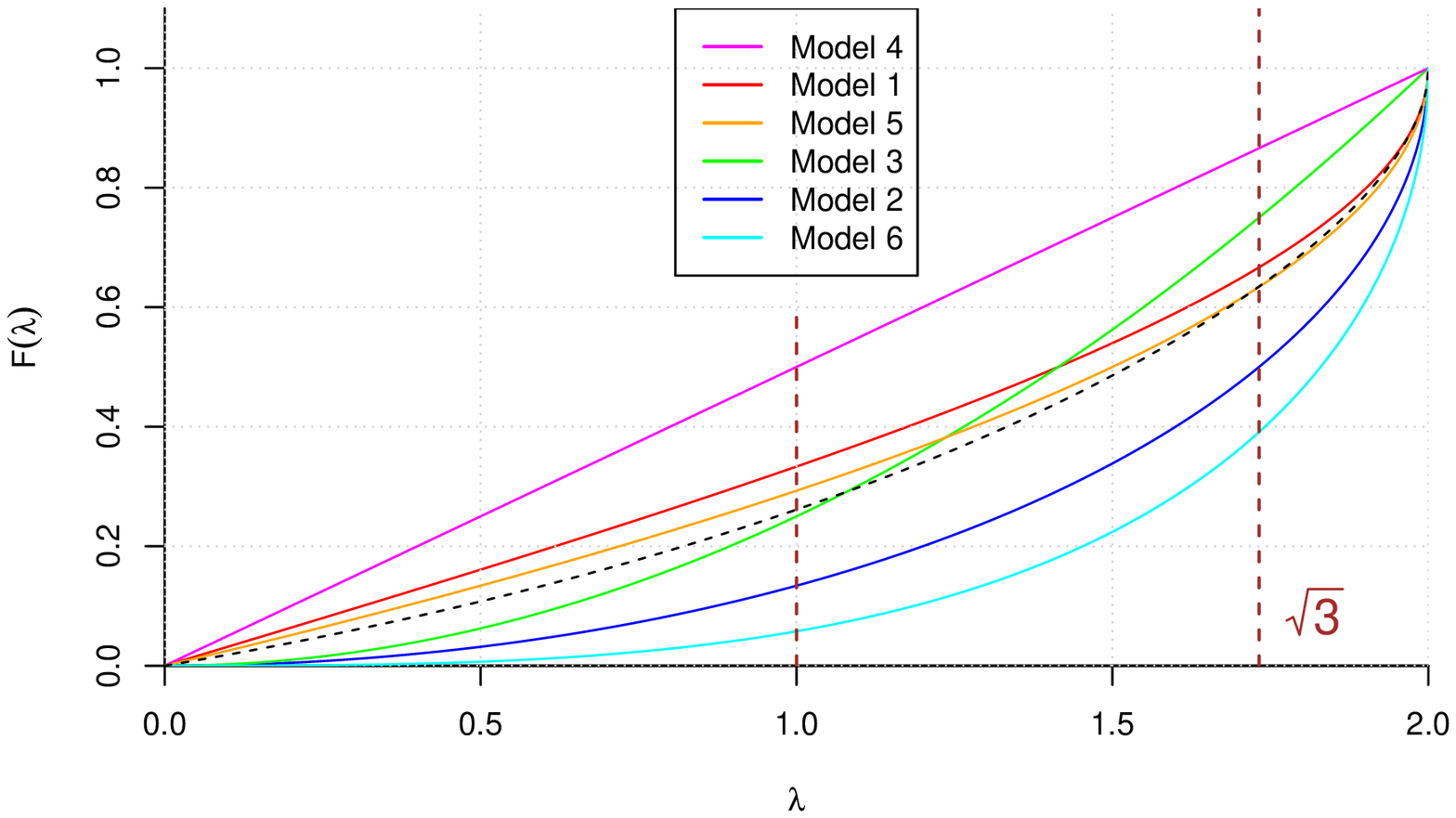,clip=,width=\linewidth}
\caption{\small \sf Probability density functions and 
cumulative function of the lengths of 
the chords in units of the radius of the circle for the
 extraction models considered in the text (the order of the models
in the legend corresponds to the decreasing order of 
$F(1\,|\,{\cal M}_i)$ and $f(0.1\,|\,{\cal M}_i)$).
The dashed lines correspond to the `averages', 
on which we shall come back in Appendix C.
} 
\label{fig:bertrand_All_F}
\end{figure}

The results obtained by the `geometric' methods 
 ${\cal M}_1$-${\cal M}_5$ are summarized in
Tab.~\ref{tab:bertrand_All_F} and in Fig.~\ref{fig:bertrand_All_F},
together with the more physical one ${\cal M}_6$,
that will be discussed in section \ref{sec:mikado}. 
In the table we have also added, for completeness, 
the expected values and the standard deviations
of the probability distributions.
More details, obtained from simulations, are shown in 
Figs.~\ref{fig:corde}-\ref{fig:corde_norot}. 
The dashed lines of  Fig.~\ref{fig:bertrand_All_F} represent
instead the arithmetic averages of the other functions, 
whose motivation will be clear in Appendix 3.
\begin{table}
\rotatebox{90}{
\begin{tabular}{c|cc|cc|cc}
Model  &  $f(\lambda)$ & $F(\lambda)$ & 
    $\mbox{E}(\lambda)$  & $\sigma(\lambda)$ &
     $\mathbf{P(\lambda\le \sqrt{3})}$ & $P(\lambda\le 1)$ \\
&&&&&& \\
\hline 
&&&&&& \\
 ${\cal M}_1$ &  {\large $\frac{1}{\pi\,\sqrt{1-(\lambda/2)^2}}$ }&
               {\large $\frac{2}{\pi}\cdot \arcsin{\frac{\lambda}{2}}$}& 
                $4/\pi$ &
                $\sqrt{2-16/\pi^2}$ & 
                 $\mathbf{2/3}$ & 
                  $1/3$ \\
&&&{\small ($\approx 1.27$)} & {\small ($\approx 0.62$)} &  
   {\small ($\approx 0.67$)} & {\small  ($\approx 0.33$)}\\
&&&&&& \\
${\cal M}_2$ &  {\large $\frac{\lambda}{4\,\sqrt{1-(\lambda/2)^2}}$ }&
               {\large $1- \sqrt{1-(\lambda/2)^2}$ } &
                 $\pi/2$ & 
                $\sqrt{8/3- \pi^2/4}$ &
                 $\mathbf{1/2}$ &
                $1-\sqrt{3}/2$\\ 
&&&{\small ($\approx 1.57$)} & {\small ($\approx 0.45$)} &  
   {\small ($= 0.5$)} & {\small  ($\approx 0.13$)}\\
&&&&&& \\
${\cal M}_3$ &   $\lambda/2$ &
                $\lambda^2/4$ &
                $4/3$ &
                 $\sqrt{2}/3$ &
                 $\mathbf{3/4}$ &
                  $1/4 $ \\
&&&{\small ($\approx 1.33$)} & {\small ($\approx 0.47$)} &  
   {\small ($= 0.75$)} & {\small  ($= 0.25$)}\\
&&&&&& \\
${\cal M}_4$ &  $1/2$ &  
               $\lambda/2$ & 
                $1$ &
                 $1/\sqrt{3}$ &
                  $\mathbf{\sqrt{3}/2}$ &
                 $1/2$ \\
&&&  & {\small ($\approx 0.58$)} &  
   {\small ($\approx 0.87$)} & {\small  ($= 0.5$)}\\
&&&&&& \\
${\cal M}_5$ & $[*]$ &  $[**]$ & 
                $4/3$ &
                 $4/(3\sqrt{5})$ &
                  $\mathbf{(3-\sqrt{3})/2}$ &
                 $1-1/\sqrt{2}$ \\
&  &&  
{\small ($\approx 1.33$)} &  {\small ($\approx 0.60$)} & 
     {\small ($\approx 0.63$)} &  {\small ($\approx 0.29$)}   \\
&&&&&& \\
${\cal M}_6$ & {\large $\frac{2}{\pi}\,\frac{(\lambda/2)^2}{\sqrt{1-(\lambda/2)^2}} $ }&
    $[***]$ & ${16}/{3\,\pi}$  & $\sqrt{3-{256}/{9\pi^2}}$ &   
   $\mathbf{{2}/{3} - {\sqrt{3}}/{2\,\pi}}$ &  ${1}/{3} - {\sqrt{3}}/{2\,\pi}$ \\
&  && {\small ($\approx 1.70$)} &  {\small ($\approx 0.34$)} &
        {\small ($\approx 0.39$)} &  {\small ($\approx 0.058$)}  \\
\hline 
&\multicolumn{6}{l}{} \\
& \multicolumn{6}{l}{[*]\ \ \ \  \, {\large $f(\lambda\,|\,{\cal M}_5)=
                 \left(\sqrt{2+\sqrt{4-\lambda^2}} + \sqrt{2-\sqrt{4-\lambda^2}}\right)/
     (4\sqrt{4-\lambda^2}) $}} \\
&\multicolumn{6}{l}{} \\
& \multicolumn{6}{l}{[**]\ \ \   {\large  $ F(\lambda\,|\,{\cal M}_5) = \frac{1}{2}\,\left(2 + 
\sqrt{2-\sqrt{4-\lambda^2}} - \sqrt{2+\sqrt{4-\lambda^2}}\,\right)$}} \\
&\multicolumn{6}{l}{} \\
& \multicolumn{6}{l}{[***]\ \  {\large $F(\lambda\,|\,{\cal M}_6) = 
\frac{2}{\pi}\,\arcsin{(\lambda/2)} - \frac{\lambda}{\pi}
\,\sqrt{1-(\lambda/2)^2} $} }
\end{tabular}
} 
\caption{{\small \sf Summary of the results obtained by the different `random'
drawing models of the chords} }

\label{tab:bertrand_All_F}
\end{table}

\newpage
\section{A chord length generator (and its implementation in R)}
\label{sec:ChordLengthGenerator}
In reality, if we are just interested in writing  
a computer program which generates the length of the chords `at random',
using one of the five methods we have seen, there is no need
to go through all the steps of the operational descriptions
of the algorithms. We can just use the probability distributions,
summarized in table \ref{tab:bertrand_All_F}. In order to do
that we need a premise, highlighted in the following 
subsection.

\subsection{A curious transformation and its practical importance}
\label{ss:InversionF}
Imagine to have a generic continuous variable $X$ whose uncertainty 
is described by the pdf $f_X(x)$ and the cumulative distribution 
$F_X(x)$ (in this subsection 
we shall use the more precise notation
introduced in footnote \ref{fn:TrasformazioneTextbook}). 
We might think at a new variable $Y$, related
to $X$ by $Y=F_X(X)$, that is
\begin{eqnarray}
y &=& g(x) = F_X(x)\,,
\end{eqnarray}
written in a way to stress that in this case $F_X()$ plays now
the role of the generic mathematical 
function $g()$, independently of its probabilistic meaning.

Making use of Eq.~\ref{eq:GeneraRuleTransf_xy}
the pdf of $Y$ can be evaluated as
\begin{eqnarray}
f_Y(y)  &=& \int_{-\infty}^{+\infty}\!\delta\left( y-F_X(x)\right)\cdot 
       f_X(x)\,dx\,.
\end{eqnarray}
Being  $F_X()$ monotonic and not-decreasing with derivative equal to $f_X()$,
 we have then 
\begin{eqnarray}
f_Y(y)  &=& \int_{-\infty}^{+\infty}\!\frac{\delta(x-x^*)}{F_X'(x^*)}\cdot 
       f_X(x)\,dx = \frac{f_X(x^*)}{f_X(x^*)} = 1\,.
\end{eqnarray}
This is a {\bf great result}, that we double check following
the same reasoning used in footnote \ref{fn:TrasformazioneTextbook}:
\begin{eqnarray*}
F_Y(y)  \equiv P(Y\le y) &=& P(Y\le F_X(x)) = 
  P(X\le F_X^{-1}(y)) \equiv   F_X\left(F_X^{-1}(y)\right) = y \\
f_Y(y) \equiv \frac{d}{dy}\,F_Y(y) &=& \frac{d}{dy}\,y = 1\,. 
\end{eqnarray*}
Independently of the pdf $f_X()$, the variable $Y$ 
defined in this way is uniformly
distributed between 0 and 1. It follows that the pdf of 
a variable defined as 
$X=F_X^{-1}(Y)$ will be $f_X(x)$. 

This observation
suggests a simple algorithm, very useful in those cases in which 
the cumulative function is {\it easy to be inverted}, 
to make a (pseudo) random number generator to produce
numbers such that our confidence on the occurrence
of $X=x$ is proportional to $f(x)$:
\begin{eqnarray}
x &=& F_X^{-1}(u)\,,\label{eq:inversione_F}
\end{eqnarray}
where  $u$ stands for the occurrence of   
a uniform (pseudo) random generator that gives
 numbers (apparently)
`at random' between 0 and 1  
(random number generators of this kind are available
in all computational environments.)

\subsection{Application to the chord problem}\label{ss:rlchords}
Fortunately we can apply this trick to all 
probability distributions of the chords found in the previous 
sections. The generic rule (\ref{eq:inversione_F}) 
gets then implemented as follows.
\begin{description}
\item{$\mathbf{\cal M}_1$:}\hspace{0.2cm} 
$\lambda = 2 \sin{\left({\pi u}/{2}\right)}$\,.
\item{$\mathbf{\cal M}_2$:}\hspace{0.2cm} $\lambda = 2 \sqrt{2 u - u^2}$\,.
\item{$\mathbf{\cal M}_3$:}\hspace{0.2cm} $\lambda = 2 \sqrt{u}$\,.
\item{$\mathbf{\cal M}_4$:}\hspace{0.2cm} $\lambda = 2 u$\,.
\item{$\mathbf{\cal M}_5$:}\hspace{0.2cm} $\lambda = 2 \sqrt{1-2 u^2+u^4}$\,.
\end{description}
And this is then, for example,
the resulting function written in the R language~\cite{R}:
\begin{verbatim}
rlchords <- function(n, meth) {
  u <- runif(n)
  switch(meth,
         2*sin(pi*u/2),
         2*sqrt(2*u-u^2),
         2*sqrt(u),
         2*u,
         2*sqrt(1-2*u^2+u^4))
}
\end{verbatim}
Issuing then the following instruction from 
an R console ('{\small $>$}' stands for the prompt) 
you can for example get an  histogram similar to the
left top one of Fig.~\ref{fig:l}:\\ 
\Rin{lambda=rlchords(100000,1); hist(lambda, nc=200, col='red')}\\
Or you can calculate sample mean and standard deviation, and fraction of 
occurrences with $\lambda$ smaller than $\sqrt{3}$
with these very simple commands\\
\Rin{mean(lambda)}\\
\Rin{sd(lambda)}\\
\Rin{length(lambda[lambda<sqrt(3)])/length(lambda)}\\
(You will get sample mean and standard deviation 
very close to the mean and standard deviation 
of the distribution {\em because it is very unlikely} 
to have something very different, as famous a theorem of
probability theory makes us feel confident.)

\newpage
\section{What should one expect from a computer drawing program?} 
\label{sec:drawing_program} 
{\small
\begin{flushright}
{\sl ``Mater artium necessitas''}
\end{flushright}
} 
\noindent
The original Bertrand problem is about {\bf drawing} chords
and not just telling numbers between 0 and 2 (taking a unitary
radius). Therefore the question we have to ask
is really to \underline{draw} the chord, by hand of by a 
computer program.\footnote{Nevertheless, the lengths 
provided by the `chord generator' presented
in the previous section do provide valid answers, since
the pdf's have been derived using some geometric rules
to produce chords and not with an abstract algorithm
to produce numbers between 0 and 2, like those 
you would get e.g. with the following R command\\
\Rin{n=10; lambda = 2*sin(runif(n, 0, pi))\^\,2 }
}
In the introduction I have 
told what I  more or less expect when I ask the practical
question, providing a sheet of paper with a pre-designed circle,
and I must say that in the last years I had no surprises
that induced me to change the model I formed in my mind. 
You may form yours with practice. 

More recently I have also asked PhD students to write
``chords generators'' with their preferred computer 
language. 
As it easy to guess, the choice goes to the algorithm
easier to implement, which for physics students is Method 1,
since they are familiar with circular motion and with 
transformations from polar to Cartesian coordinates. 

The other methods are somehow tedious because, if taken 
{\it literally}, they require several steps with formulae
not used everyday, that one needs to derive. For example
Method 2, requires {\it literally}: 1) to choose a radius
at random; 2) to chose a point on the radius; 
3) find the equation of the line orthogonal to the radius
in that point; 4) find the interceptions of the line 
with the circle. Indeed -- I have to confess with some shame --
the first time I was playing with the Bertrand problem,
I was implementing in R these detailed procedures. 
When during this year course I tried to make an Android app to draw 
`random' chords on a circle, using App Inventor\,\cite{AI2},
I was horrified by the formulae I had to `write' with that
tool. So I initially implemented only Method 1. After a while
I also implemented Method 2, but \underline{not} 
following the procedures described above. The trick was
to extract a point along the horizontal diameter, 
thus coinciding with the abscissa, and operate then
a random rotation. In that way it was possible  
to reuse somehow the `blocks' (the graphical 
programming elements shown e.g. in  Fig.~\ref{fig:AppInventorM2})
developed for Method 1. 

The rest of this section is devoted to simulation issues,
showing how to avoid pedantic procedures and without pretending 
that the suggested algorithms are the `best' in some sense
that should be better defined. (My suggestion to students
is that the for everyday use the `fastest' algorithm
is the one that they write more rapidly and understand better --
unless you need it for special purposes, it is a waste of time 
to spend several hours of your life to write a piece of program 
that provides the result in microseconds instead then in 
tens, hundreds or even thousands of milliseconds). 
 
\subsection{Model 1 (${\cal M}_1$)}
As stated above, this is the one that appears the simplest
(to implement in a program) 
to physics students, to most colleagues and to myself. 
Here is how it appears in R (\code{n} is the number of chords, 
that should be set in a previous command).\\
\Rin{ph1 <- runif(n, 0, 2*pi)}\\
\Rin{ph2 <- runif(n, 0, 2*pi)}\\
\Rin{p1 <- cbind(cos(ph1), sin(ph1))}\hspace{1.0cm}{\tt \# [1]}\\
\Rin{p2 <- cbind(cos(ph2), sin(ph2)}\\
\Rin{l <- sqrt( (p2[,1]-p1[,1])\^\,2 + (p2[,2]-p1[,2])\^\,2)}\\
The result is a `vector'  \code{l} of \code{n} lengths of chords
(in units of the radius). Plus we have the matrices of 
interception points (each raw is a point).

\subsection{Model 2 (${\cal M}_2$)}
\begin{figure}
\begin{center}
\epsfig{file=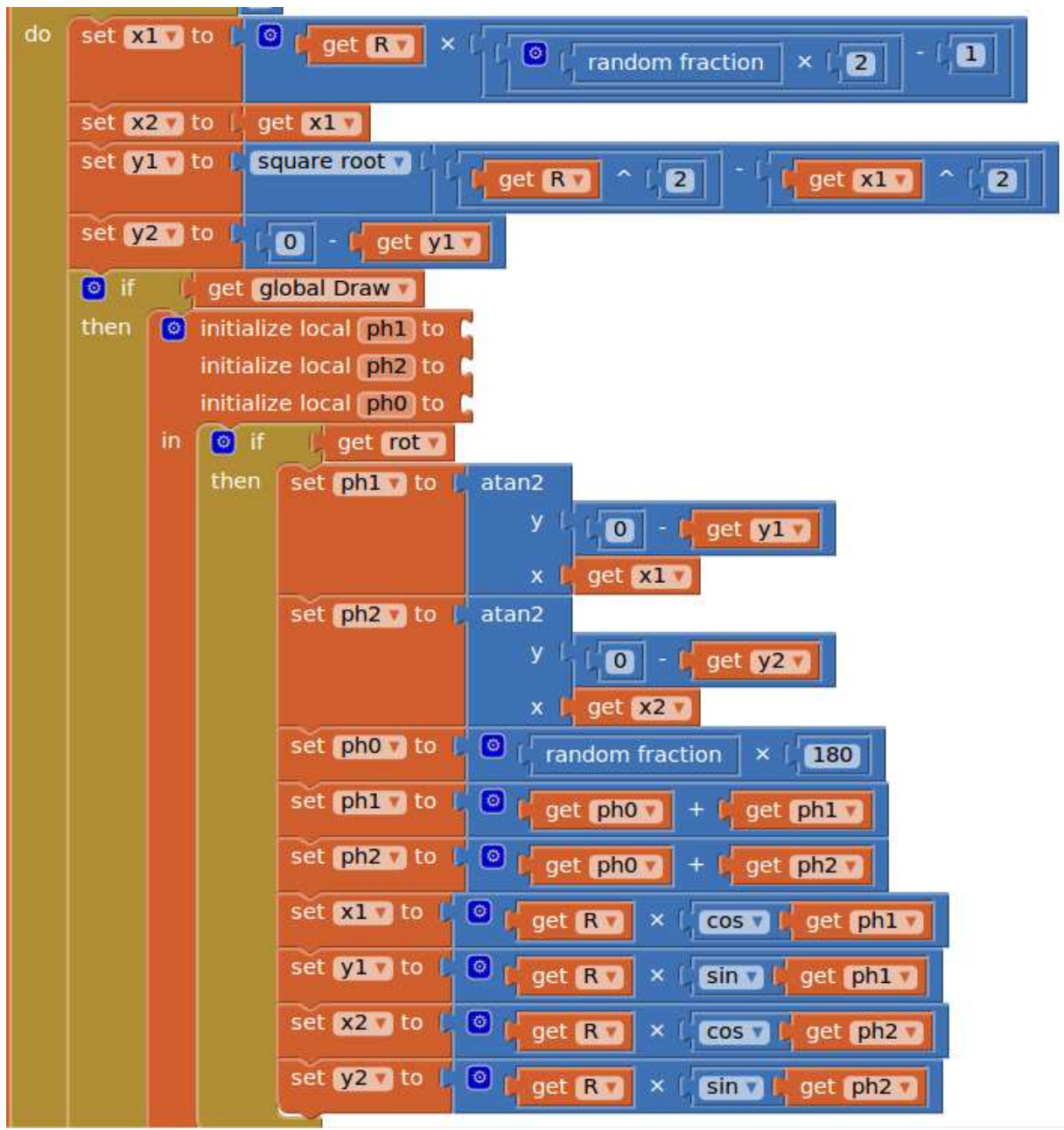,clip=,width=1.05\linewidth} 
\end{center}
\caption{\small \sf App Inventor blocks for the core of Method 2. 
(The minus sign of the {\tt y} argument of {\tt atan2} is due
to the fact that  in App Inventor the $y$ coordinate 
inside a  `canvas' 
is upside down, being the origin in the top left corner, while
the `heading' angle follows the usual convention. Note
also how, contrary to other scientific 
libraries, the trigonometric functions use degrees.)}
\label{fig:AppInventorM2}
\end{figure}

In this case we start extracting $x_1=x_2$ between 
$-1$ and $1$ and evaluate the corresponding ordinates
on the circumference, i.e.\\ 
\Rin{p1 <- p2 <- runif(n, -1, 1)}\\
\Rin{p1 <- cbind(p1, sqrt(1-p1\^\,2))}\hspace{1.0cm}{\tt \# [2]}\\
\Rin{p2 <- cbind(p2, -sqrt(1-p2\^\,2))}\\
Then we define a random rotation angle and add
it to the polar angles calculated from the points:\\
\Rin{phr <- runif(n, 0, pi)}\\
\Rin{ph1 <- phr + atan2(p1[,2], p1[,1])}\\
\Rin{ph2 <- phr + atan2(p2[,2], p2[,1])}\\
At this point we can reuse exactly the last three lines
of code of the previous method, i.e. starting from the
line tagged by ``{\tt \# [1]}''.
The resulting App Inventor blocks are shown 
in Fig.~\ref{fig:AppInventorM2}.

\subsection{Model 3 (${\cal M}_3$)}\label{ss:simulation_M3}
\begin{figure}
\begin{center}
\epsfig{file=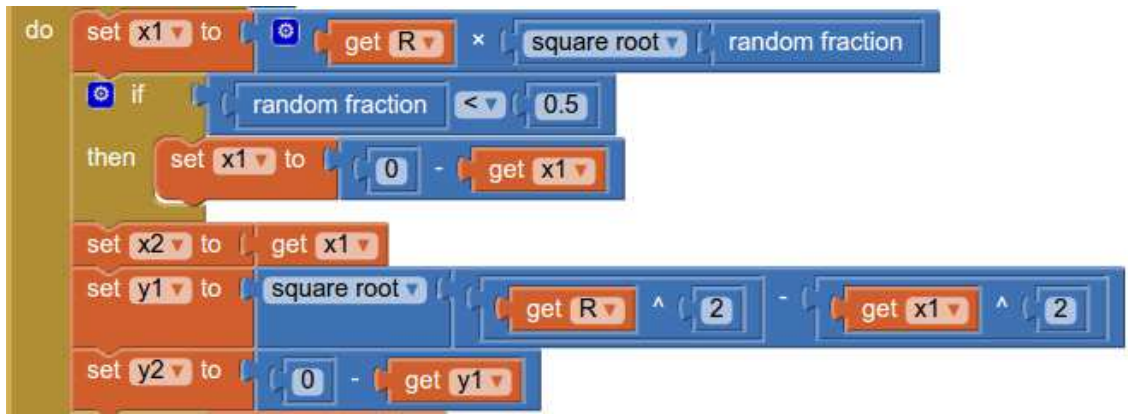,clip=,width=\linewidth} 
\end{center}
\caption{\small \sf App Inventor blocks to turn Method 2 into 
Method 3 (see text).}
\label{fig:AppInventorM3}
\end{figure}

To chose a point uniformly inside the circle we could
extract uniformly $x$ and $y$ between $-1$ and $1$
and discard the points which are outside the 
circle of radius 1. But we can use of the previous code
(or App Inventor blocks) if we extract a point along the radius
with pdf $f(\rho) = 2\rho$, as we have learned in subsection 
\ref{ss:Meth3}, making use of the technique learned in 
subsection \ref{ss:InversionF}. \\
\Rin{rho <- sqrt(runif(n))}\\
But, in order to reuse the previous code, we have to invert 
at random (with probability 1/2) the sign of this numbers.
Technically this can be done in R creating a vector
of random $-1$'s and $1$'s obtained by a binomial generator
and multiplying it,  element by element, with the vector {\tt rho}. 
Thus our starting abscissas will be\\
\Rin{p1 <- p2 <-  rho * ( rbinom(n, 1, 0.5)*2 - 1 )}\\
After this, we continue exactly as in the second line
of R code of the previous method, 
tagged by ``{\tt \# [2]}''. The 
implementation of this variation in App Inventor
is shown in Fig.~\ref{fig:AppInventorM3}.

\subsection{Model 4 (${\cal M}_4$)}
\begin{figure}[!b]
\begin{center}
\epsfig{file=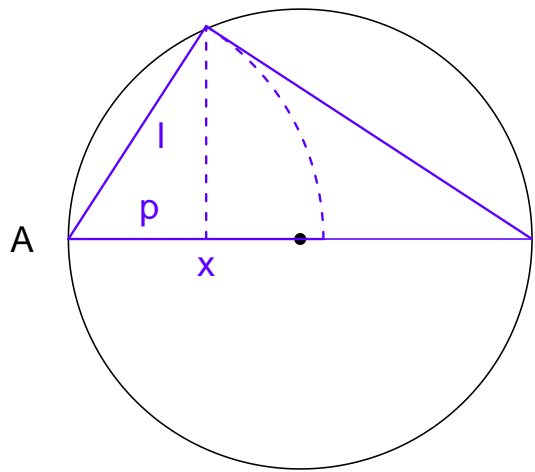,clip=,width=0.455\linewidth}
\end{center}
\caption{\small \sf A sketch which show how to calculate the abscissa of the 
the interception ($x$) by simple geometry (see text). 
} 
\label{fig:bertrand_meth4_Euclide}
\end{figure}

Also in the case of the fourth method, we can reuse
the code written for Method 2, without having to calculate
the intersections of two circles. In fact the first
endpoint is $(-R,0)$, while the second can be easily 
found using elementary geometry.
With the help of 
Fig.~\ref{fig:bertrand_meth4_Euclide} 
we can recognize a useful rectangular triangle 
and then make use of a famous theorem 
of  geometry. The projection $p$
is then $l^2/2R$ and then the abscissa $x_2$ 
of the intersection is equal to $x_2=p-R=l^2/2R-R$. 
The corresponding ordinate is therefore
$y_2=\sqrt{R^2-x_2^2}$. Then we can rotate `at random'
the points as done above, when we described Method 2. 
These are the two lines of R code that replace the 
\underline{first} line of code above\\
\Rin{p1 <- rep(-1, n)}\\
\Rin{p2 <- runif(n, 0, 2)\^\,2/2 - 1}\\
and from the second line, tagged by ``{\tt \# [2]}'', 
it will be all the same.

\subsection{Model 5 (${\cal M}_5$)}
The difference with the previous model is that 
now the abscissa of the two points (before rotation!)
are the same, while the ordinates are one the opposite
opposite of the other, as in Model 2. We have then \\
\Rin{p2 <- p1 <- runif(n, 0, 2)\^\,2/2 - 1}\\
and then we continue from the second line of the 
code of Model 2,  tagged by ``{\tt \# [2]}''.\mbox{}\\  \mbox{}

\begin{figure}[h]
\begin{center}
\epsfig{file=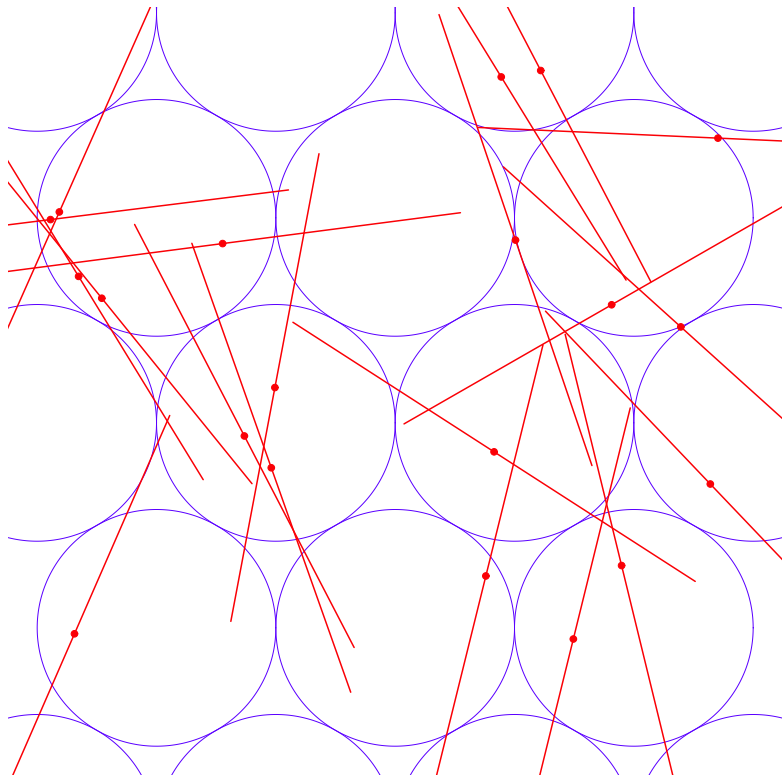,clip=,width=0.67\linewidth}
\end{center}
\caption{\small \sf  Mikado sticks thrown at random 
on a pattern of circles.} 
\label{fig:mikado_1}
\end{figure}
\mbox{}\vspace{-1.2cm}
\section{A physical model: throwing mikado 
sticks on a pattern of circles}\label{sec:mikado}
The authors of Ref.~\cite{DiPorto} claim to have found 
a `conclusive physical solution' to the Bertrand problem, but I have strong 
doubts that they have ever tried 
to implement their model in a real experiment. 
Something which seems to me more realistic is the kind
of game sketched in Fig.~\ref{fig:mikado_1}:
\begin{figure}
\begin{center}
\epsfig{file=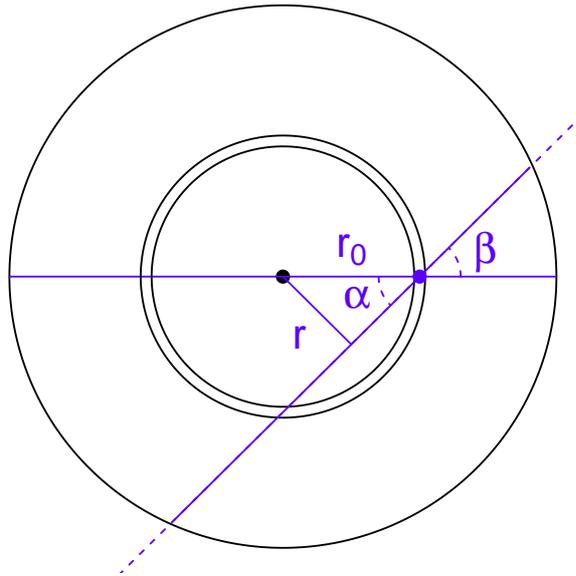,clip=,width=0.5\linewidth}
\end{center}
\vspace{-0.3cm}
\caption{\small \sf Sketch of the mikado experiment to
show how to evaluate the distance of the chord from the center from
the position of the center of the stick and its orientation (see text).} 
\label{fig:bertrand_mikado_1}
\end{figure}
a pattern of circles\,\footnote{In Fig.~\ref{fig:mikado_1}
all circles have the same size, but this is not a necessary 
requirement, as it will be clear in a while.} 
on a table, or on the floor,
on which we throw mikado sticks, or toothpicks, or needles or something
similar. The only, obvious, conditions is on the minimum
length of the sticks, that has to be at least twice the 
maximum diameter of the circles.\footnote{Also this condition
is not necessary, if we think to prolong the stick in either
end by a ruler to draw the chords. And, finally, it is not even required
that the reference point, needed to decide inside which 
circle the chord has to be drawn, has to coincide 
with the center of the stick.
The formulation in the text is, or at least so seems to me,
the easiest to be implemented in a real `game', 
similar to the famous ``Buffon's needle''.}

If we throw `ad random' the sticks, somehow towards the center
of the pattern in order to avoid complications with 
boundary conditions, 
we expect their centers and their orientations `uniformly'
distributed (the former in the plane, the latter
in angle w.r.t. a given direction).
We consider only sticks whose reference point, marked somehow, is inside 
one circle and consider the resulting chords defined
by the intersection of the stick with the circumference
of that circle.
As we can see from Fig.~\ref{fig:mikado_1}, the efficiency is rather high 
(and, as a byproduct, we can try to estimate empirically 
the area of the regions between circles, but this is 
a different story\ldots).

Evaluating the pdf of the expected chords length might
be complicated, but fortunately we can make use of some 
of our previous results. Let us indicate now with $r_0$ the distance
from the center of the stick to the center of the circle
and with  $r$ the distance
of the chord from the center of the circle
(see Fig.~\ref{fig:bertrand_mikado_1}). 
The respective quantities in units of 
$R$ are then  $\rho_0=r_0/R$  and  $\rho=r/R$. 
 
By the hypothesis inherent to the throwing 
mechanism the center of the stick is
uniformly distributed in the circle. This implies,
as we already know, that high values of $\rho_0$ are more
probable than small ones, namely
$f(\rho_0\,|\,{\cal M}_6)=2\rho_0$. The fractional distance 
$r$ is related to $r_0$ by  $r=r_0\sin\alpha$, where
the angle $\alpha$, defined in the 
construction in Fig.~\ref{fig:bertrand_mikado_1}, 
is uniformly distributed 
between 0 and $\pi/2$.  
These are then our starting conditions\,\footnote{An alternative way
would be to use the angle $\beta$ defined in 
Fig.~\ref{fig:bertrand_mikado_1}, ranging
between $0$ to $\pi$, with $f(\beta\,|\,{\cal M}_6) =1/\pi$. 
The angle $\alpha$ inside the rectangular triangle will be equal to
$\beta$ if $\beta$ is smaller than $\pi/2$, and $\pi-\beta$ elsewhere.
The relation (\ref{eq:rho_da_rho0}) would then 
be replaced by $\rho = \rho_0\,\sin\beta$
and the integral (\ref{eq:int_delta_d_alpha}) replaced 
on the equivalent one in $d\beta$ between 0 and $\pi$,
with the factor $2/\pi$ in the integrand replaced by 
 $1/\pi$,  {\em apparently} leading to results differing by
a factor of 2. This apparent contradiction is resolved
noted that transformation rule of the Dirac delta has now two roots,
$\beta^*_1 = \arcsin(\rho/\rho_0)$ and 
$\beta^*_2 = \pi - \arcsin(\rho/\rho_0)$. 
But, since $|\cos\beta^*_1| = |\cos\beta^*_2|$, we have two 
identical contributions, thus exactly compensating
the missing factor 2.}
\begin{eqnarray}
\rho &=& \rho_0\,\sin\alpha \label{eq:rho_da_rho0}\\
f(\rho_0\,|\,{\cal M}_6) &=& 2\rho_0\hspace{1.7cm}(0\le \rho_0\le 1) \\
f(\alpha\,|\,{\cal M}_6) &=& \frac{1}{\pi/2} = \frac{2}{\pi}\hspace{0.7cm}(0\le \alpha \le \frac{\pi}{2})\,.
\end{eqnarray}
We can then calculate the probability distribution 
of $\rho$ and  make then use of the reasoning applied in Method 3
to evaluate the probability distribution of $\lambda$.  
Indeed, for a given $\rho_0$, the
pdf of $\rho$, conditioned by the value of $\rho_0$, 
will be given by
\begin{eqnarray}
f(\rho\,|\,{\cal M}_6,\rho_0) &=& \int_{0}^{\pi/2}\!\delta(\rho-\rho_0\,\sin\alpha)\cdot f(\alpha)\,d\alpha  \label{eq:int_delta_d_alpha}\\
&=& \int_{0}^{\pi/2}\frac{\delta(\alpha-\alpha^*)}{\rho_0\,\cos{\alpha^*}}
\cdot  \frac{2}{\pi}\,d\alpha\,,
\end{eqnarray} 
with $\alpha^* = \arcsin(\rho/\rho_0)$, from which it follows
\begin{eqnarray}
f(\rho\,|\,{\cal M}_6,\rho_0) &=& \frac{2}{\pi\,\rho_0\sqrt{1-(\rho/\rho_0)^2}}\\
 &=& \frac{2}{\pi\,\sqrt{\rho_0^2-\rho^2}}\,.
\end{eqnarray} 
Having $f(\rho\,|\,{\cal M}_6,\rho_0)$ and 
$f(\rho_0\,|\,{\cal M}_6)$ we can then evaluate 
 $f(\rho\,|\,{\cal M}_6)$ as
\begin{eqnarray}
f(\rho\,|\,{\cal M}_6) &=& \int_\rho^1 
f(\rho\,|\,{\cal M}_6,\rho_0)\cdot f(\rho_0\,|\,{\cal M}_6)\,d\rho_0\,,
\end{eqnarray} 
in which we have to pay attention to the condition
$\rho\le \rho_0$. The result is 
\begin{eqnarray}
f(\rho\,|\,{\cal M}_6) &=& \int_\rho^1 
\frac{2}{\pi\,\sqrt{\rho_0-\rho^2}} \cdot 2\rho_0 \,d\rho_0\\
 &=& \int_\rho^1 
\frac{4\,\rho_0}{\pi\,\sqrt{\rho_0^2-\rho^2}} \,d\rho_0\\
&=& \frac{4}{\pi}\,\sqrt{1-\rho^2}\,.\label{eq:pdf_rho_M6}
\end{eqnarray} 
Having calculated the pdf of the distances of the chords
from the center of the circle, we continue as in 
Eq.~(\ref{eq:trasf_M3_inizio}), thus obtaining
\begin{eqnarray}
f(\lambda\,|\,{\cal M}_6) &=& \int_0^1
\!\delta\left(\lambda -  2 \sqrt{1-\rho^2}\right)\cdot 
\frac{4}{\pi}\,\sqrt{1-\rho^2} \,d\rho \\
&=& \int_0^1
\!\frac{\delta\left(\rho-\rho^*\right)}
{\left.2\rho^2/\sqrt{1-\rho^2}\right|_{\rho=\rho^*}}
\cdot 
\frac{4}{\pi}\,\sqrt{1-\rho^2} \,d\rho \\
&=& \frac{4/\pi\,\sqrt{1-{\rho^*}^2}}{2\rho^*/\sqrt{1-{\rho^*}^2}} 
= \frac{2}{\pi}\,\frac{1-{\rho^*}^2}{\rho^*}\,,
\end{eqnarray}
with the usual $\rho^*=\sqrt{1-(\lambda/2)^2}$. 
We get finally 
\begin{eqnarray}
f(\lambda\,|\,{\cal M}_6) &=& 
\frac{2}{\pi}\,\frac{(\lambda/2)^2}{\sqrt{1-(\lambda/2)^2}}\\
F(\lambda\,|\,{\cal M}_6) &=& 
\frac{2}{\pi}\,\arcsin{(\lambda/2)} - \frac{\lambda}{\pi}
\,\sqrt{1-(\lambda/2)^2}\,, \label{eq:M6F_lambda}
\end{eqnarray}
\begin{figure}
\begin{tabular}{c}
 {\large $[\,{\cal M}_6\,]$ }\\
\epsfig{file=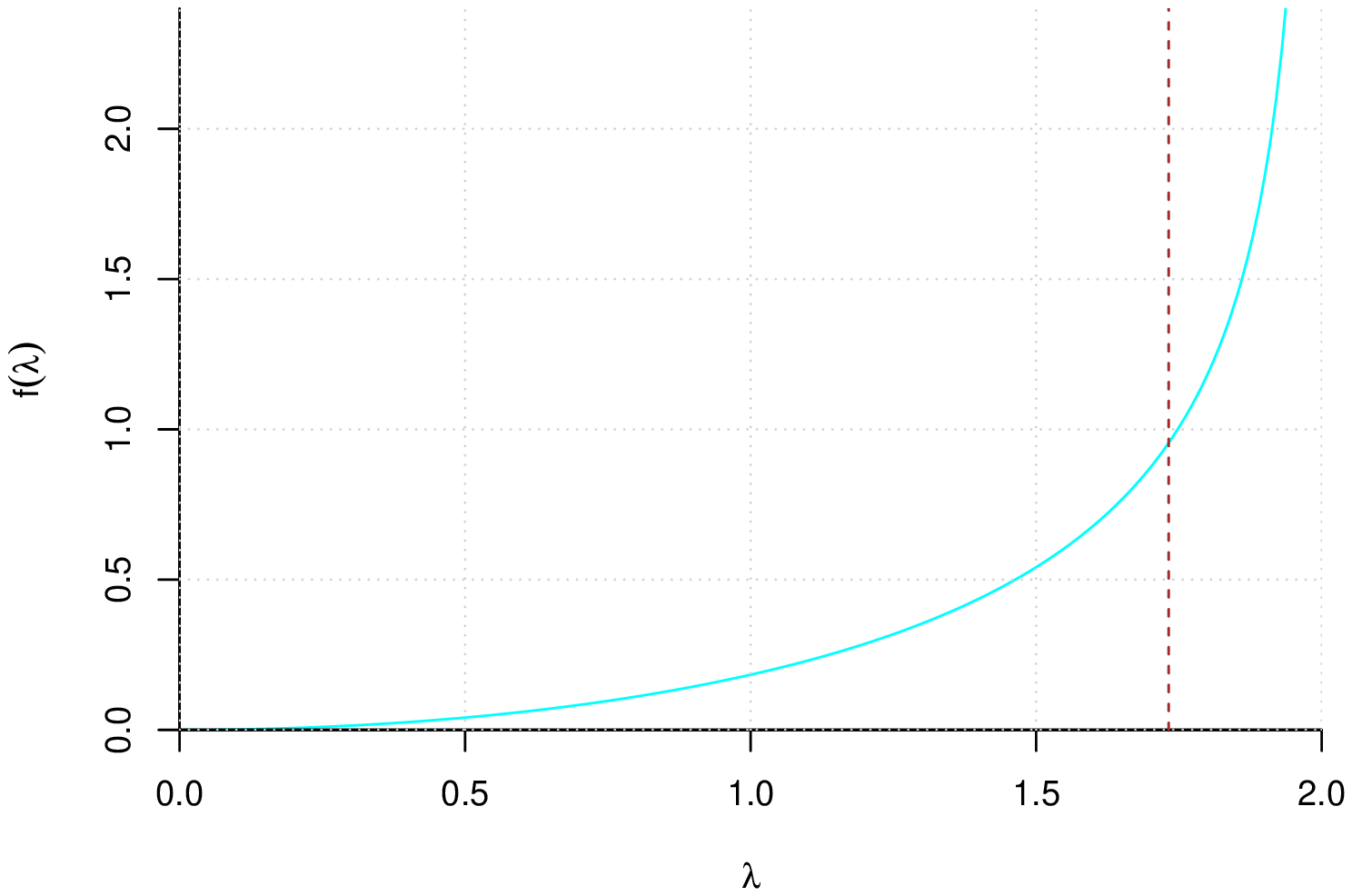,clip=,width=\linewidth} \\
\epsfig{file=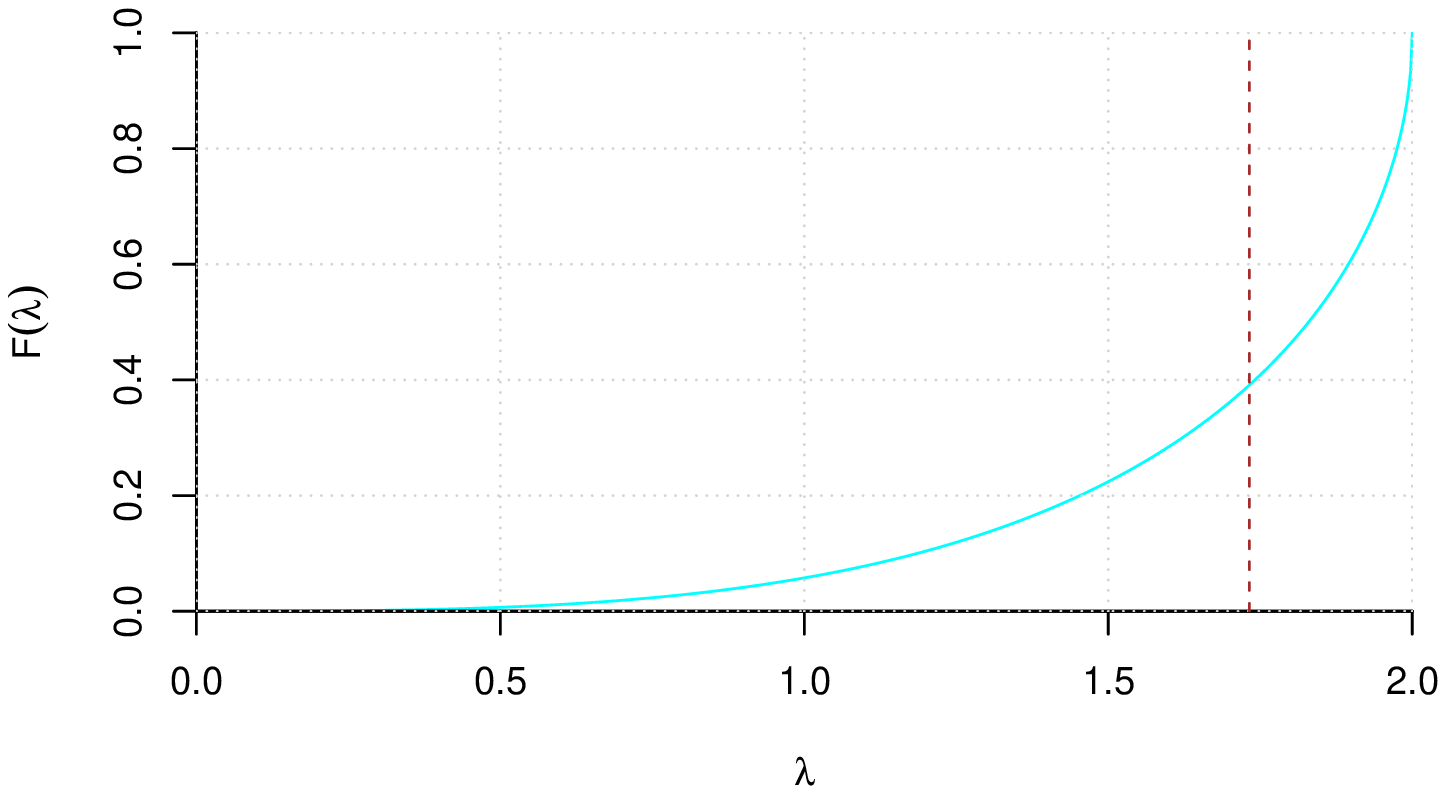,clip=,width=\linewidth}
\end{tabular}
\caption{\small \sf Probability distribution of $\lambda=l/R$ of the chords
generated with Method 6 (`mikado'). 
The dashed vertical line indicates $\lambda=\sqrt{3}$.}
\label{fig:f_F_meth6}
\end{figure}
shown in Fig.~\ref{fig:f_F_meth6} and 
from which we can calculate the usual indicators
\begin{eqnarray}
\mbox{E}(\lambda\,|\,{\cal M}_6) &=& \frac{16}{3\,\pi} 
         \approx  1.70\\
\sigma(\lambda\,|\,{\cal M}_6)  &=& \sqrt{3-\frac{256}{9\,\pi^2}}
\approx 0.34 \\
P(\lambda \le \sqrt{3}\,|\,{\cal M}_6) &=&
\frac{2}{3} - \frac{\sqrt{3}}{2\,\pi} \approx 0.39 \\
P(\lambda \le 1\,|\,{\cal M}_6) &=&
\frac{1}{3} - \frac{\sqrt{3}}{2\,\pi} \approx 0.058\,. 
\end{eqnarray}

\subsection{Remarks on simulation}
We have seen in section \ref{sec:ChordLengthGenerator}
how to write chord length generators for the different
methods without having to go through the steps of drawing the chords,
while we have learned in section \ref{sec:drawing_program} 
how to reuse the code to draw the chords. In this case the situation
seems a bit more complicate, but there are fortunately
some ways out.

\subsubsection{Random chord length generator}
\begin{figure}
\begin{tabular}{cc}
\epsfig{file=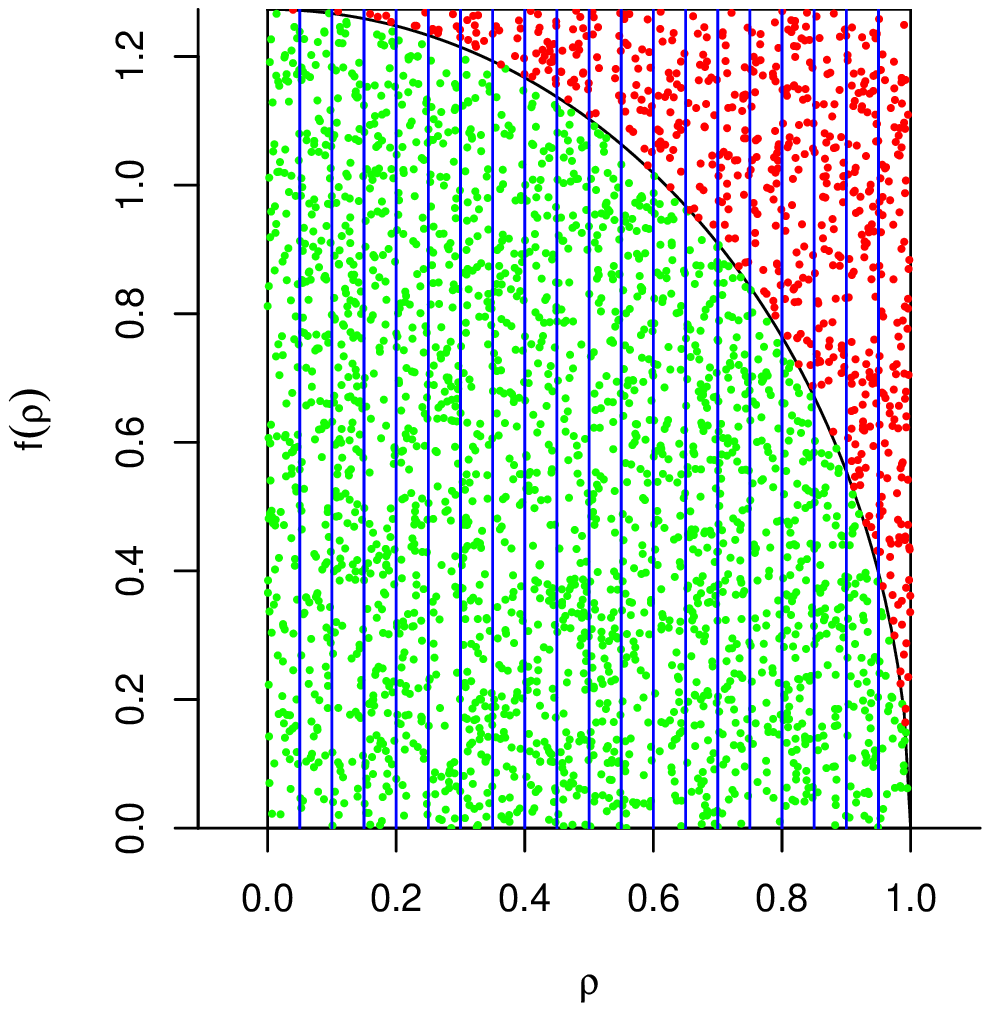,clip=,width=0.49\linewidth} &
\epsfig{file=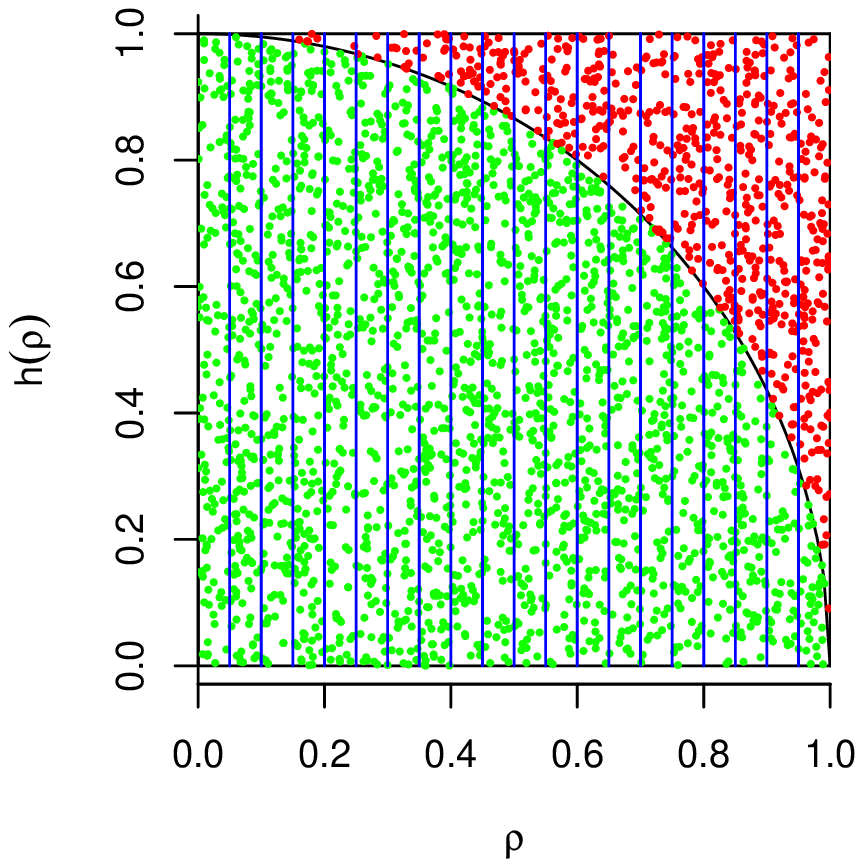,clip=,width=0.49\linewidth} 
\end{tabular}
\caption{\small \sf Extracting $\rho$ according to its
pdf using the hit/miss Monte Carlo technique.
(But we shall make direct extractions inside the
quarter of circle! See text.)}
\label{fig:hit-miss}
\end{figure}

The generators described in subsection \ref{ss:rlchords}
were based on the trick of inverting 
the cumulative distribution. However Eq.~(\ref{eq:M6F_lambda}) is not easily
invertible and we cannot use that trick.\footnote{One might
think to use the {\it hit/miss} method, described 
in the text in the following lines.
But this is also
not possible to be used since $f(\lambda)$ is divergent for 
$\lambda\rightarrow 2$, although its integral is finite.}
 We could then go
one step behind, and try to generate the values of $\rho$,
normalized distance of the chords from the center, and then 
calculate, from each $\rho$, the corresponding $\lambda$. 
But also in this case
the cumulative distribution of interests is not invertible, 
being equal to 
\begin{eqnarray}
F(\rho,|\,{\cal M}_6)&=& \frac{2}{\pi}\,\left(\rho\,\sqrt{1-\rho^2}
+\arcsin\rho\right)\,.
\end{eqnarray}
We could make use of the {\em hit/miss} method, simply
introduced with the help of Fig.~\ref{fig:hit-miss}.
The left plot shows (black curve) the pdf $f(\rho)$ inside
a rectangle defined by the range of $\rho$ and the range
of its pdf. We then throw `at random' points uniformly
inside the rectangle and mark in green those which
fall between $f(\rho)$ and the abscissa, in red those 
which fall `outside'. We draw then some vertical lines
to identify `slides' in $\rho$. We see that the number 
of the green point in each slice is proportional to
 $f(\rho)$ calculated in the middle of the slide. 
We can imagine then the slides very thin, with the usual
procedures done in calculus, to convince ourselves that
the probability that a green point falls inside a 
slide is proportional to $f(\rho)$. This is a well known 
technique to make extractions according to a given
pdf, at the expenses of some inefficiency, which in our
case is tolerable (just the fraction of red points over the total). 
It becomes instead not tolerable if the pdf is very peaked
somewhere and assumes very small values in the rest of
the range of the variable. Or it becomes impossible to be used 
when the pdf diverges and the rectangle become infinite 
high, as it would be with $f(\lambda\,|\,{\cal M}_6)$.

To conclude, in our case the hit/miss technique would 
be appropriate, but we can do even better, if we observe that
the factor $4/\pi$ in Eq.~(\ref{eq:pdf_rho_M6}) 
is irrelevant for the reasoning. 
If we drop it, we are left with $\sqrt{1-\rho^2}$, which describe
a circle in the first quadrant, as shown in the right plot of
Fig.~\ref{fig:hit-miss}, with the ordinate indicated now by $h(\rho)$,
being a generic `height' and not a pdf. 
But we  already know how to extract points uniformly
inside a circle without having to throw points in the square 
circumscribed! What we need is just to limit the extraction in 
a quarter of the circle, and this can be easily achieved 
limiting the polar angle between 0 and $\pi/2$.
Here is directly the R command to generate a single value of $\rho$,
followed by the corresponding value of $\lambda$\\
\Rin{rho=sqrt(runif(1))*cos(runif(1,0,pi/2))}\\
\Rin{lambda <- 2 * sqrt(2 - rho\^\,2)}\\
and finally in a single step, with $n$ values:\\
\Rin{lambda <- 2 * sqrt(2 - runif(n)*cos( runif(n,0,pi/2)\^\,2 )}\\
Here is then the new version of our generator with 
all six models implemented:
\begin{verbatim}
rlchords <- function(n, meth) {
  u <- runif(n)
  switch(meth,
         2*sin(pi*u/2),
         2*sqrt(2*u-u^2),
         2*sqrt(u),
         2*u,
         2*sqrt(1-2*u^2+u^4),
         2*sqrt(1-u*cos(runif(n, 0, pi/2))^2))
}
\end{verbatim}

\subsubsection{Random chord drawer}
\begin{figure}
\begin{tabular}{|c|c|}
\hline
\epsfig{file=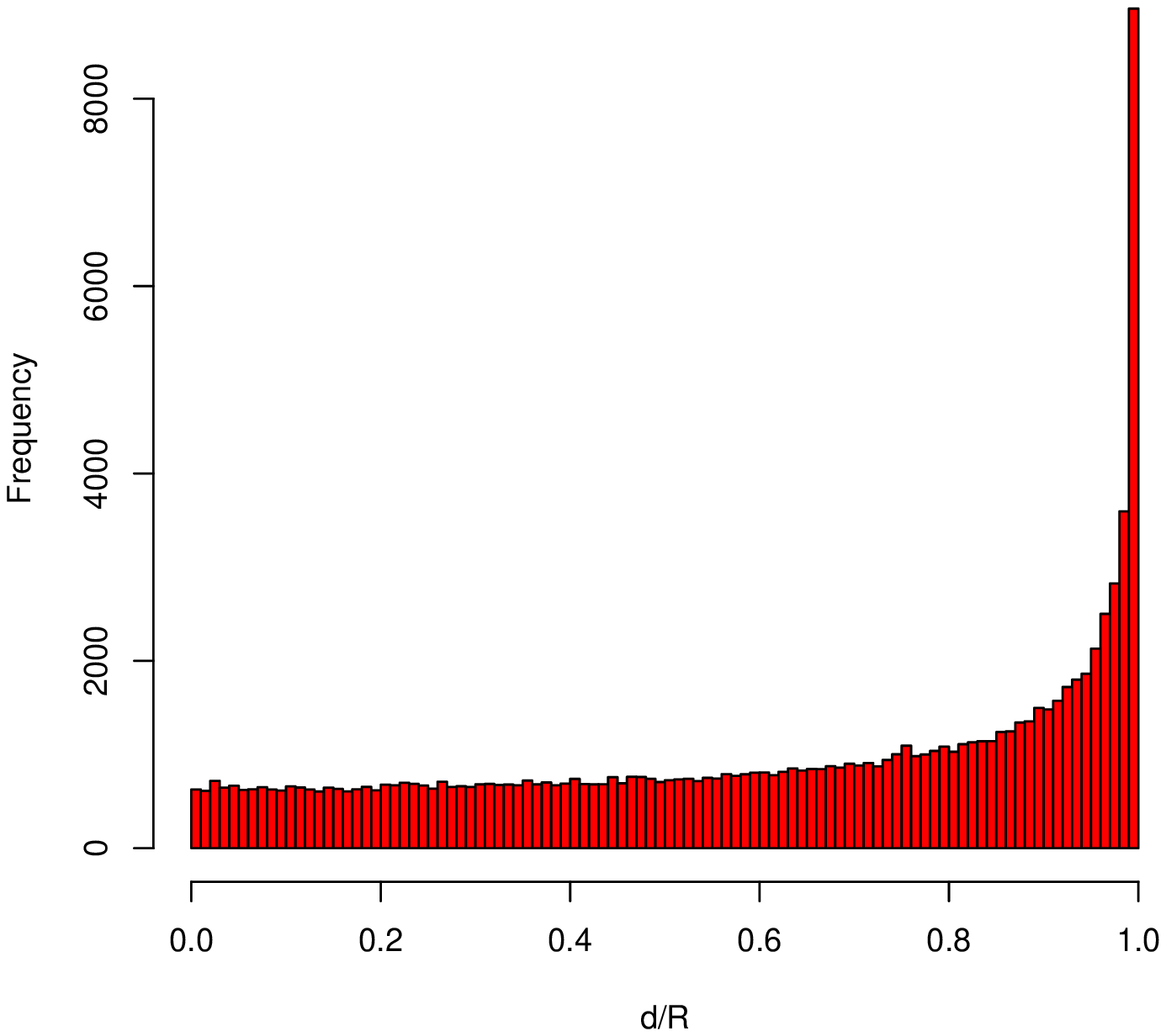,clip=,width=0.46\linewidth} &
\epsfig{file=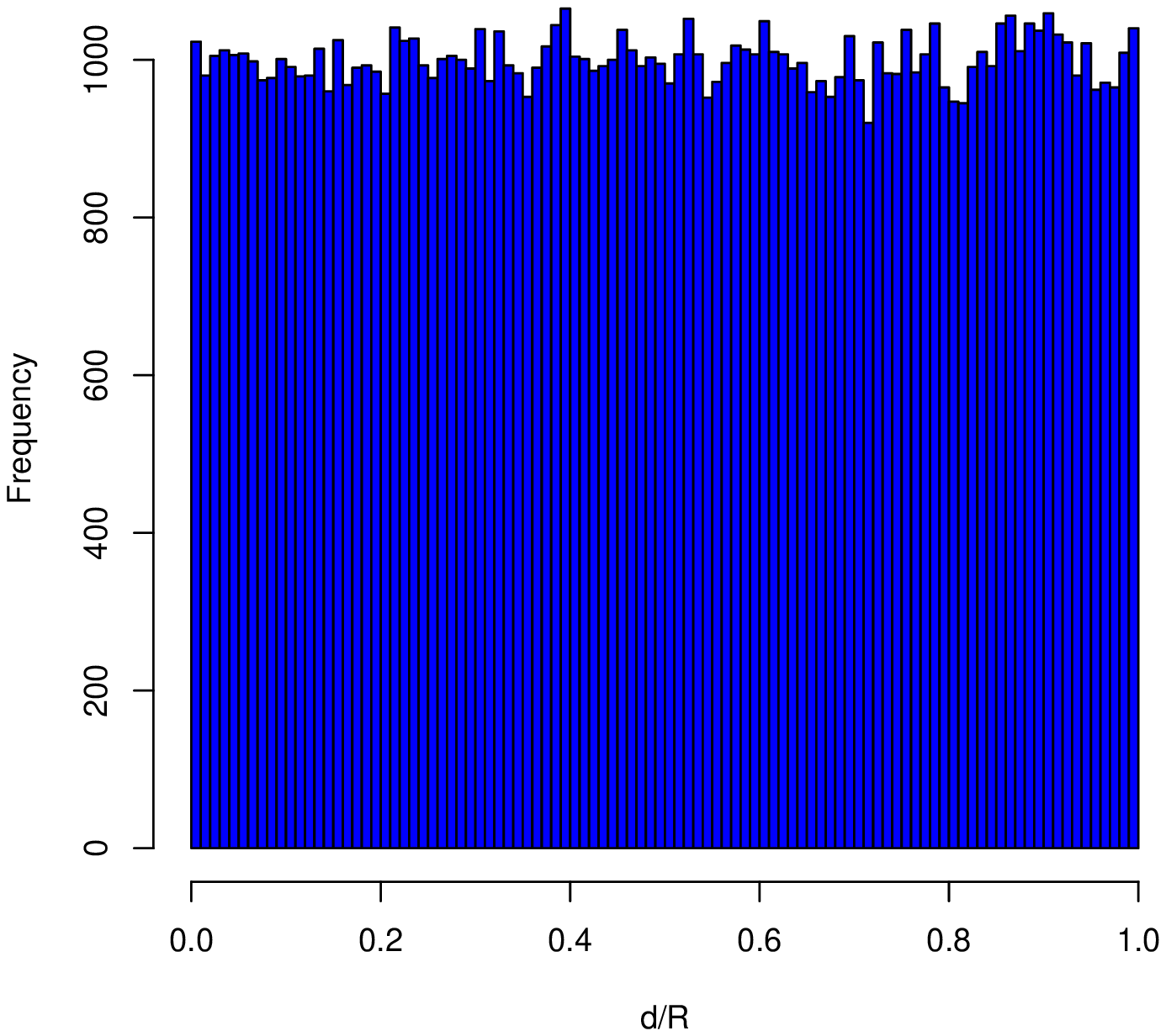,clip=,width=0.46\linewidth} \\
\hline
\epsfig{file=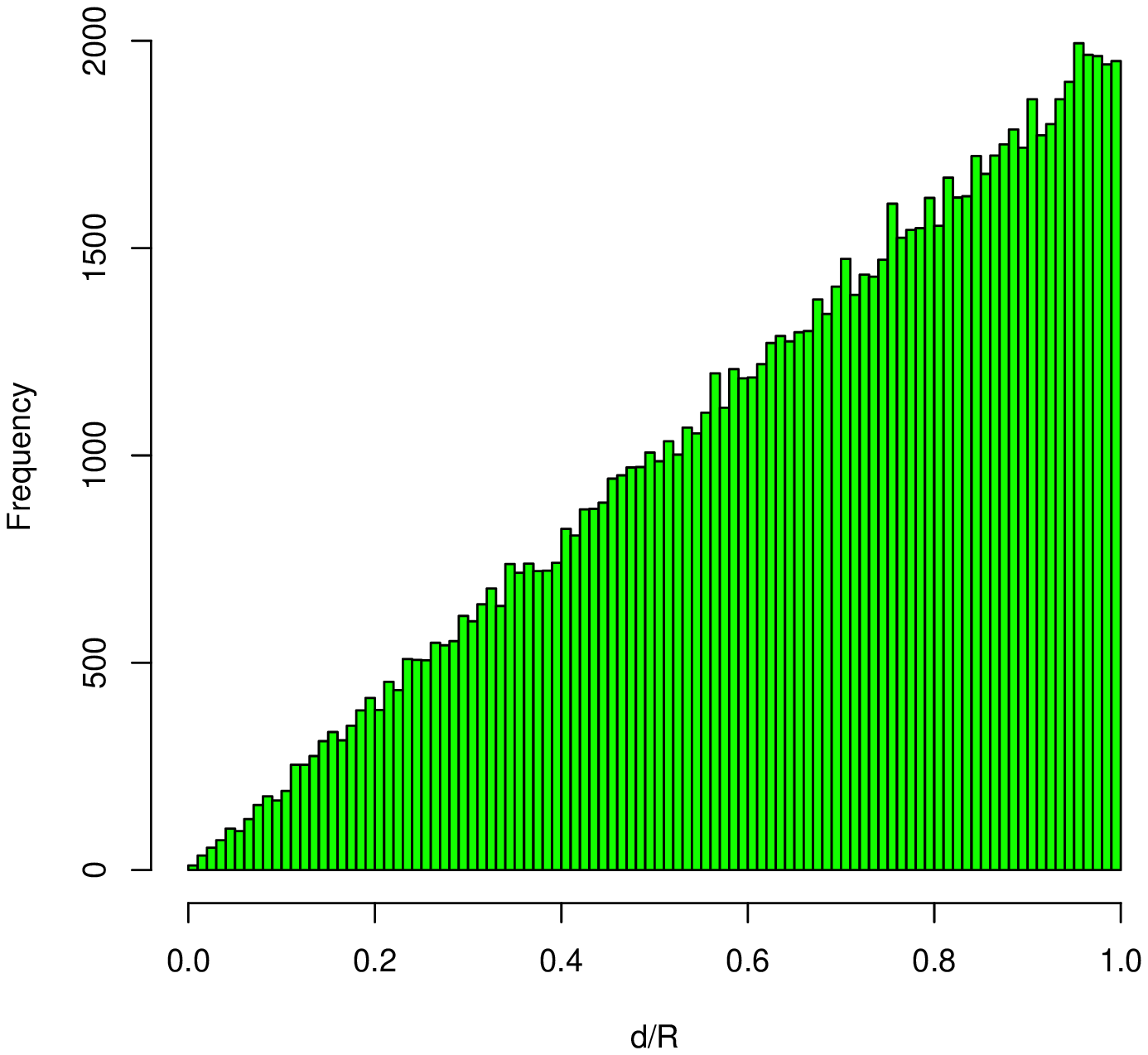,clip=,width=0.46\linewidth} &
\epsfig{file=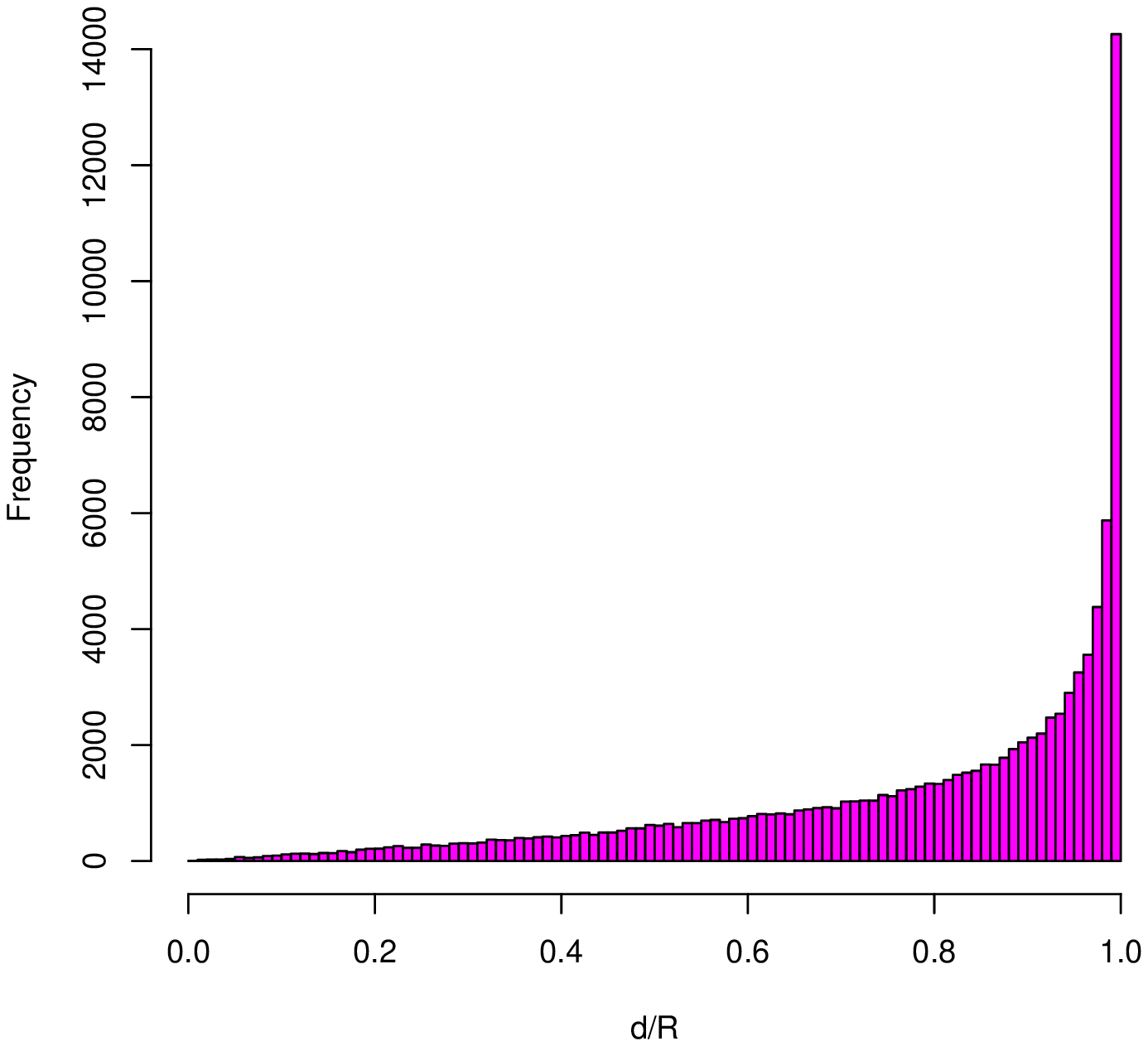,clip=,width=0.46\linewidth}\\
\hline
\epsfig{file=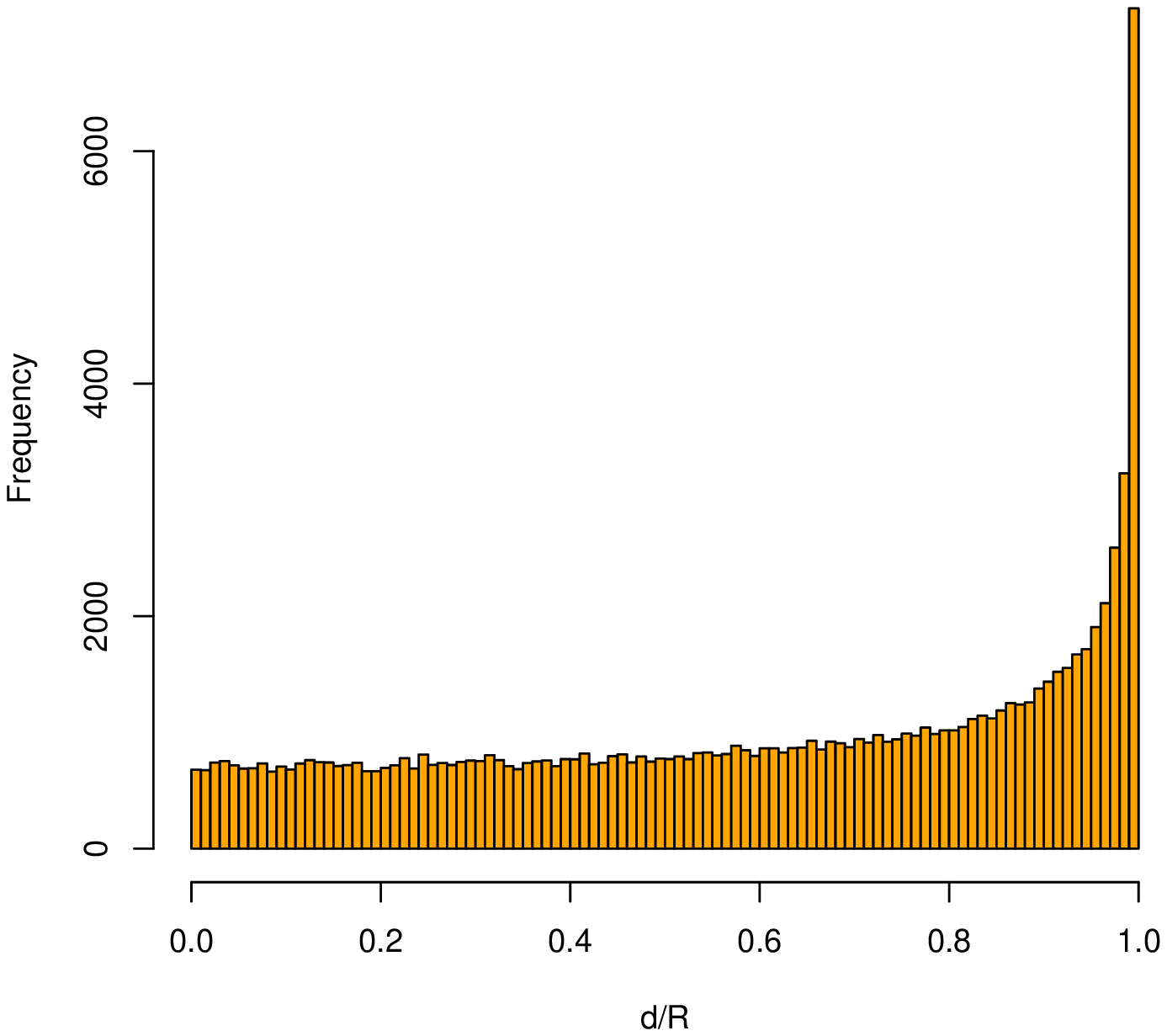,clip=,width=0.46\linewidth} &
\epsfig{file=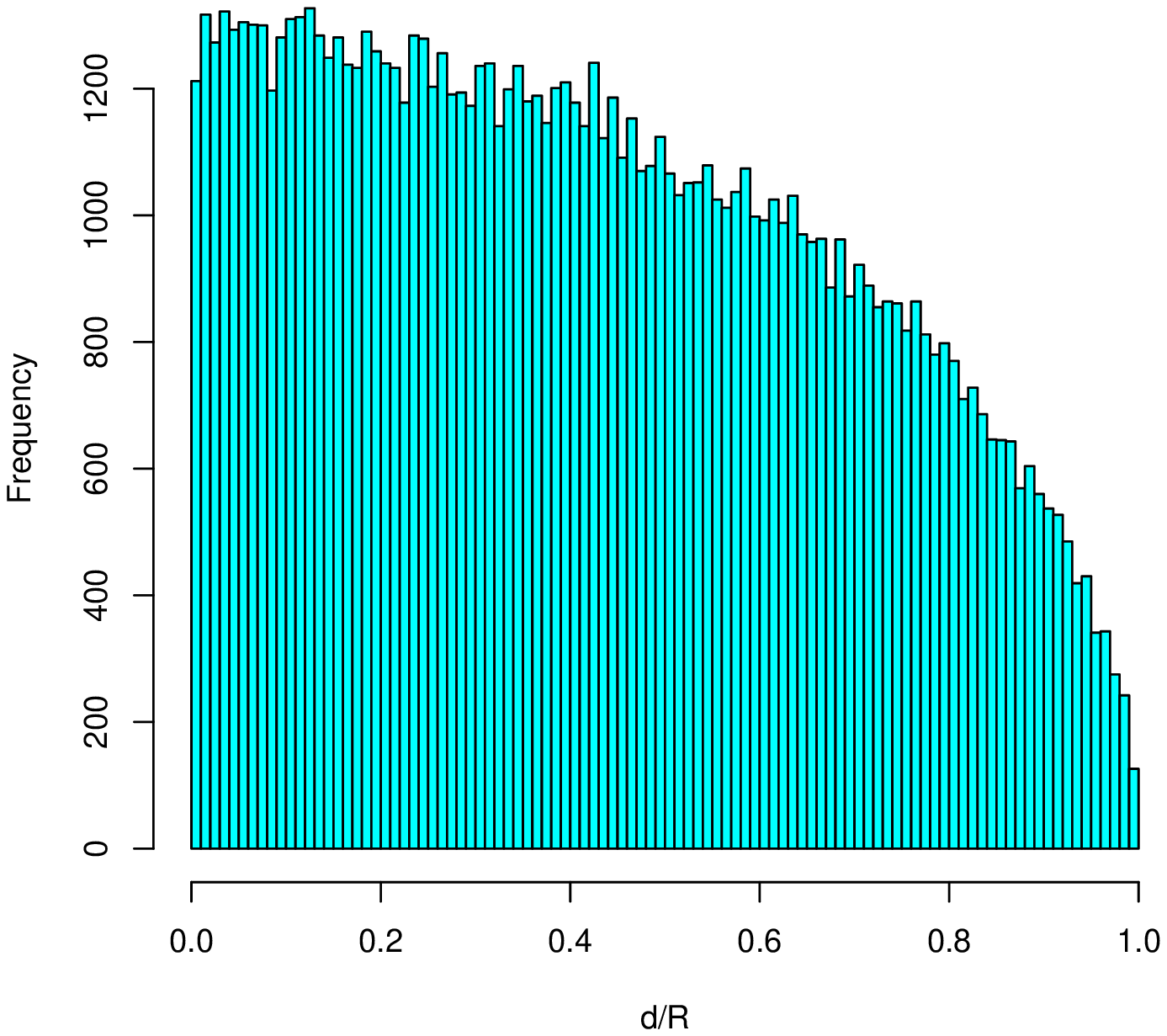,clip=,width=0.46\linewidth}\\
\hline
\end{tabular}
\caption{\small \sf Distribution of the distance  of the chords
from center of the circle
in samples produced with the various methods.}
\label{fig:r}
\end{figure}

Sampling $\rho$ is the key to simulate chords 
inside the circle selected by the mikado reference point,
allowing us to reuse the code written for other methods.
We can then easily extend what we have done
in section \ref{sec:drawing_program}. All what
we have to do is to replace in the code
for Model 3 (subsection \ref{ss:simulation_M3}) 
the line\\
\Rin{rho <- sqrt(runif(n))}\\
by\\
\Rin{rho <- sqrt(runif(n, 0, 1))*cos(runif(n, 0, pi/2))}\\ 
and the game is over. Results from simulations are shown
in Figs.~\ref{fig:corde}-\ref{fig:corde_norot}.

\section{Conclusions}
Arguing, in abstract terms, about 
{\em the} solution of the Bertrand problem
-- not a paradox! -- is as scientific as the
Byzantine debates about the sex of angels.
Nevertheless, the question can still have a sense if 
framed in a practical contest, 
asking people to draw, by hand or with 
the help of computer graphics, 
\underline{a} chord and betting on the result.

Several random drawing methods have been analyzed
in great detail, in order to show several issues
related to the evaluation of the probability distributions
of functions of variables whose pdf was assumed
and to the drawing simulation. 

Of the six methods analyzed, five can be classified
as `geometrical', two of which are the `classical' solutions.
There was no intention to invent abstruse methods and indeed
the two ``with ruler and compass'' algorithms
seem to me even more `natural' (or at least more suited
to a class of people) that the more famous, `classical' Method 3.

Finally, the only sound experiment I could think about 
(but perhaps I miss of fantasy)
leads to a solution different than the claimed 
`conclusive' physical solutions of the `paradox'.\cite{Jaynes_Bertrand,DiPorto}
\begin{center}
{\em \large --- Il faut cultiver notre jardin ---}
\end{center}

\newpage{}

\section*{Appendix A: Distributions of 
chord distances from the center of the circle}
Out of the six methods analyzed above, one (${\cal M}_2$)
is defined in terms of the distance between chord
and center of the circle, while in two others
(${\cal M}_3$ and ${\cal M}_6$)
we have learned that that distance is a useful quantity
to simplify the solution of the problem. The three
pdf's of the normalized distances in these three cases are
\begin{eqnarray}
f(\rho\,|\,{\cal M}_2) &=& 1\\
f(\rho\,|\,{\cal M}_3) &=& 2\,\rho\\
f(\rho\,|\,{\cal M}_6) &=& \frac{4}{\pi}\,\sqrt{1-\rho^2}\,,
\end{eqnarray}
while we have shown in Fig.~\ref{fig:r} the results of the simulation
for all six methods. For completeness, 
let us do the exercise to calculate
the pdf $f(\rho)$ also for the other three methods. This is in fact
easy, at least in principle,  
because there a simple geometric relation
between the length of a chord and its distance from the center,
since the chord and the radii to its endpoints 
identify an equilateral triangle. 
The only technical problem is that of getting a closed expression 
for the pdf. Let us remind here the relations between 
length $\lambda$ of the chord and distance $\rho$ 
from the center, both in units of the radius.
\begin{eqnarray}
\rho &=& \sqrt{1-(\lambda/2)^2} \\
\lambda &=& 2\,\sqrt{1-\rho^2}\,,
\end{eqnarray}
which we shall use in the following subsections. In fact,
using our transformation rule with
the Dirac delta, $f(\rho)$ is given by
\begin{eqnarray}
f(\rho \,|\,{\cal M}_i) &=& \int_{0}^{2}\!
\delta\left(\rho-  \sqrt{1-(\lambda/2)^2}\right) \cdot
f(\lambda \,|\,{\cal M}_i)\,d\lambda \\
 &=& \int_{0}^{2}\!
\delta\left(\rho-  \sqrt{1-(\lambda/2)^2}\right) 
\frac{\delta(\lambda -\lambda^*)}
     {\left|\frac{d}{d\lambda}\,\sqrt{1-(\lambda/2)^2}\right|_{\lambda=\lambda^*}}
\cdot f(\lambda \,|\,{\cal M}_i)\,d\lambda\nonumber\\ 
&=& \frac{4\,\sqrt{1-(\lambda^*/2)^2}}{\lambda^*}\cdot
 f(\lambda^* \,|\,{\cal M}_i) \nonumber\\
&=& \frac{2\,\rho}{\sqrt{1-\rho^2}}\cdot 
f(\lambda^* \,|\,{\cal M}_i) 
\end{eqnarray}
being $\lambda^*=2\,\sqrt{1-\rho^2}$.

Here are the results for the remaining three models:
\begin{eqnarray}
f(\rho \,|\,{\cal M}_1) &=& \frac{2\,\rho}{\sqrt{1-\rho^2}}\cdot 
\frac{1}{\pi\,\sqrt{1-(\lambda^*/2)^2)}} \\
    &=&   \frac{2\,\rho}{\sqrt{1-\rho^2}}\cdot \frac{1}{\pi\,\rho}
\nonumber \\
&=& \frac{2}{\pi}\frac{1}{\sqrt{1-\rho^2}}
\end{eqnarray}
\begin{eqnarray}
f(\rho \,|\,{\cal M}_4) &=& \frac{2\,\rho}{\sqrt{1-\rho^2}}\cdot 
\frac{1}{2} \\
&=& \frac{\rho}{\sqrt{1-\rho^2}} \\
f(\rho \,|\,{\cal M}_5) &=& \frac{2\,\rho}{\sqrt{1-\rho^2}}\cdot 
\frac{\sqrt{2+2\,\sqrt{1-{(\lambda^*/2)}^2}} +\sqrt{2-2\,\sqrt{1-{(\lambda^*/2)}^2}} }
     {4\times 2\,\sqrt{1-{(\lambda^*/2)}^2}} \\
&& \nonumber\\
&=& \frac{2\,\rho}{\sqrt{1-\rho^2}}\cdot  
\frac{\sqrt{2+2\rho}+\sqrt{2-2\rho}}{8\,\rho} \\
&& \nonumber\\
&=& \frac{\sqrt{2+2\rho}+\sqrt{2-2\rho}}{4\,\sqrt{1-\rho^2}}\,, 
\end{eqnarray}
where the form $\sqrt{4-{\lambda^*}^2}$ have been
written as $2\sqrt{1-(\lambda^*/2)^2}$, in which we easily
recognize $2\rho$.

\begin{figure}
\begin{center}
\epsfig{file=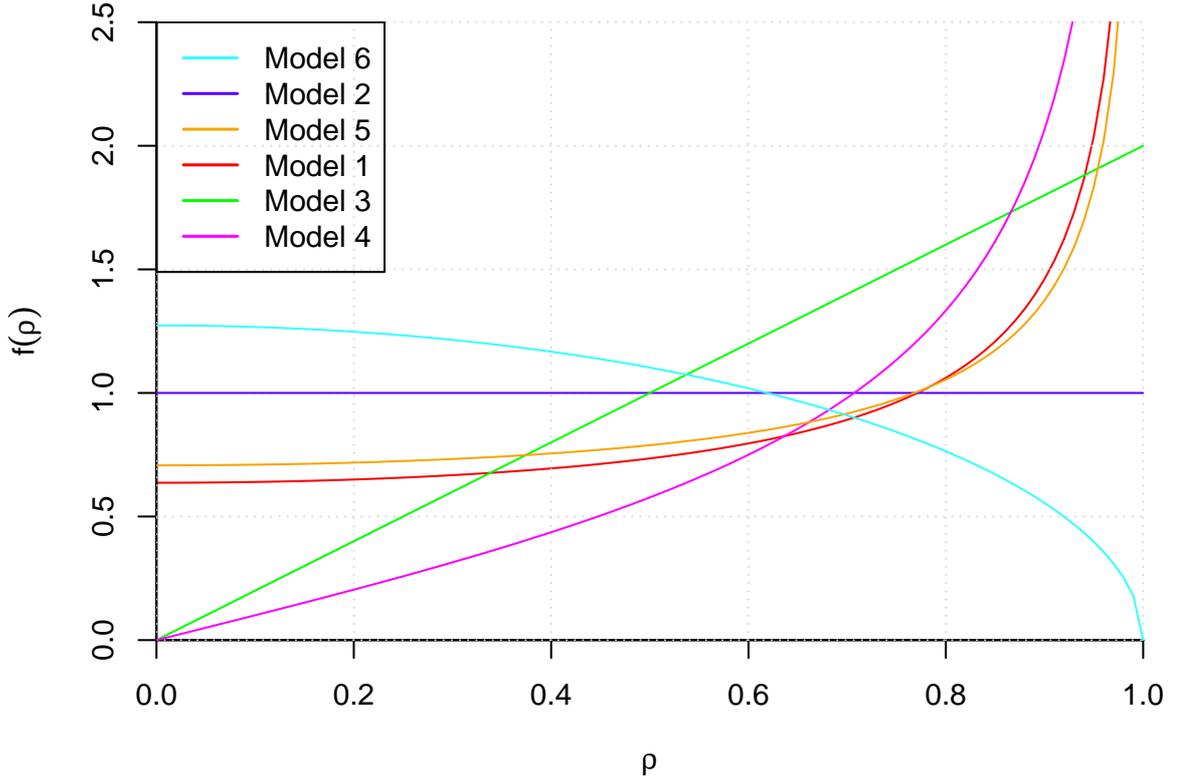,clip=,width=\linewidth}
\end{center}
\caption{\small \sf Distribution of the distance  of the chords
from center of the circle. 
(The order of the models
in the legend corresponds to the decreasing order of the pdf's
for small values of $\rho$.)
}
\label{fig:distr_rho}
\end{figure}

The results are summarized in Fig.~\ref{fig:distr_rho}.
As it happened for the pdf's of $\lambda$, some of them
are divergent for $\rho\rightarrow 1$, but it has been 
checked that all integrals from 0 to 1 are finite and 
indeed yield 1, as it must be.

\section*{Appendix B: A chord length generator based on the Beta distribution}
If it makes no sense to talk about the `correct', or even the `best',
solution in abstract terms, it has also little sense to make an inventory
of all possible solutions. The reason on the six ones shown here,
let us repeat it again,
was the following:
\begin{itemize}
\item models ${\cal M}_1$, ${\cal M}_2$ and ${\cal M}_3$
 are the `classical' ones, i.e. those proposed by Bertram himself;
\item model ${\cal M}_4$ is as simple as the other three
      (and certainly, in my opinion, much more natural to think
      about than  ${\cal M}_3$), justified by the custom 
      of a ruler and compass geometer;
\item model ${\cal M}_5$ is a variation of ${\cal M}_4$, 
      in which the endpoints of the chord are defined by the 
      two intersections
      of the  pencil lead when the compass is rotated clockwise
      and anticlockwise;
\item finally, model ${\cal M}_6$ was an attempt to think about
      of a chord drawing game, in analogy to Buffon's needle.
\end{itemize}
In all cases (even in the sixth one, that could be indeed played
in practice), 
in order to refer to real cases, the question has been turned into
the result of a generator that a `student' 
(or anyone else who can write a computer program)
might write. 
And, although at the very beginning one would mainly think about 
to ${\cal M}_1$, and perhaps ${\cal M}_2$, once the 
`student' starts thinking about the problem and learns 
how to reuse programming code, 
we have to be {\em afraid} (always think to bets!)
that more sophisticated extraction models could be implemented.

Finally, once the student realizes that what matters is the
distribution of the normalized distance, 
shown in the previous appendix, the number of methods 
which can be easily implemented immediately diverges. 
One only needs an easy generator 
of a quantity in the range between 0 and 1.
The easiest possibility which offers
quite some variability in shapes
is provided by the Beta distribution,
defined as 
\begin{equation}
f(x\,|\,\mbox{Beta}(r,s))=\frac{1}{\beta(r,s)}x^{r-1}(1-x)^{s-1}
\hspace{0.6cm}\left\{\!\begin{array}{l}  r,\,s > 0 \\
   0\le x\le 1 \,.  \end{array}\right.\,,
\label{eq:distr_beta}
\end{equation}
in which the denominator, equal to 
$$\beta(r,s)=\int_0^1 x^{r-1}(1-x)^{s-1}\,\mbox{d}x\,,$$
 defines the {\em beta function}. 
\begin{figure}
\begin{center}
\begin{tabular}{|c|c|}\hline 
& \\
\multicolumn{1}{|l|}{{\bf A)} {\small $r=s=$\,{\bf 1}, 1.1 e 0.9}} & 
\multicolumn{1}{l|}{{\bf B)} {\small $r=s=$\,{\bf 2}, 3, 4, 5}} \\ 
\epsfig{file=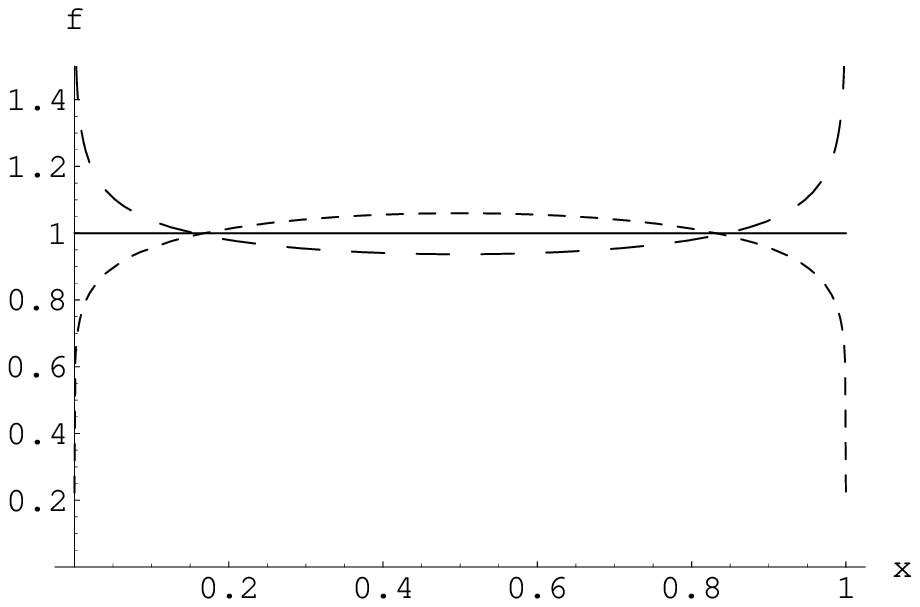,width=0.4\linewidth,clip=} &
\epsfig{file=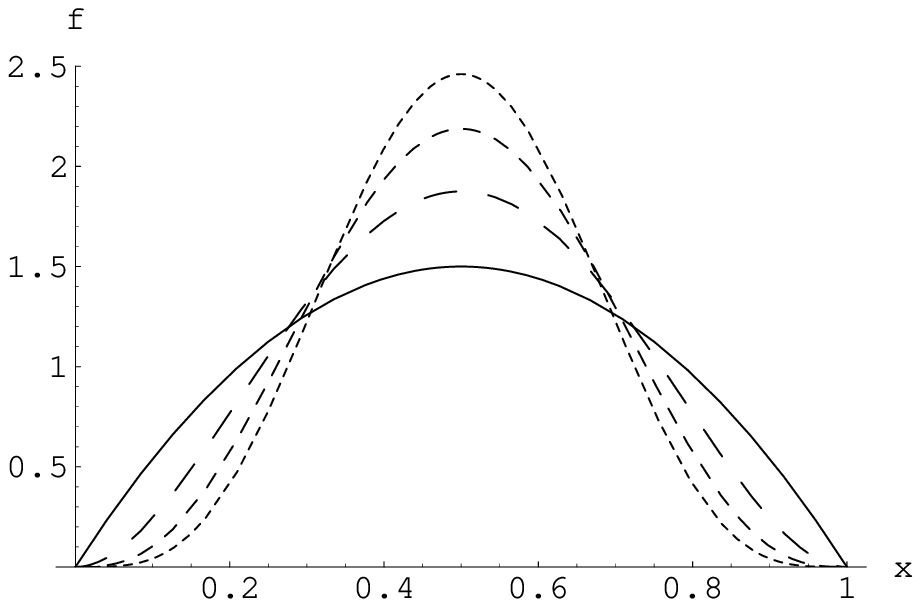,width=0.4\linewidth,clip=} \\ \hline 
& \\
\multicolumn{1}{|l|}{{\bf C)} {\small $r=s=$\,{\bf 0.8}, 0.5, 0.2, 0.1}} & 
\multicolumn{1}{l|}{{\bf D)} {\small $r=0.8$; $s=$\,{\bf 1.2}, 1.5, 2, 3}}\\
\epsfig{file=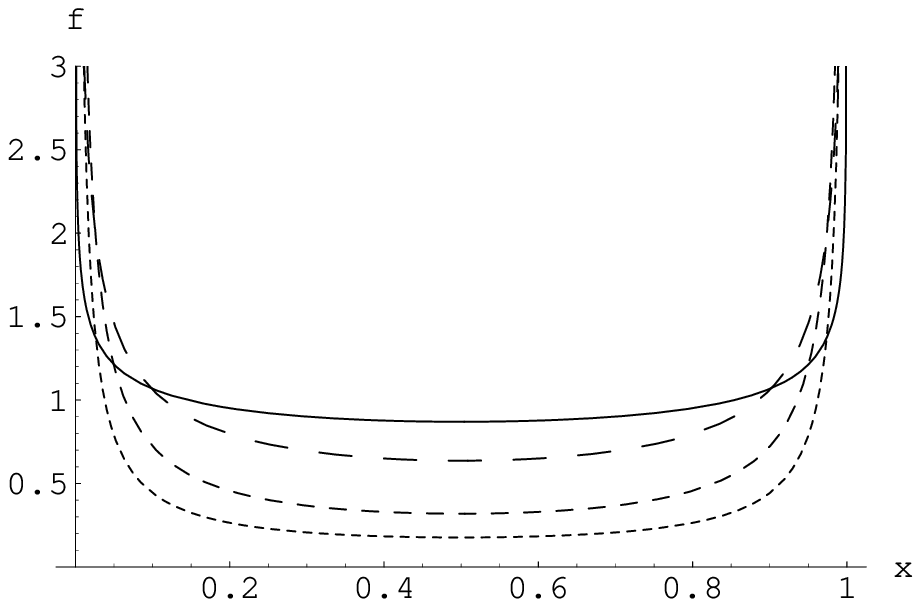,width=0.4\linewidth,clip=} &
\epsfig{file=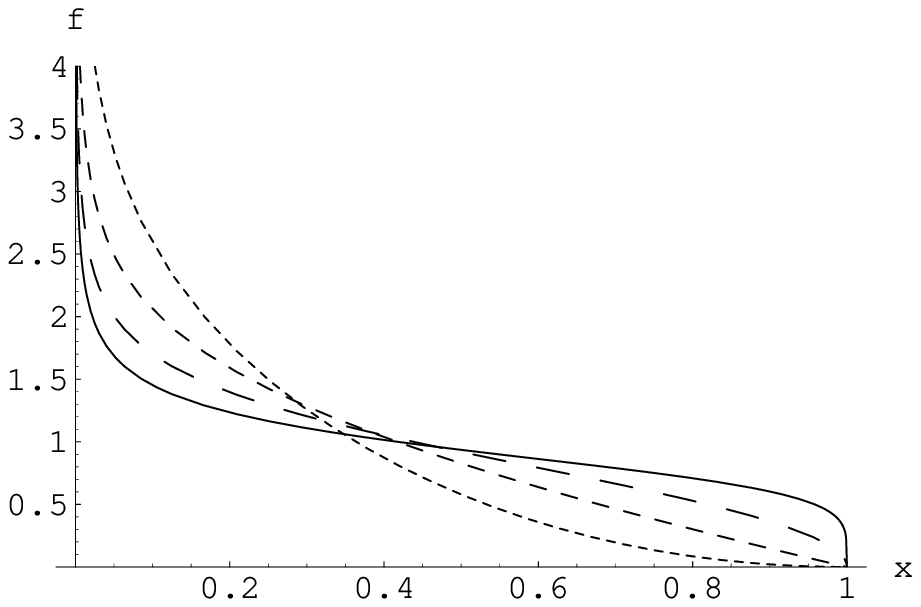,width=0.4\linewidth,clip=} \\ \hline 
& \\
\multicolumn{1}{|l|}{{\bf E)} {\small $s=0.8$; $r=$\,{\bf 1.2}, 1.5, 2, 3}} &
\multicolumn{1}{l|}{{\bf F)} {\small $s=2$; $r=$\,{\bf 0.8}, 0.6, 0.4, 0.2}}\\
\epsfig{file=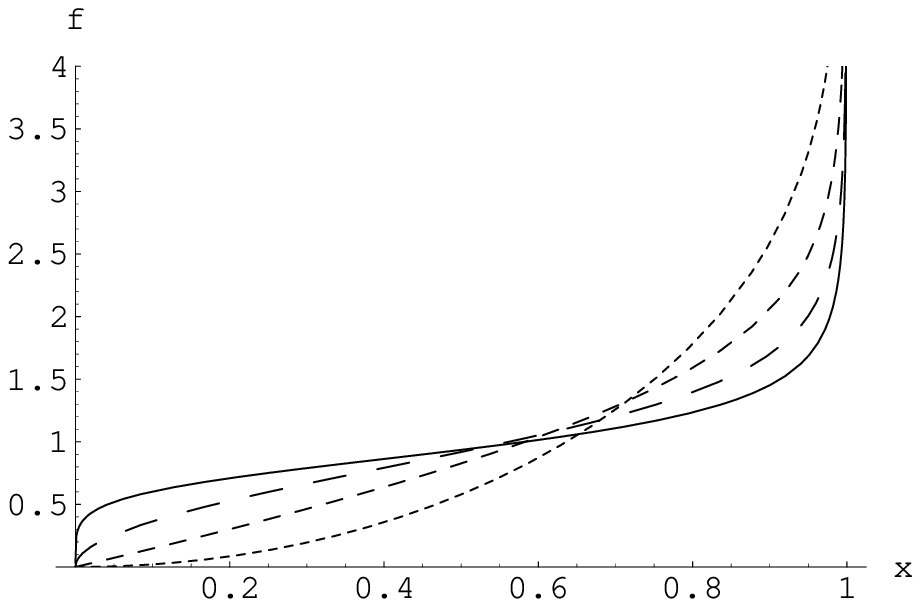,width=0.4\linewidth,clip=} &
\epsfig{file=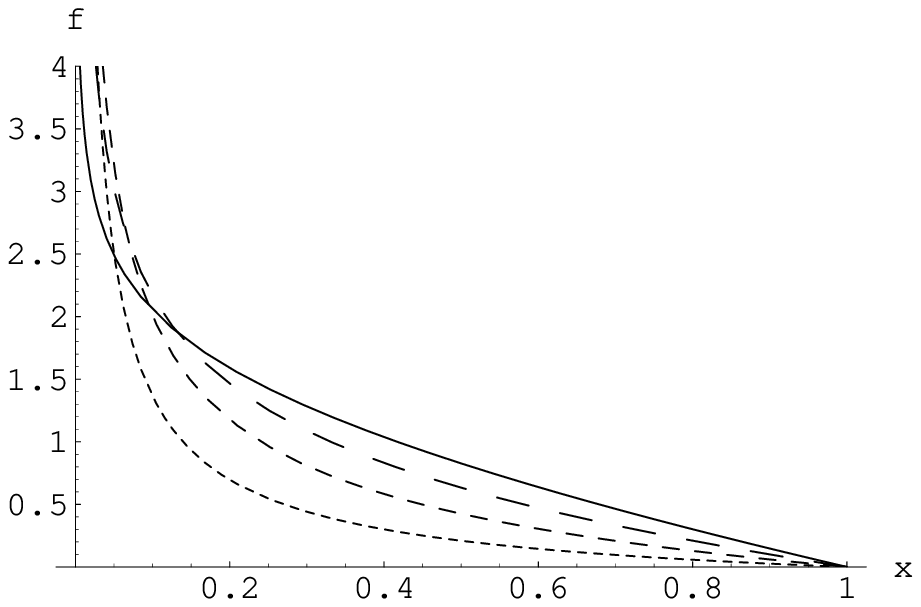,width=0.4\linewidth,clip=} \\ \hline 
& \\
\multicolumn{1}{|l|}{{\bf G)} {\small $(r,\,s)=$ (3,\,5), ({\bf 5,\,5}),
(5,\,3)}} & 
\multicolumn{1}{l|}{{\bf H)} {\small $(r,\,s)=$ (30,\,50), ({\bf 50,\,50}),
(50,\,30)}} \\
\epsfig{file=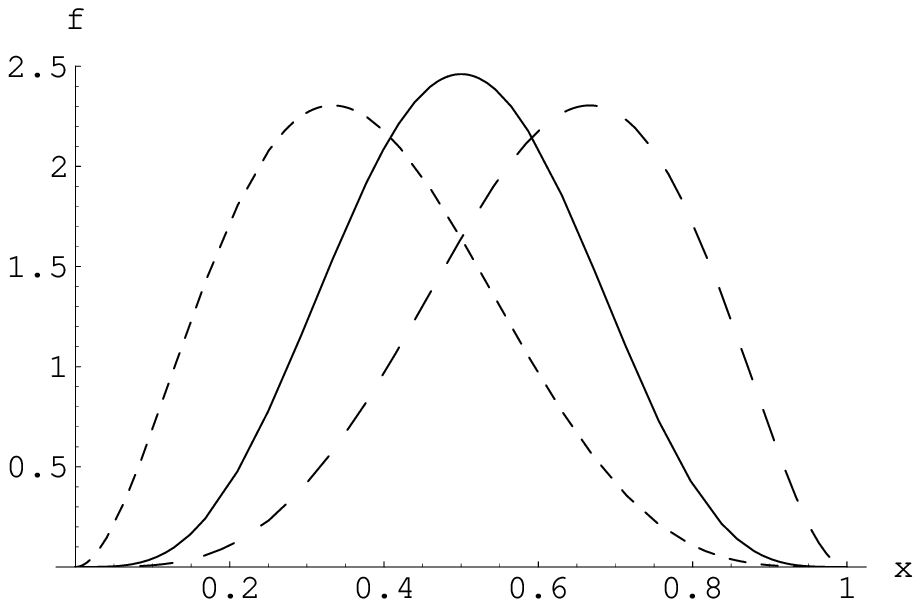,width=0.4\linewidth,clip=} &
\epsfig{file=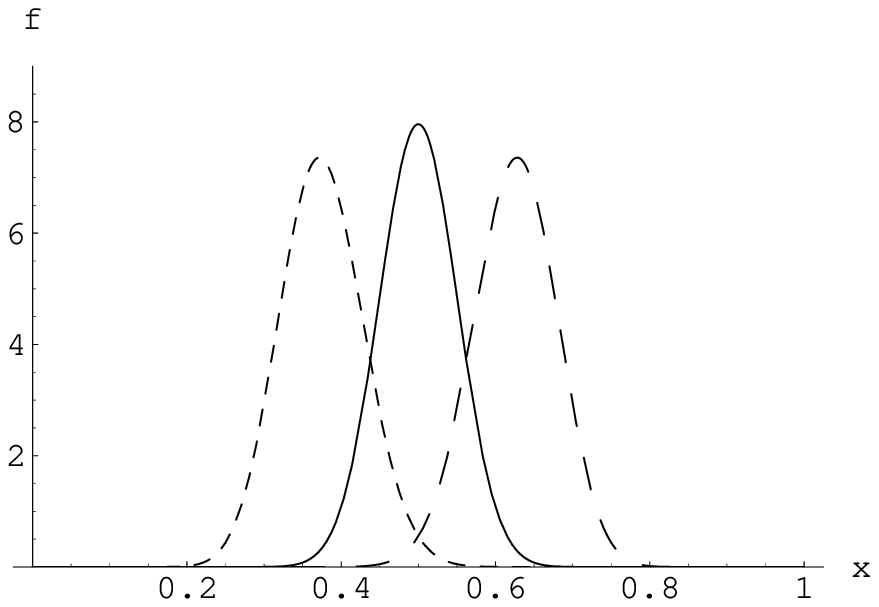,width=0.4\linewidth,clip=} \\ \hline
\end{tabular}
\end{center}
\caption{\small Examples of Beta distributions for some 
values of  $r$ and $s$. 
The parameters in bold refer to continuous curves.}
\label{fig:betas}
\end{figure}
Examples of the beta distributions are shown in 
Fig.~\ref{fig:betas} (I refrain from showing the resulting
chord distributions\ldots). Here it is just how a chord length generator
could be implemented in R:\\
\Rin{rlchordsBeta <- function(n, r, s) 2*sqrt(1-rbeta(n, r, s)\^\,2)}\\
And here it is how to recover very easily the simulations with Methods 2 
and 3:\\
\Rin{n=100000; hist(rlchordsBeta(n, 1, 1), nc=200, col='blue')}\\
\Rin{hist(rlchordsBeta(n, 2, 1), nc=200, col='green')}\\
Other fancy distributions are left to the fantasy of the reader.

\section*{Appendix C: Turning the Bertrand problem into an inferential-predictive
game}
{\small
\begin{flushright}
{\sl ``I play at {\em \'ecart\'e} with a gentleman whom I know to be perfectly  honest.}\\
{\sl What is the chance that he turns up the king?}\\
{\sl It is 1/8. This is a problem of the {\em probability of effects}.}\\
{\sl \ldots}\\
{\sl I play with a gentleman whom I do not know.}\\
{\sl He has dealt ten times, and he has turned the king up six times.}\\
{\sl What is the chance that he is a sharper?}\\
{\sl This is a problem in the probability of causes.}\\
{\sl It may be said that it is the essential problem of the experimental method''}\\
(Henry Poincaré)
\end{flushright}
} 
\noindent
The Bertrand problem can be rephrased in the terms of the causes and the effects
of the above quote. If we assume a  {\em cause} (a precise random
drawing model) we are uncertain about the {\em effects} (the chord length).
But we might also be uncertain about the model. And usually
we do not consider {\em all} possible models equally likely,
where ``all model'' only means ``all models of which we can think about''.
Usually  there will be models in which we believe
more and models in which we believe less, and we then 
rank the possibilities with a degree of belief, or probability. 

As a consequence, the uncertainty 
about the occurrence of each possible effect (the length of a chord, rounded
somehow) has to take into account the probability of each value of the length,
 given the model, and the probability of the model itself. 
This is done using the following well known rule of probability theory:
\begin{eqnarray}
f(\lambda\,|\,I) &=& \sum_i f(\lambda\,|\,{\cal M}_i,I)\cdot P({\cal M}_i\,|\,I)\,,
\end{eqnarray}
where the background information $I$ has been 
made here explicit to remind that probability is always conditional probability.
A similar formula applies to the probability that the length of the chord
is smaller than any given value, that is
\begin{eqnarray}
F(\lambda\,|\,I) &=& \sum_i F(\lambda\,|\,{\cal M}_i,I)\cdot P({\cal M}_i\,|\,I)\,.
\end{eqnarray}
In both cases we have a weighted average, with weights
equal to the probabilities of the models. 
Figure  \ref{fig:bertrand_All_F} shows with dashed lines
the case in which the six models ${\cal M}_1$-${\cal M}_6$
are {\em initially} considered equally likely. 
In particular, for the original Bertrand question we have,
in units of the radius,
\begin{eqnarray}
P(\lambda\le\sqrt{3}\,|\,I) &=& 
\sum_i P(\lambda\le \sqrt{3}\,|\,{\cal M}_i,I)\cdot P({\cal M}_i\,|\,I)\,.
\label{eq:updatePlambdaleLtr}
\end{eqnarray}
Let us now think to a different question . 
We assume we know a chord length, 
generated by one of the possible models. Our problem
does not concern any longer the length of that chord,
assuming we are happy with the provided precision, 
but rather what algorithm was used to generate it. 
A problem that Poincaré would have classified as 
{\em the essential problem of the experimental method}
is then to {\em guess}, in quantitative terms, which model
was used. 

Let us imagine, for example, that the extracted chord
has a length, in our customary units  of the radius,  of 0.1, 
that is $\lambda_1=0.1$. Analyzing Fig.~\ref{fig:bertrand_All_F}
we tend to exclude with great certainty models 2, 3 and 6,
attributing instead the result most likely to ${\cal M}_4$.
But we would not believe this model much more than 
${\cal M}_1$ or ${\cal M}_5$. 
If instead we had (very) good reasons, as we have learned, 
to exclude   ${\cal M}_4$ and  ${\cal M}_5$, we were practically sure
that $\lambda_1=0.1$ came from ${\cal M}_1$. 
As a consequence, our expectation about the length of a second chord
will be that resulting from model ${\cal M}_1$ 
preferred by the {\em experimental evidence}. 

Imagine instead the `curious' (but not unlikely) 
situation in which: model 6 was for some reasons
excluded; we considered the other five models initially equally likely; 
the result was $\lambda_2=1.4$.  In this case 
the experimental information would practically irrelevant, 
and this data will not 
update our opinion concerning the drawing model 
(see Fig.~\ref{fig:bertrand_All_F} to figure out the reason).

If we continue the extractions (always using the same generator!)
we keep updating the probability of the different models
and the probability density function of the length of the
{\em next} chord extracted, or, to make the game simpler,
the probability that the next $\lambda$ will be equal or
smaller than $\sqrt{3}$, that we recalculate each time from
Eq.~(\ref{eq:updatePlambdaleLtr}). Put in this terms the problem
is similar to the six box one discussed in Ref.~\cite{GdA_AJP}, 
each box having a different content of black and white balls.
Indeed we have even the same number of `causes'. The main difference
is that, instead of having only two outcomes
(black and white), we have now all possible values
between 0 and 2, although discretized. This discretization
yields to another important difference. While
in the six box problem
some of the causes could eventually  be
`falsified', i.e. their probability could become exactly zero
(a black ball cannot come from a box containing only white balls,
and the other way around), in the chord problem 
falsification is impossible.\footnote{Some of the pdf's
go to zero for $\lambda\rightarrow0$. However the 
probability in a finite interval around zero
is different from zero because the pdf are larger than zero for 
$\lambda>0$.}
Given the similarity of the inferential and predictive games
I refer then to \cite{GdA_AJP} for a mini introduction
to probabilistic inference needed to analyze our problem.

Probability theory teaches us how to update the odds, i.e. the
ratio of probabilities, with a rule that it is convenient 
to rewrite in our case as
\begin{eqnarray}
\frac{P({\cal M}_i\,|\,\lambda_n,I)}
     {P({\cal M}_j\,|\,\lambda_n,I)} &=& 
\frac{P(\lambda_n\,|\,{\cal M}_i,I)}
     {P(\lambda_n\,|\,{\cal M}_j,I)} \times
\frac{P({\cal M}_i\,|\,\lambda_{n-1},I)}
     {P({\cal M}_j\,|\,\lambda_{n-1},I)}\,,    
\label{eq:sequential_bayes}
\end{eqnarray}
where $\lambda_n$ is the length at the $n$-th
extraction, starting from $n=0$, that is `no extraction'
and then $P({\cal M}_i\,|\,\lambda_{0},I)$ stand for the {\em priors}.
As we see in (\ref{eq:sequential_bayes}), the {\em posterior}
from the $n$-th inference becomes the prior
of the $(n+1)$-th. Moreover, we are not considering the pdf of $\lambda$,
but the probability to obtaining a number from our chord generator,
rounded somehow. For example, if we round at two significant digits
and we get $\lambda_1=1.23$, the probabilities of interest 
will be\footnote{This rule can be applied also at the
edges, since the pdf is null outside the range of the variable,
as we shall see in the R code, in we shall pay attention
that the cumulative function is 0 for $\lambda$ 
below 0 and 1 for $\lambda$ above 2.}
\begin{eqnarray}
P(\lambda_1\,|\,{\cal M}_i,I) &=& 
\int_{1.225}^{1.235}\!\!f(\lambda\,|\,{\cal M}_i,I)\,d\lambda\\
&=& F(1.235\,|\,{\cal M}_i,I) -  F(1.225\,|\,{\cal M}_i,I)\,.
\end{eqnarray}
In this case we shall make the problem more realistic
and avoid the singularities of some of the pdf'
(and also be more precise in the calculation of the probability,
in the region in which the pdf's are not enough linear).

This is the R code we need to calculate the cumulative distributions
for the different models
\begin{verbatim}
F.lambda <- function(l, mod) {
   if(l<= 0) 0
   else if (l>=2) 1
   else switch(mod,
               2/pi*asin(l/2),
               1 - sqrt(1-l^2/4),
               l^2/4,
               l/2,
               (2+sqrt(2-sqrt(4-l^2))-sqrt(2+sqrt(4-l^2)))/2,
               2/pi*asin(l/2) - l/pi*sqrt(1-l^2/4) )
}
\end{verbatim}
(Note how the functions reports 0 for arguments smaller than 0,
and 1 for arguments larger than 2).

Then we need a function to calculate the odds between 
the different hypotheses. We can calculate them with respect to
a reference one, from which all others can be easily calculated. 
Here it is (the default reference model was chosen the one producing a 
flat distribution of $\lambda$, but, as we shall see,
the precise default value is irrelevant):
\begin{verbatim}
chordsModelsBF <- function (l, last.digit=0.01, ref.mod=4) {
  bf <- rep(1,6)
  hld <- last.digit/2
  for (mod in c(1:6)) {
    bf[mod] = (F.lambda(l+hld, mod) - F.lambda(l-hld, mod)) /
               (F.lambda(l+hld, ref.mod) - F.lambda(l-hld, ref.mod))  
  }
  return(bf)
}
\end{verbatim}
As we see from the list of the arguments, we have to pass
also the last digits. The functions return a vector of values. 
For example\\
\Rin{chordsModelsBF(1.23)}\\
\Rout{0.8073569 0.7799414 1.2300000 1.0000000 0.8058272 0.6107326}\\
in which we see that this value provides a slight evidence in favor of
${\cal M}_3$, as clear from Fig.~\ref{fig:bertrand_All_F}, 
while 1.95 would favor, in the order, ${\cal M}_6$ and 
 ${\cal M}_5$, while ${\cal M}_4$ is the less favored: \\
\Rin{chordsModelsBF(1.95)}\\
\Rout{2.868580 4.393492 1.950000 1.000000 3.166248 5.454361}\\
Let also try to extreme values, leaving the evaluation
of the results to the reader (results are rounded
to facilitate the reading).\\
\Rin{round(chordsModelsBF(0), 7)}\\
\Rout{0.6366204 0.0012500 0.0025000 1.0000000 0.5003129 0.0000027}\\
\Rin{round(chordsModelsBF(2), 1)}\\
\Rout{18.0 28.3  2.0  1.0 20.0 36.0}

\begin{figure}
\begin{center}
\epsfig{file=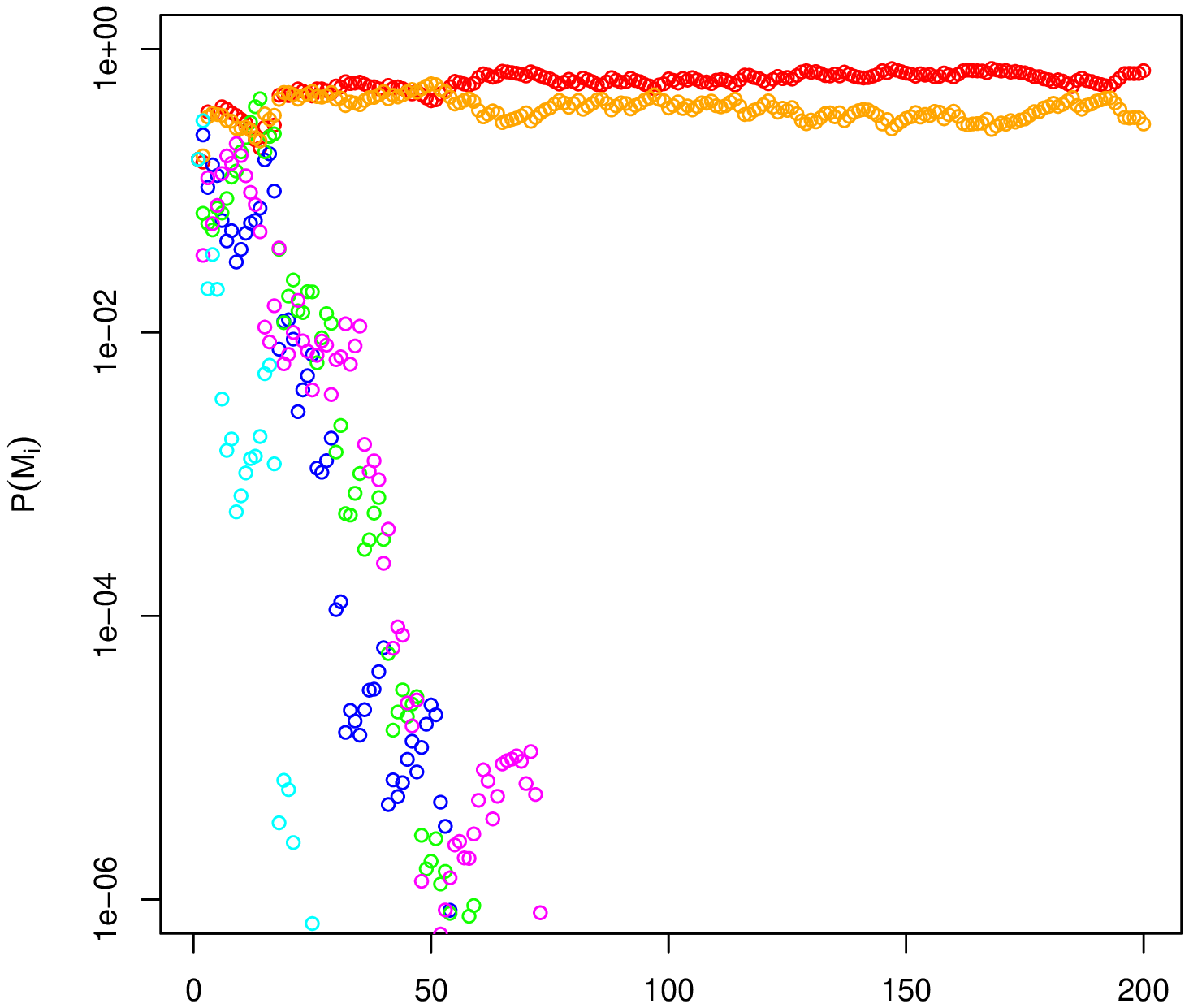,clip=,width=\linewidth} \\
\epsfig{file=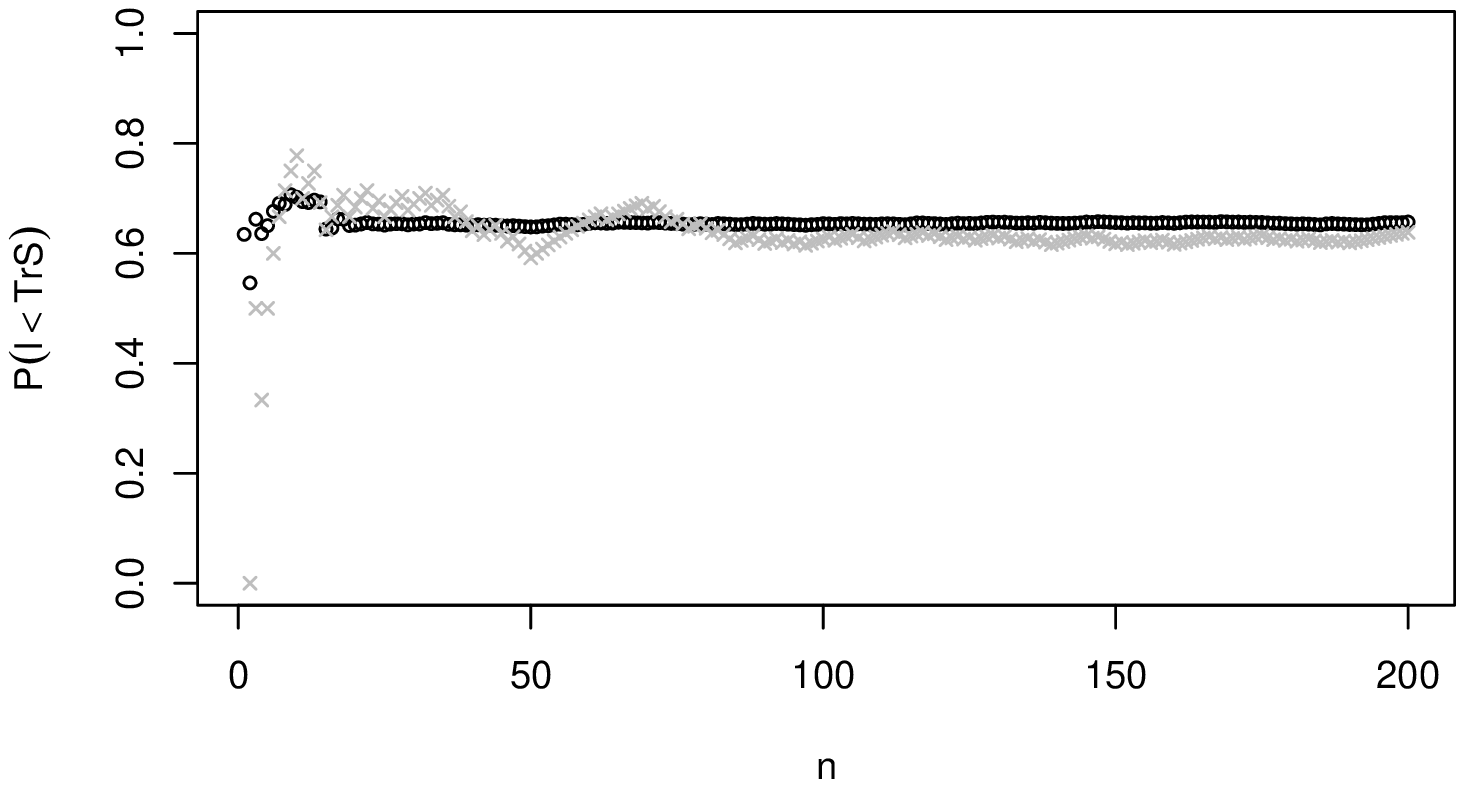,clip=,width=\linewidth} 
\end{center}
\caption{\small \sf Probability of `causes' (above) and probability
of an `effect' (below), the last compared with the relative 
frequency of occurrence (see text). [True model: ${\cal M}_1$]}
\label{fig:bayes_bertrand_M1}
\end{figure}

\begin{figure}
\begin{center}
\epsfig{file=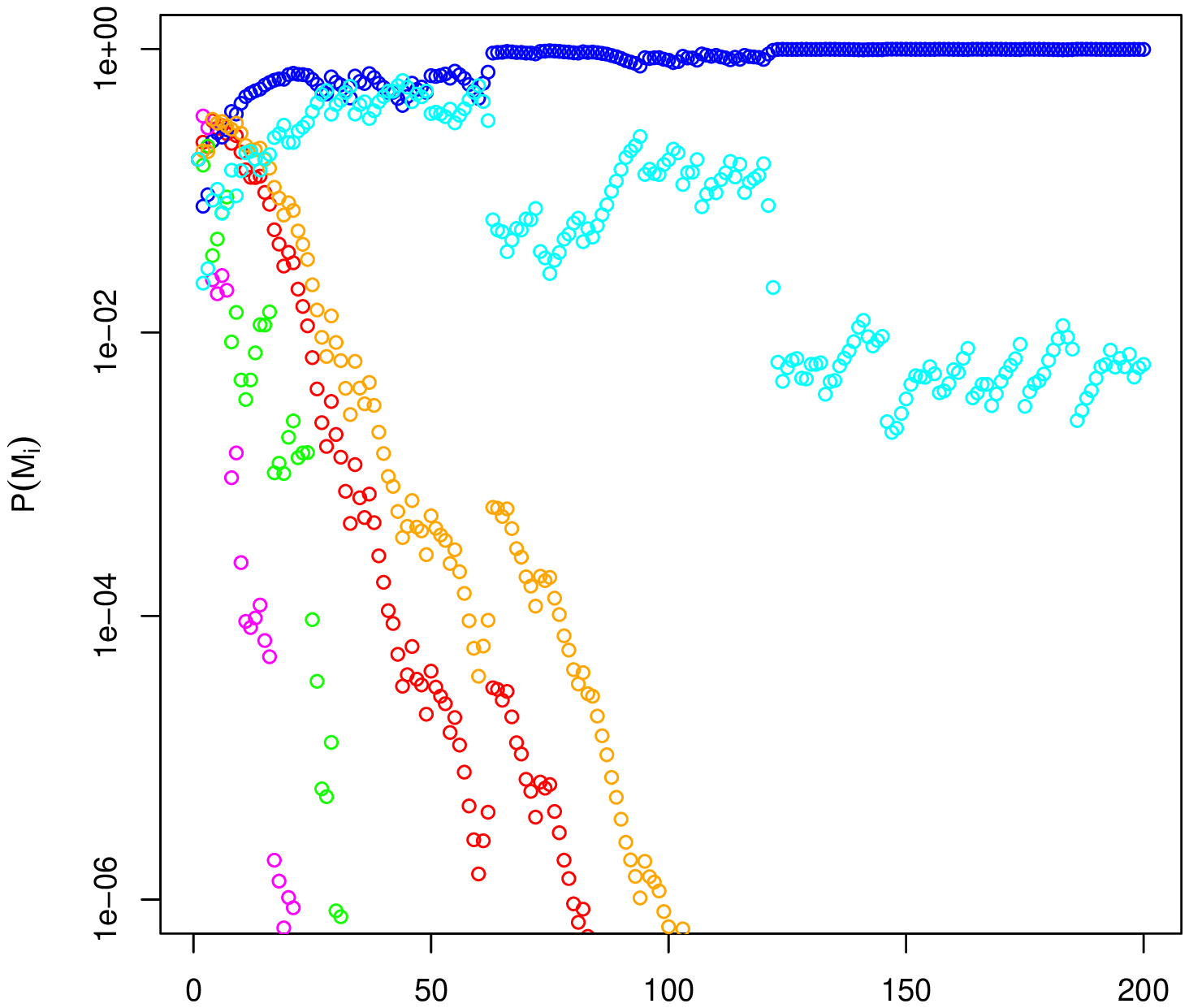,clip=,width=\linewidth} \\
\epsfig{file=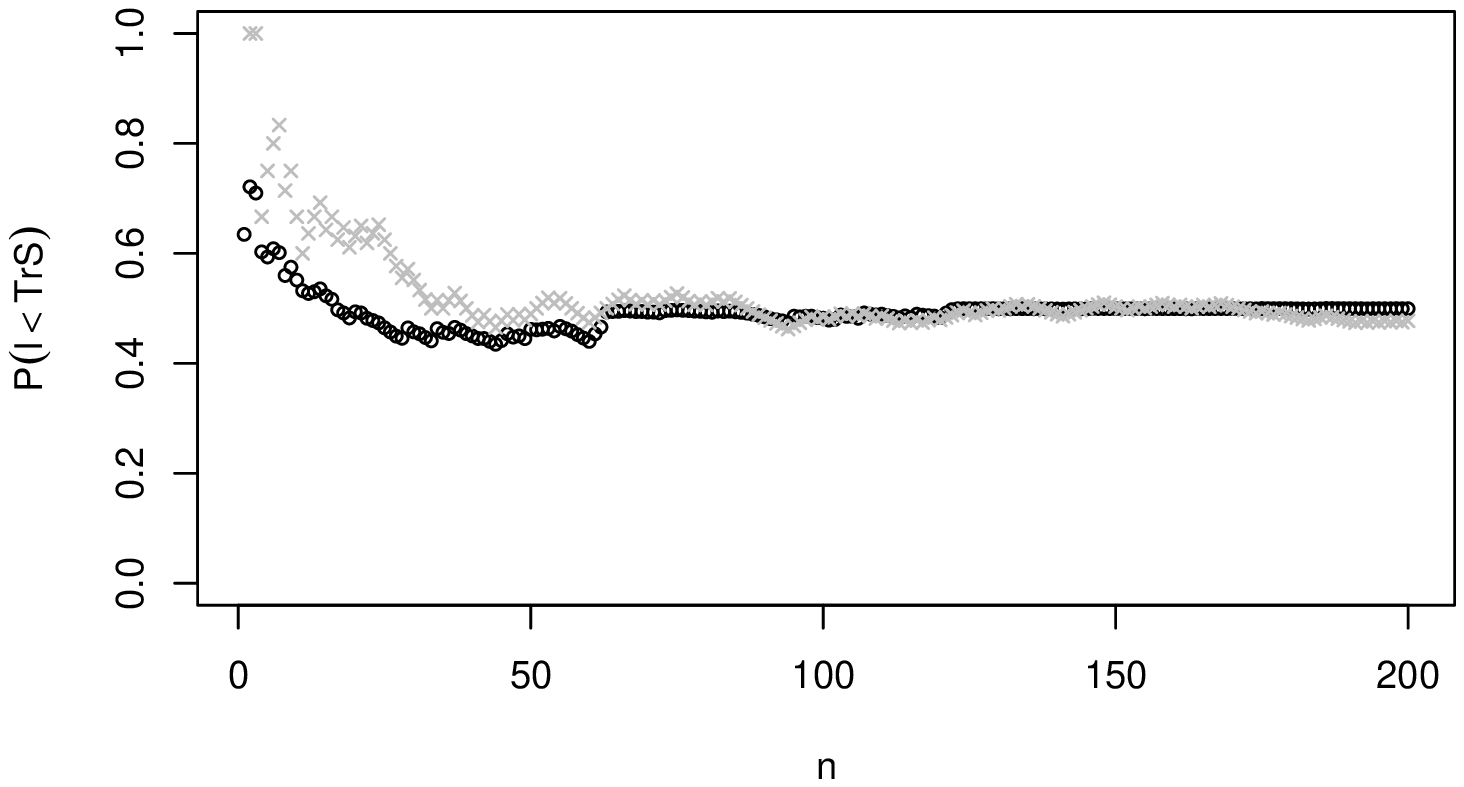,clip=,width=\linewidth} 
\end{center}
\caption{\small \sf Probability of `causes' (above) and probability
of an `effect' (below), the last compared with the relative 
frequency of occurrence (see text). [True model: ${\cal M}_2$]}
\label{fig:bayes_bertrand_M2}
\end{figure}

\begin{figure}
\begin{center}
\epsfig{file=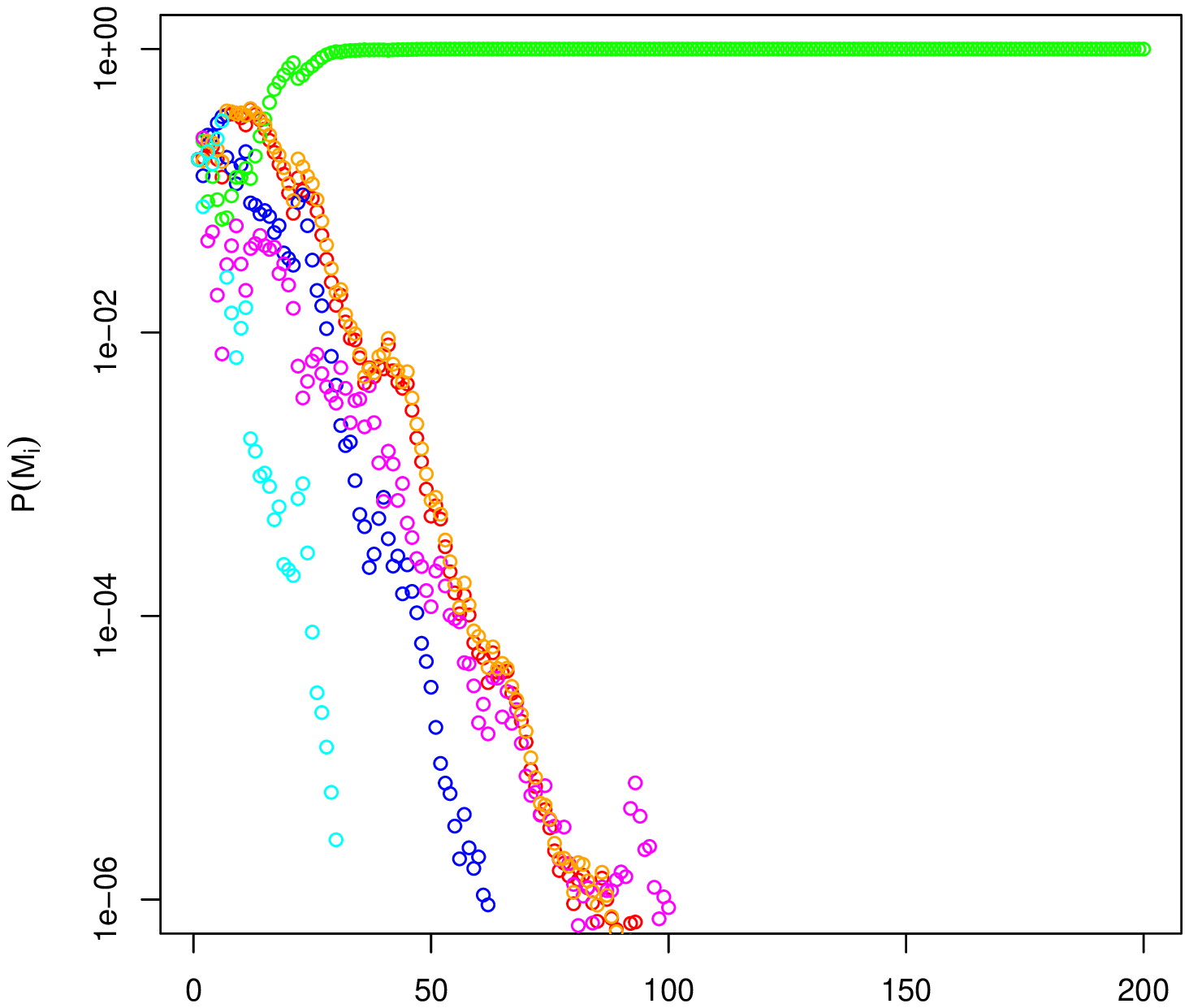,clip=,width=\linewidth} \\
\epsfig{file=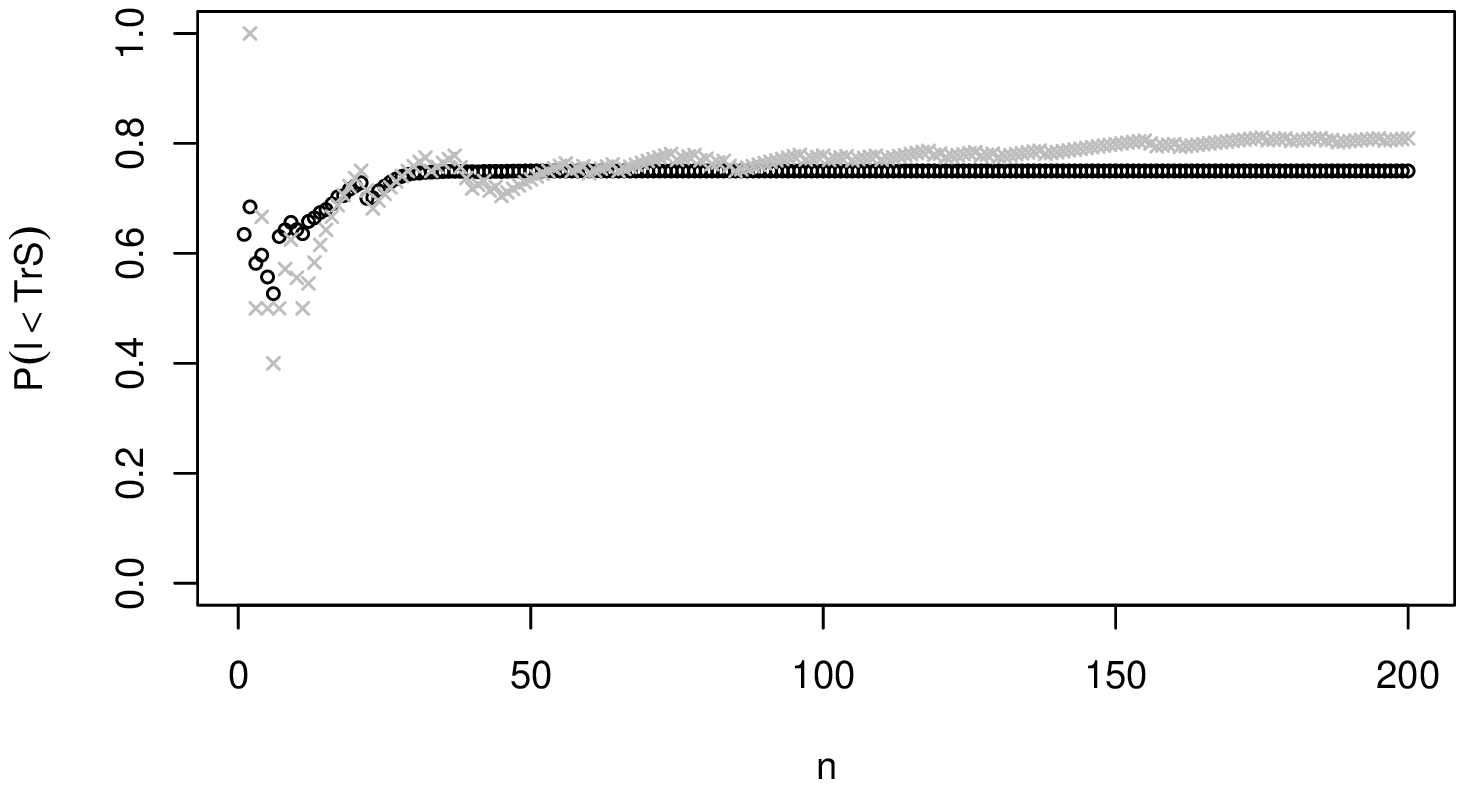,clip=,width=\linewidth} 
\end{center}
\caption{\small \sf Probability of `causes' (above) and probability
of an `effect' (below), the last compared with the relative 
frequency of occurrence (see text). [True model: ${\cal M}_3$]}
\label{fig:bayes_bertrand_M3}
\end{figure}

\begin{figure}
\begin{center}
\epsfig{file=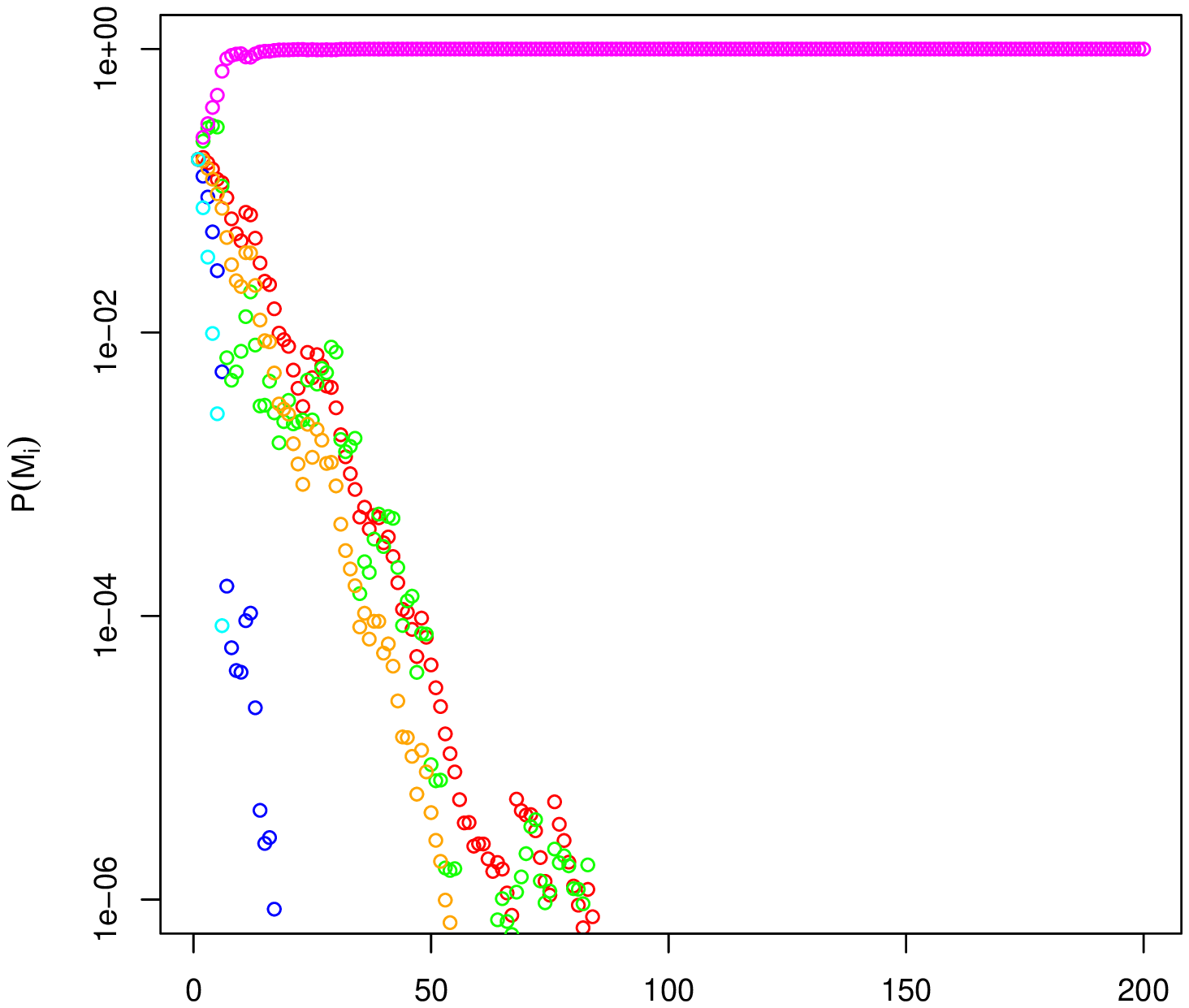,clip=,width=\linewidth} \\
\epsfig{file=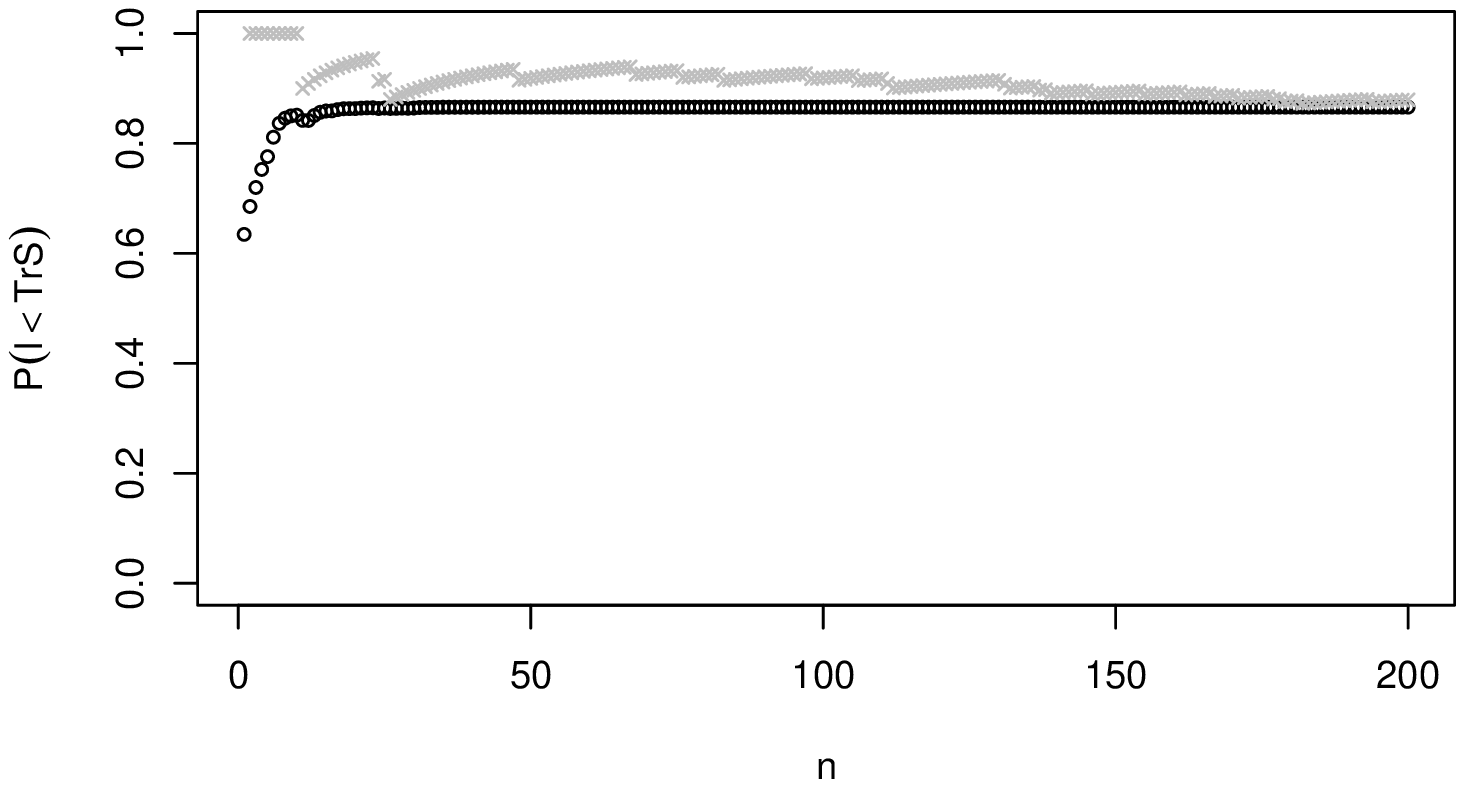,clip=,width=\linewidth} 
\end{center}
\caption{\small \sf Probability of `causes' (above) and probability
of an `effect' (below), the last compared with the relative 
frequency of occurrence (see text). [True model: ${\cal M}_4$]}
\label{fig:bayes_bertrand_M4}
\end{figure}

\begin{figure}
\begin{center}
\epsfig{file=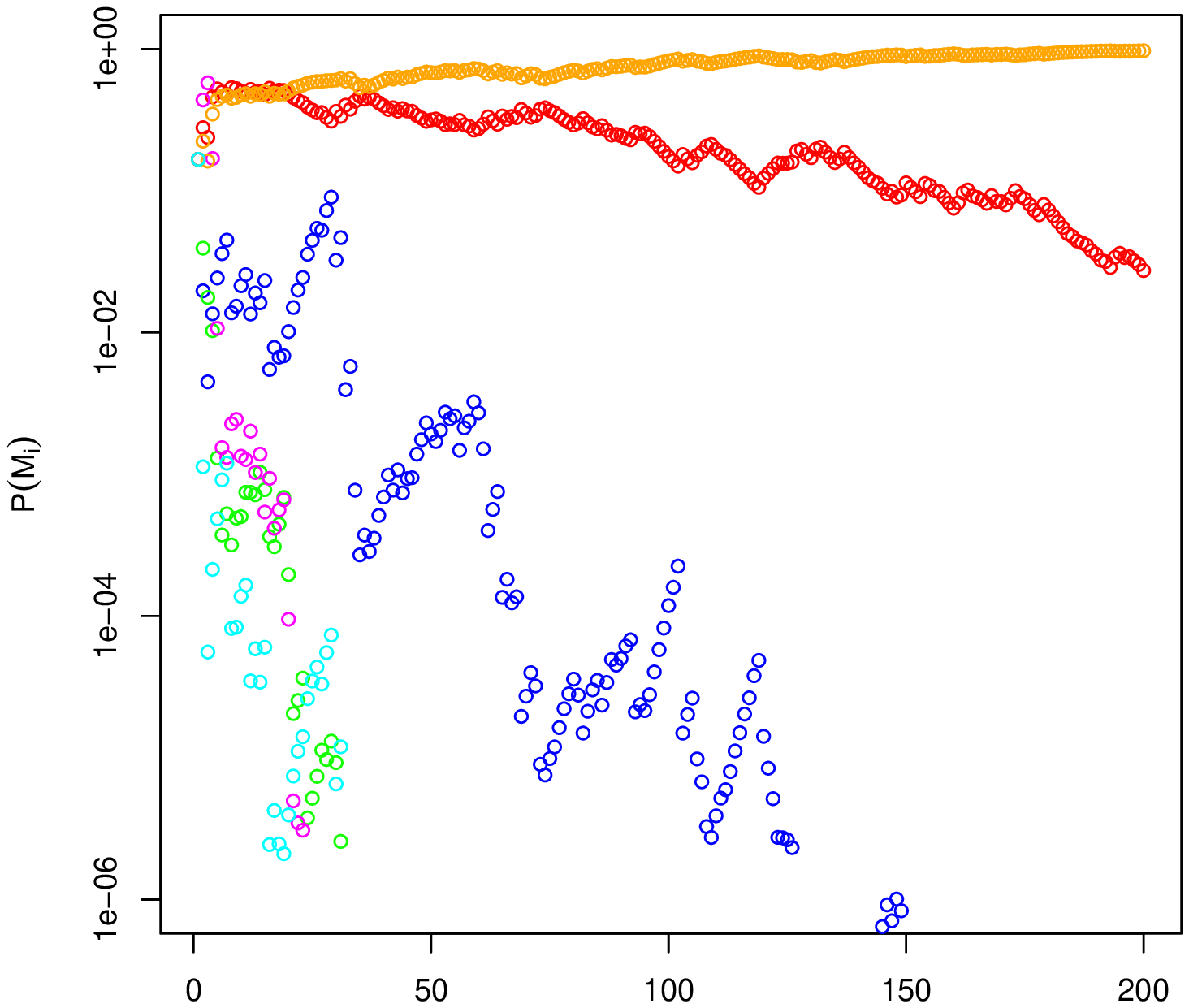,clip=,width=\linewidth} \\
\epsfig{file=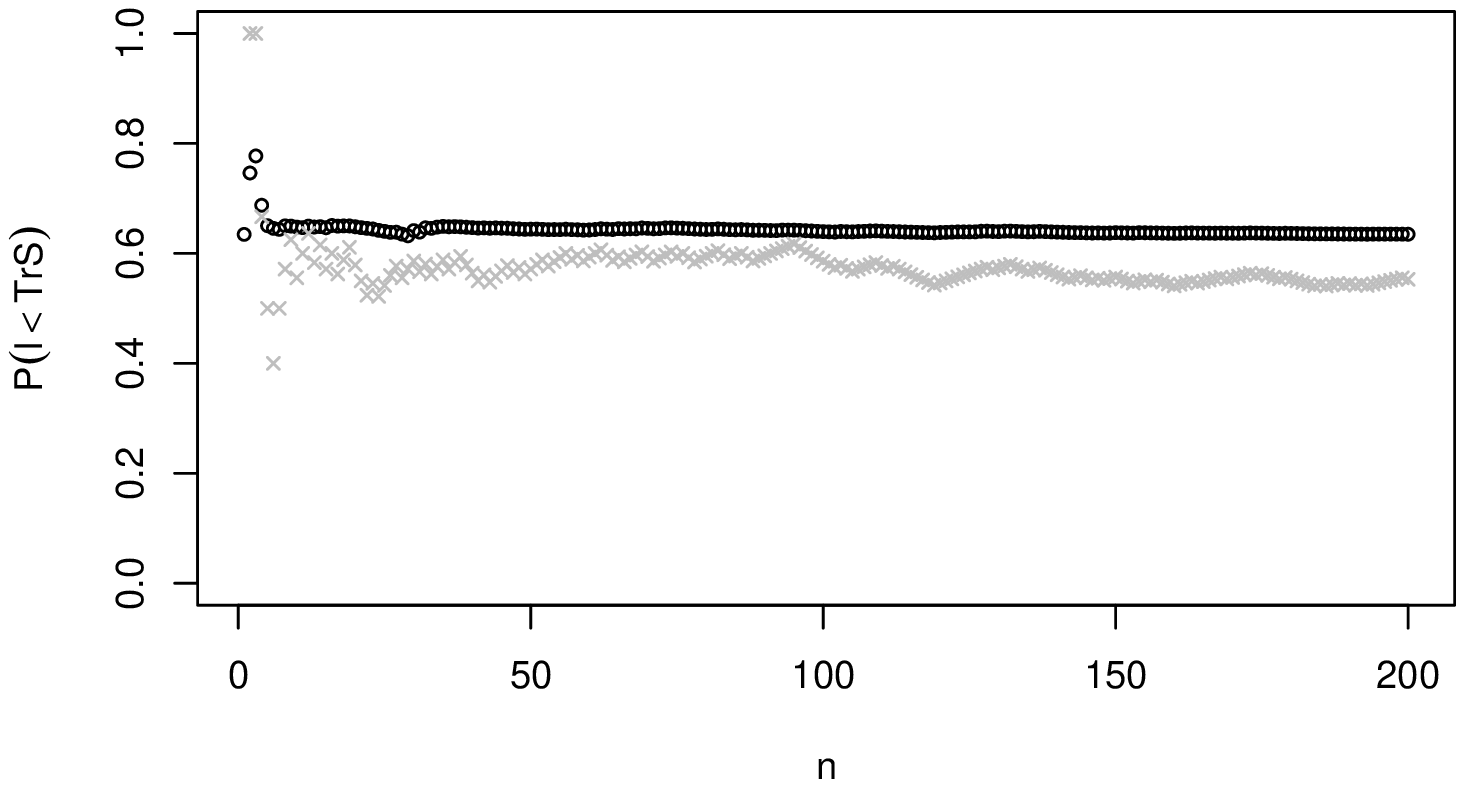,clip=,width=\linewidth} 
\end{center}
\caption{\small \sf Probability of `causes' (above) and probability
of an `effect' (below), the last compared with the relative 
frequency of occurrence (see text). [True model: ${\cal M}_5$]}
\label{fig:bayes_bertrand_M5}
\end{figure}

\begin{figure}
\begin{center}
\epsfig{file=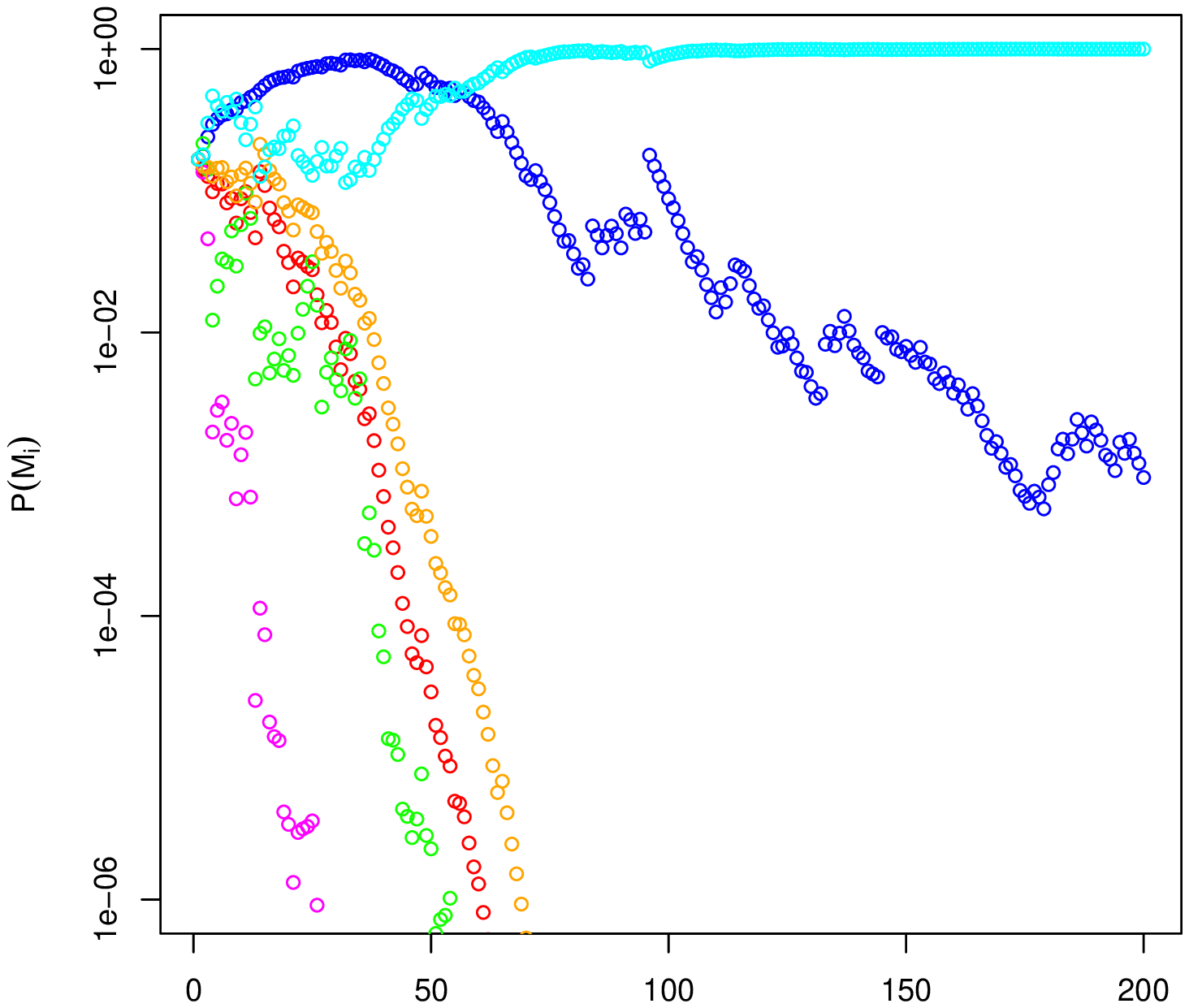,clip=,width=\linewidth} \\
\epsfig{file=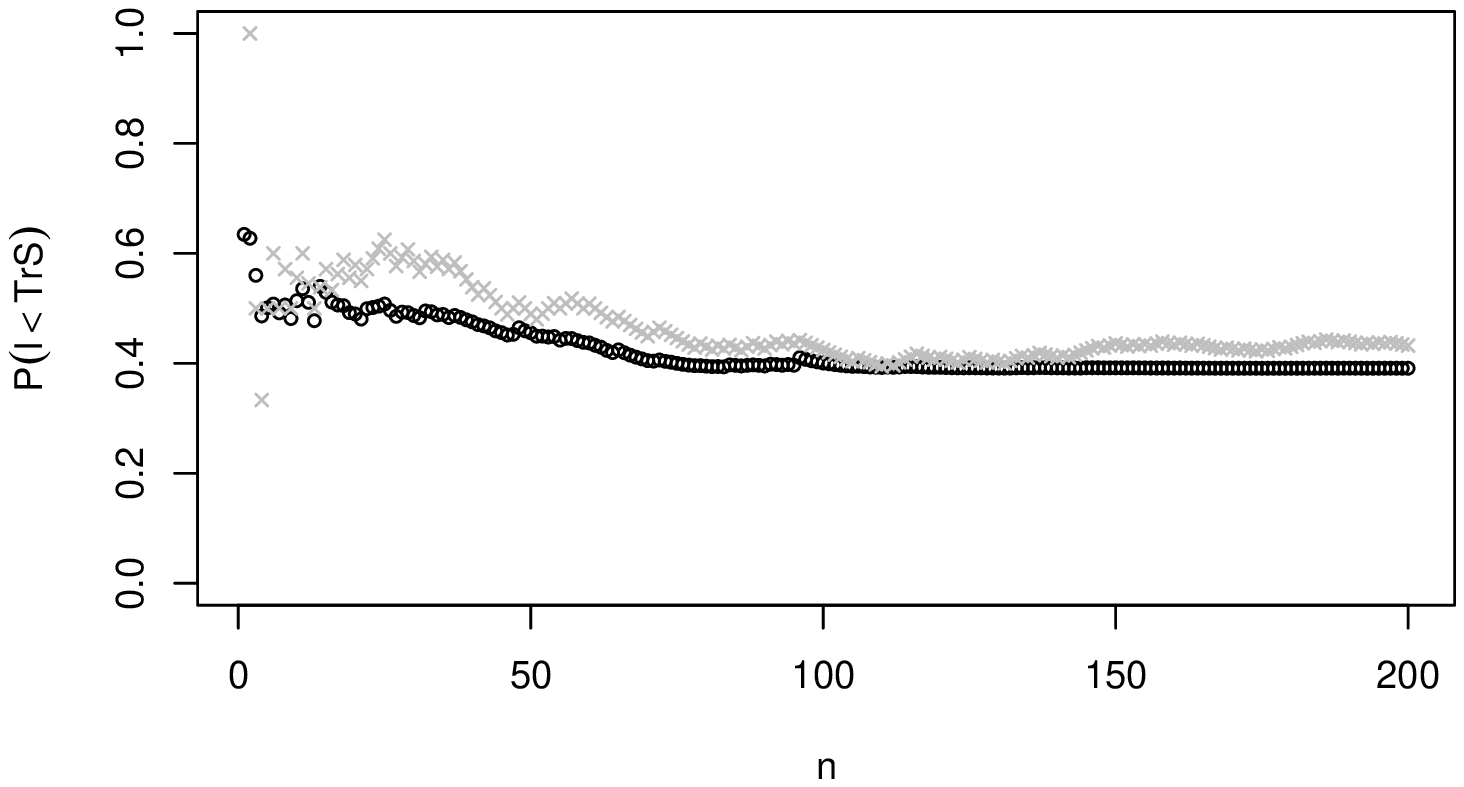,clip=,width=\linewidth} 
\end{center}
\caption{\small \sf Probability of `causes' (above) and probability
of an `effect' (below), the last compared with the relative 
frequency of occurrence (see text). [True model: ${\cal M}_6$]}
\label{fig:bayes_bertrand_M6}
\end{figure}

At this point we are ready to make the simulations. 
This is the the rest of the code to make the extractions
and to update accordingly the various probabilities:
\begin{verbatim}
rmod <- sample(1:6)[1]

odds <- matrix(rep(1, 6), c(1,6))
probs <- matrix(rep(1/6, 6), c(1,6))
bf <- matrix(rep(NA, 6), c(1,6))
pLEsqrt3.Mi <- c(2/3, 1/2, 3/4, sqrt(3)/2, (3-sqrt(3))/2, 2/3-sqrt(3)/(2*pi) )
pLEsqrt3 <- sum(probs[1,]*pLEsqrt3.Mi)
fLEsqrt3 <- NA

n <- 200
decimal.digits <- 2
last.digit <- 10^-decimal.digits
l <- NA
for (i in 2:n) {
  l[i] <- round(rlchords(1, rmod), decimal.digits)
  #print(ref.mod)
  bf <- rbind(bf, chordsModelsBF(l[i], last.digit, ref.mod) )
  odds <- rbind(odds, odds[i-1,]*bf[i,])
  ref.mod <- which(odds[i,]==max(odds[i,]))[1]
  odds[i,] = odds[i,] / odds[i,ref.mod]
  probs <- rbind(probs, odds[i, ]/sum(odds[i,]))
  pLEsqrt3[i] <- sum(probs[i,]*pLEsqrt3.Mi)
  fLEsqrt3[i] <- length(l[l<=sqrt(3)])/i
}
\end{verbatim}
Then this is how to plot the histories of the probabilities
of the models:
\begin{verbatim}
plot(probs[,1], ylim=c(10^-6,1), xlab='', ylab=expression(P(M[i])),
     log='y', col='red', cex=0.9)
points(probs[,2], col='blue', cex=0.9)
points(probs[,3], col='green', cex=0.9)
points(probs[,4], col='magenta', cex=0.9)
points(probs[,5], col='orange', cex=0.9)
points(probs[,6], col='cyan', cex=0.9)
\end{verbatim}
Another interesting plot concerns the probability that the 
{\em next} chord will have $\lambda\le\sqrt{3}$, which was
stored, step by step, in the vector {\tt pLEsqrt3}, compared
to the relative frequency of occurrence of such a condition
in the previous steps:
\begin{verbatim}
plot(pLEsqrt3, ylim=c(0,1), col='black', 
     xlab='n', ylab=expression(P(l<TrS)), cex=0.7)
points(fLEsqrt3, ylim=c(0,1), pch=4, cex=0.7, col='gray')
\end{verbatim}
The results are
shown in Fig.~\ref{fig:bayes_bertrand_M1}-\ref{fig:bayes_bertrand_M6},
with the models identified by the same color code
used in the previous figures.
In particular we see that there are model easy to
be identified and other more difficult, as it can be understood
from the plots in  Fig.~\ref{fig:bertrand_All_F}. As far as the comparison
between the probability of the next effect calculated from
probability theory (black circles) 
and that evaluated from the past frequency (gray), we see
that the former is more stable and rapidly converging to the 
correct one, corresponding to the model about which we become
practically sure.  

\end{document}